\documentclass[11pt]{article}
\usepackage{amsfonts,amssymb,amsmath,amsthm,dsfont,graphicx,hyperref,mathrsfs,euscript,enumitem,bm,soul}
\usepackage[title, titletoc]{appendix}
\usepackage[onehalfspacing]{setspace}
\usepackage{caption,subcaption,comment}
\hypersetup{pdfborder = {0 0 0},colorlinks=true,linkcolor=blue,urlcolor=blue,citecolor=blue}
\usepackage{natbib}

\usepackage[top=1in,bottom=1in,left=1in,right=1in]{geometry}

\newtheorem{thm}{Theorem}

\newtheorem{assumption}{Assumption}

\theoremstyle{definition}
\newtheorem{remark_tmp}{Remark}

\allowdisplaybreaks[1]

	\DeclareMathOperator*{\argmin}{arg\,min}
	\DeclareMathOperator*{\argmax}{arg\,max}
		
\renewcommand{\P}{\mathbb{P}}
\newcommand{\E}{\mathbb{E}}
\newcommand{\V}{\mathbb{V}}
\newcommand{\Cov}{\mathbb{C}\mathrm{ov}}
\newcommand{\I}{\mathds{1}}
\newcommand{\cval}{\mathfrak{c}}

\newcommand{\bb}{\mathbf{b}}
\newcommand{\bD}{\mathbf{D}}
\newcommand{\bN}{\mathbf{N}}
\newcommand{\bQ}{\mathbf{Q}}
\newcommand{\bw}{\mathbf{w}}

\newcommand{\bbeta}{\boldsymbol{\beta}}

\newcommand{\bgamma}{\boldsymbol{\gamma}}

\newcommand{\bxi}{\boldsymbol{\xi}}

\newcommand{\bSigma}{\boldsymbol{\Sigma}}

\newcommand{\bXi}{\boldsymbol{\Xi}}

\usepackage{marginnote}

\begin{document}

\title{On Binscatter\thanks{We especially thank Jonah Rockoff and Ryan Santos for detailed, invaluable feedback on this project. We also thank two Coeditors, four anonymous referees, Raj Chetty, Michael Droste, John Friedman, Andreas Fuster, Paul Goldsmith-Pinkham, Andrew Haughwout, Ben Hyman, Randall Lewis, David Lucca, Stephan Luck, Xinwei Ma, Ricardo Masini, Emily Oster, Filippo Palomba, Jesse Rothstein, Jesse Shapiro, Boris Shigida, Rocio Titiunik, Seth Zimmerman, Eric Zwick, and seminar participants at various seminars, workshops and conferences for helpful comments and discussions. Oliver Kim, Ignacio Lopez Gaffney, Shahzaib Safi, and Charles Smith provided excellent research assistance. Cattaneo gratefully acknowledges financial support from the National Science Foundation through grants SES-1947805, SES-2019432, and SES-2241575. Feng gratefully acknowledges financial support from the National Natural Science Foundation of China (NSFC) through grants 72203122, 72133002, and 72250064. The views expressed in this paper are those of the authors and do not necessarily reflect the position of the Federal Reserve Bank of New York or the Federal Reserve System. Companion general-purpose software and complete replication files are available at \url{https://nppackages.github.io/binsreg/}.}
\bigskip }
\author{Matias D. Cattaneo\thanks{Department of Operations Research and Financial Engineering, Princeton University.} \and
	    Richard K. Crump\thanks{Macrofinance Studies, Federal Reserve Bank of New York.} \and
	    Max H. Farrell\thanks{Department of Economics, UC Santa Barbara.} \and 
	    Yingjie Feng\thanks{School of Economics and Management, Tsinghua University.}}
\maketitle

\begin{abstract}
    Binscatter is a popular method for visualizing bivariate relationships and conducting informal specification testing. We study the properties of this method formally and develop enhanced visualization and econometric binscatter tools. These include estimating conditional means with optimal binning and quantifying uncertainty. We also highlight a methodological problem related to covariate adjustment that can yield incorrect conclusions. We revisit two applications using our methodology and find substantially different results relative to those obtained using prior informal binscatter methods. General purpose software in \texttt{Python}, \texttt{R}, and \texttt{Stata} is provided. Our technical work is of independent interest for the nonparametric partition-based estimation literature.
\end{abstract}


\thispagestyle{empty}
\clearpage

\doublespacing
\setcounter{page}{1}
\pagestyle{plain}

\pagestyle{plain}

\section*{Introduction}

The classical scatter plot is a fundamental visualization tool in data analysis. Given a sample of bivariate data, a scatter plot displays all $n$ data points at their coordinates $(x_i, y_i)$, $i =1, \ldots, n$. By plotting every data point, one obtains a visualization of the joint distribution of $y$ and $x$. When used prior to regression analyses, a scatter plot allows researchers to assess the functional form of the regression function, the variability around this conditional mean, and recognize unusual observations, bunching, or other anomalies or irregularities.

Classical scatter plots, however, have several limitations and have fallen out of favor. For example, with the advent of larger data sets, the cloud of points becomes increasingly dense, rendering scatter plots uninformative. Even for moderately sized but noisy samples it can be difficult to assess the shape and other properties of the conditional mean function. Further, with increasing attention paid to privacy concerns, plotting the raw data may be disallowed completely. Another important limitation of the classical scatter plot is that it does not naturally allow for a visualization of the relationship of $y$ and $x$ while controlling for other covariates, which is a standard goal in social sciences.

Binned scatter plots, or binscatters, have become a popular and convenient alternative tool in applied microeconomics for visualizing bivariate relations \citep[see][and references therein, for an overview of the literature]{Starr-Goldfarb_2020_SMJ}. A binscatter is made by partitioning the support of $x$ into a modest number of bins and displaying a single point per bin, showing the average outcome for observations within that bin. This makes for a simpler, cleaner plot than a classical scatter plot, but it does not present the same information. While a scatter plot allows one to display the entirety of the data, a binscatter shows only an estimate of the conditional mean function. A binned scatter plot is therefore not an exact substitute for the classical scatter plot, but it can be used to judge functional form, provide a qualitative assessment of features such as monotonicity or concavity, and guide later regression analyses. Handling additional covariates correctly is a particularly subtle issue.

In this paper we introduce a suite of formal and visual tools based on binned scatter plots to restore, and in some dimensions surpass, the visualization benefits of the classical scatter plot. We deliver a fully featured toolkit for applications, including estimation of conditional mean functions, visualization of variance and precise quantification of uncertainty, and formal tests of substantive hypotheses such as linearity or monotonicity. Our toolkit allows for characterizing key features of the data without struggling to parse the dense cloud of large data sets or sharing identifying information of individual data points. As a foundation for our results we deliver an extensive theoretical analysis of binscatter and related partition-based methods. We also highlight a prevalent methodological problem related to covariate adjustment present in prior binscatter implementations, which can lead to incorrect estimates and visualizations of the conditional mean, in both shape and support. We demonstrate how incorrect covariate adjustment in binscatter applications can mislead practitioners when assessing linearity or other hypothesized parametric or shape specifications of the unknown conditional mean. 

The concept of a binned scatter plot is simple and intuitive: divide the data into $J < n$ bins according to the covariate $x$, often using empirical ventiles, and then calculate the average outcomes among observations with covariate values lying in each bin. The final plot shows the $J$ points $(\bar{x}_j, \bar{y}_j)$, the sample averages for units with $x_i$ falling within the $j$th bin ($j=1,2,\dots,J$). Further, by plotting only averages, discrete-valued outcomes are easily accommodated. The result is a figure which shares the conceptual appeal, visual simplicity, and \emph{some} of the utility of a classical scatter plot. 

In a binned scatter plot the $J$ points are then used to visually assess the bivariate relation between $y$ and $x$. Because each of the $J$ points in a binned scatter plot shows a conditional average, i.e., the average outcome given that $x_i$ falls into a specific bin, using the plot to examine the conditional mean is intuitive. The primary use is assessing the shape of this mean function: whether the relationship is linear, monotonic, convex, and so forth. In applications, a roughly linear binscatter often precedes a linear regression analysis. Indeed, we provide formal results which justify such an approach in a principled, valid way. 

\begin{figure}[!ht]
\caption{\small{\bf Illustration of Binned Scatter Plots.} This figure illustrates the construction of a binned scatter plot using data from \cite{Akcigitetal2021_QJE}. The dependent variable is the log number of patents per state per year, and the independent variable is the log of the marginal net of tax rate for 90th percentile earners. No control variables are included.}
\label{fig:intro}
\begin{subfigure}[t]{0.5\columnwidth}
\centering	
\caption{\it Raw Scatter Plot}
\includegraphics[trim={0cm 1cm 0cm 1cm}, clip, scale=.275]{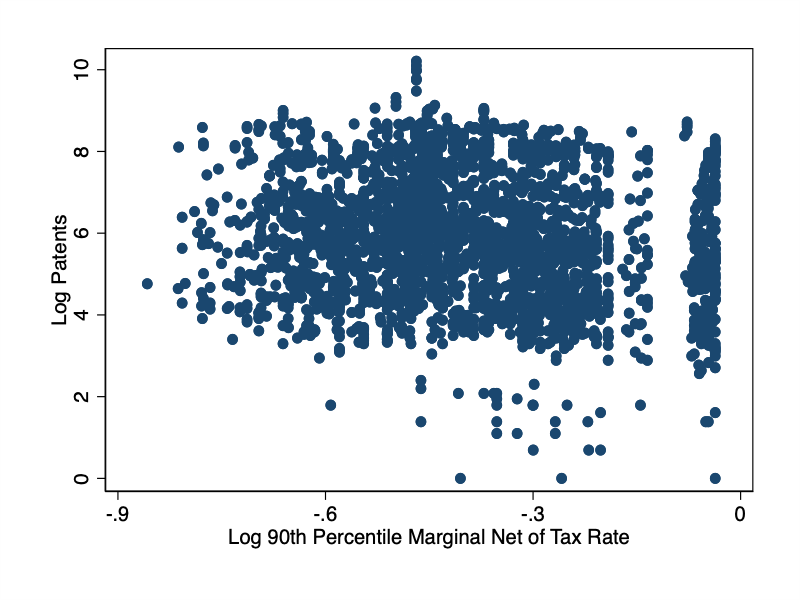}
\end{subfigure}\hfill
\begin{subfigure}[t]{0.5\columnwidth}
\centering	
\caption{\it Bin Cutoffs and Binscatter}
\includegraphics[trim={0cm 1cm 0cm 1cm}, clip, scale=.275]{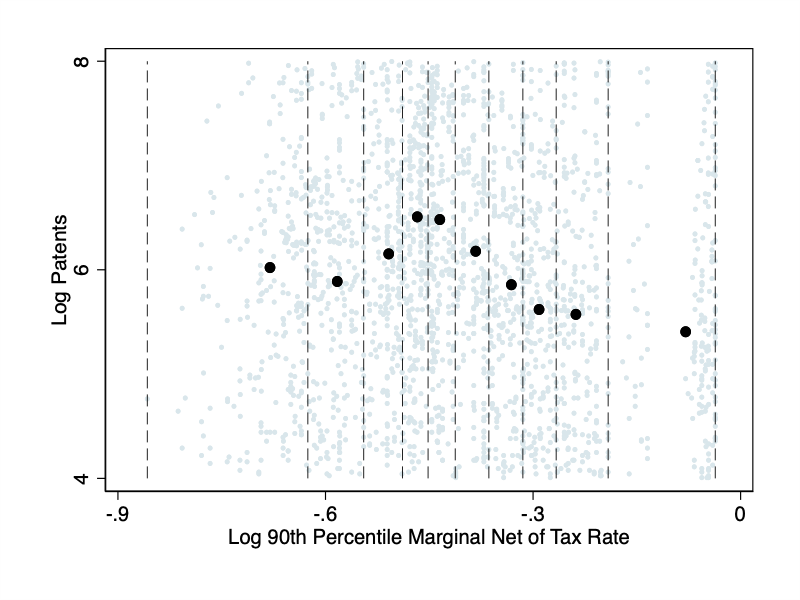}
\end{subfigure}\\[.5 ex]
\begin{subfigure}[t]{0.5\columnwidth}
\centering	
\caption{\it Binscatter Plot}
\includegraphics[trim={0cm 1cm 0cm 1cm}, clip, scale=.275]{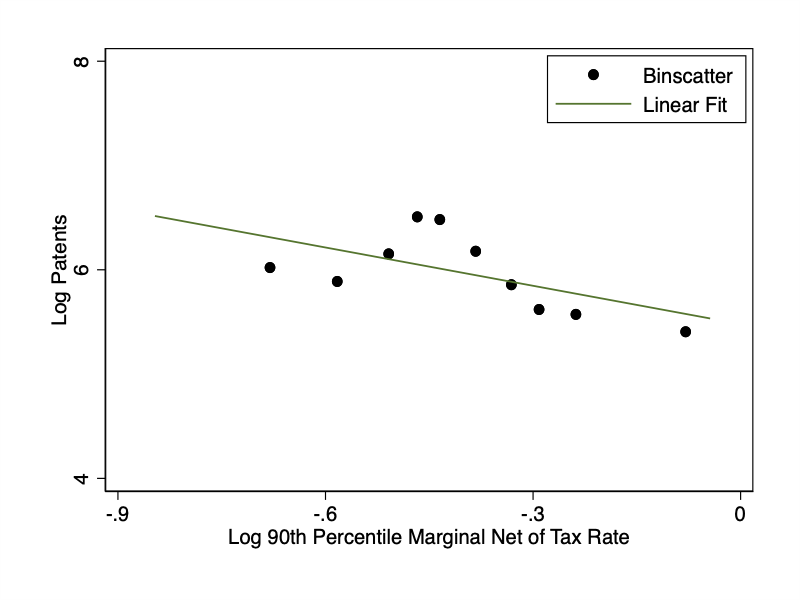}
\end{subfigure}\hfill
\begin{subfigure}[t]{0.5\columnwidth}
\centering	
\caption{\it Conditional Mean Point Est.}
\includegraphics[trim={0cm 1cm 0cm 1cm}, clip, scale=.275]{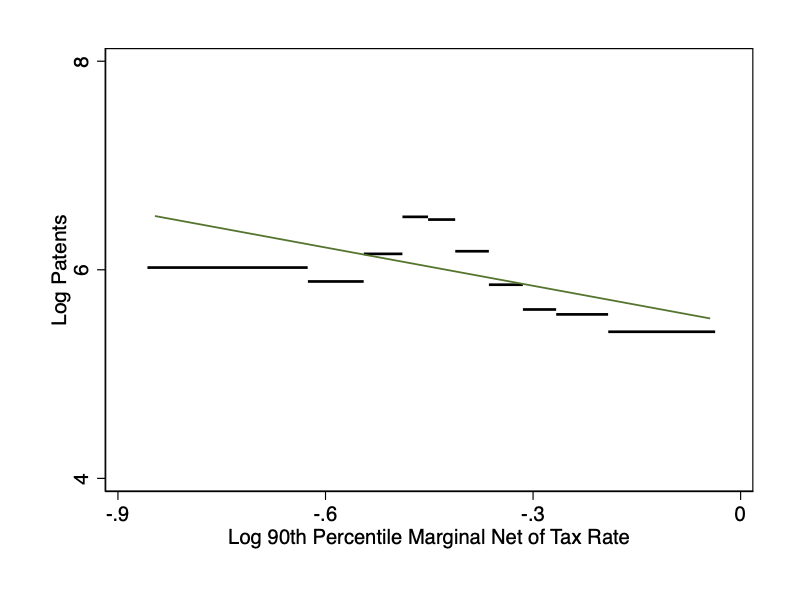}
\end{subfigure}\\[.5 ex]
\end{figure}

Figure \ref{fig:intro} shows an example of this construction using the data from \citet[AGNS hereafter]{Akcigitetal2021_QJE}. This recent paper will serve as a running example throughout the text to illustrate our main ideas and results using real data. AGNS study the effect of corporate and personal taxes on innovation in the United States over the twentieth century. Figure \ref{fig:intro}(a) presents a raw scatter plot of log patents and the variable of interest, transformed marginal tax rates.\footnote{The authors use the logarithm of one minus the 90th percentile marginal tax rate or, equivalently, the logarithm of the 90th percentile marginal net of tax rate. This transformed variable implies that a positive relation between $y$ and $x$ implies that higher marginal tax rates are associated with lower quantity of innovation.} Despite a sample size of about 3,000 observations it is difficult to draw any inferences about the data from the scatter plot. (Section \ref{sec:applications} studies a much larger data set.) Figure \ref{fig:intro}(b) shows a binned scatter plot being constructed, with the raw data in the background, and \ref{fig:intro}(c) isolates the binscatter, and overlays a linear regression fit. Graphs like \ref{fig:intro}(c) are often found in empirical papers. An important note is that although the binned scatter plot invites the viewer to ``connect the dots'' smoothly, the actual estimator is piecewise constant, as shown explicitly in Figure \ref{fig:intro}(d). Though graphically distinct, this is formally identical to the dots in Figure \ref{fig:intro}(c). Figure \ref{fig:intro} also highlights the fact that although the averaging is useful for evaluating the conditional mean, it masks other features of the conditional distribution which may be important to the subsequent analysis. This presents a clear limitation to the usefulness of binscatter methods for visualization and analysis. Note how much information is lost in moving from Figure \ref{fig:intro}(b) to \ref{fig:intro}(c). Our later inference tools help to remedy this limitation by augmenting the binned scatter plot with formal uncertainty quantification.

\begin{figure}[!ht]
\caption{\small{\bf Covariate Adjustment.} This figure illustrates the role of covariate adjustment in the construction of binned scatter plots using data from \cite{Akcigitetal2021_QJE}. The dependent variable is the log number of patents per state per year, and the independent variable is the log of the marginal net of tax rate for 90th percentile earners. The additional control variables are the lagged corporate tax rate, lagged population density, personal income per capita, and R\&D tax credits, along with state and year fixed effects. The left plot is inspired by Figure I(A) in \cite{Akcigitetal2021_QJE} using 50, rather than 100, bins (when the corrected covariate adjustment is used there is insufficient variation in the variable of interest to feasibly accommodate the larger choice of bins). The middle plot is a correctly scaled version of the left plot. The right plot presents the binned scatter plot using the correct covariate adjustment approach. Binscatter estimates are based on weights of each state's 1940 population count.}
\captionsetup{justification=centering}
\label{fig:intro-covs}
\begin{subfigure}[t]{0.333\columnwidth}
\centering	
\caption{\it Standard Binscatter\\(Incorrect Scale)}
\includegraphics[trim={0cm 1cm 0cm 1cm}, clip, scale=.2]{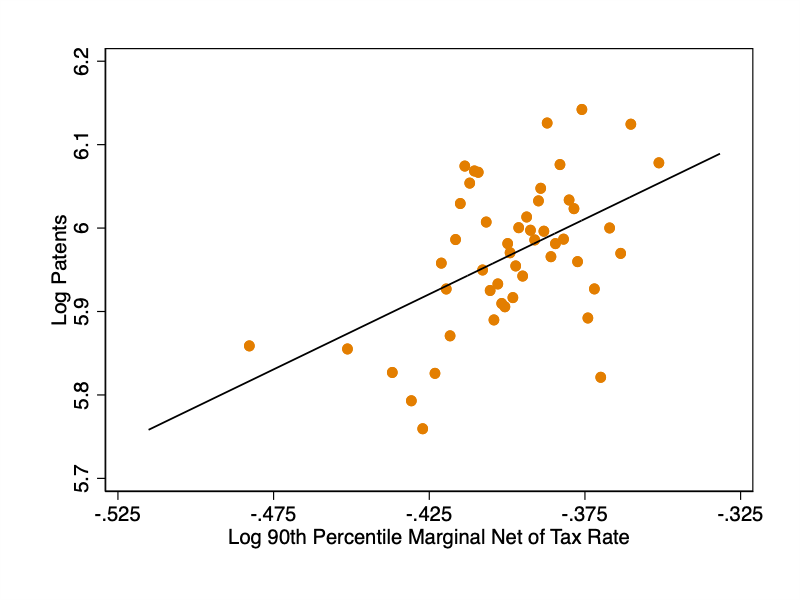}
\end{subfigure}
\begin{subfigure}[t]{0.333\columnwidth}
\centering	
\caption{\it Standard Binscatter\\(Correct Scale)}
\includegraphics[trim={0cm 1cm 0cm 1cm}, clip, scale=.2]{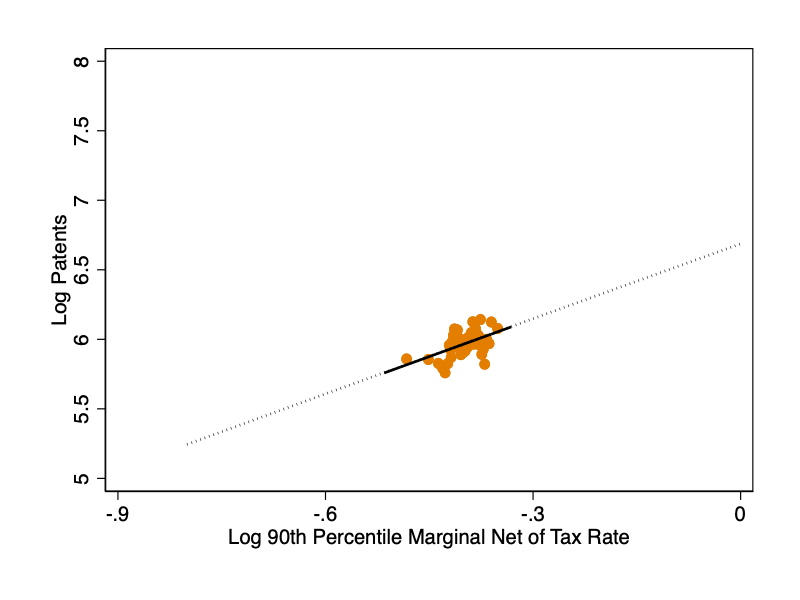}
\end{subfigure}
\begin{subfigure}[t]{0.333\columnwidth}
\centering	
\caption{\it Corrected Binscatter\\(Correct Scale)}
\includegraphics[trim={0cm 1cm 0cm 1cm}, clip, scale=.2]{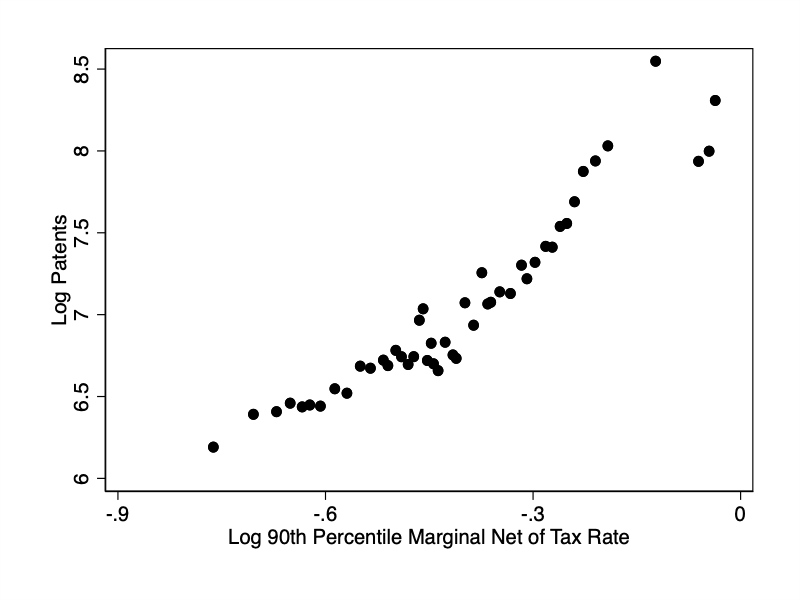}
\end{subfigure}
\end{figure}

It is common practice to use additional control variables and fixed effects when constructing a binscatter. The standard plots, like Figure \ref{fig:intro}(c), will often be made after ``controlling'' for a set of covariates. This turns out to be a subtle issue, as the controls affect the visualization as well as the degree of uncertainty. Even the common practice of adding a regression line to a binned scatter plot is not straightforward to do correctly. We highlight important methodological and theoretical problems with the commonly used practice of first ``residualizing out'' additional covariates before constructing a binscatter. This is only formally justified when the true function is linear. Instead we show that the shape and support of the conditional mean can be incorrect when employing common practice. Figure \ref{fig:intro-covs} shows the practical importance of this issue by revisiting AGNS. Their benchmark specifications study the relation between log patents and marginal tax rates utilizing a rich set of control variables including fixed effects (see Table II and Figure I in AGNS). In their macro-level approach, the authors show that higher taxes negatively affect the quantity of innovation. Figure \ref{fig:intro-covs}(a) is inspired by Figure I(A) in the original paper. Comparing the $x$ axis to the raw scatter plot of Figure \ref{fig:intro}(a) we see the distortion of the support. Figure \ref{fig:intro-covs}(b) is the correctly scaled plot in the original paper; it is essentially uninformative about the shape of the mean. Finally, Figure \ref{fig:intro-covs}(c) shows the corresponding results using our corrected covariate-adjustment approach.

We provide an array of results and tools for binned scatter plots aimed at improving their empirical application. We improve on the estimation of conditional mean functions and provide tools for quantifying uncertainty. To facilitate our analysis, we first demonstrate that a binscatter is a nonparametric estimator, and we provide a modeling framework that enables formal analysis, allowing us to deliver new, more powerful methods and to resolve conceptual and implementation issues. We clarify precisely the parameters of interest in applications, both for visualization and formal inference. Our framework centers around a partially linear model, wherein we show how to control for additional variables in a principled and interpretable way, and discuss why prior implementations are not recommended. 

Within our framework, we also discuss the choice of the number of bins, $J$. We elucidate how the choice of $J$ relates to the interpretation of the binscatter plot and its role in nonparametric estimation. When we use a binscatter to recover the conditional mean function we must assume $J$ grows with the sample size as is standard in semi- and nonparametric inference. In this case, we provide data-driven methods for an optimal choice of $J$. We can also consider a fixed, user-chosen $J$, which may yield a simple and appealing visualization of a coarsened version of the conditional mean. For example, selecting $J=10$ has a natural interpretation of comparing average outcomes in different deciles of the distribution of $x_i$. Our results also apply in this case.

We then turn to uncertainty quantification. For visualization, we provide confidence bands that capture the uncertainty in estimating the conditional mean or other functional parameters of interest. A confidence band is a region that contains the entire function with some pre-set probability, just as a confidence interval covers a single value, and is thus the proper tool for assessing uncertainty about the regression function. Confidence bands can be used to visually assess the plausibility of parametric functional forms, such as linearity. Confidence bands partly restore the uncertainty visualization capability of the classical scatter plot by capturing how certain we are about the functional form of the conditional mean. Further, our confidence bands are explicitly functions of the conditional heteroskedasticity in the underlying data. Delivering a valid confidence band requires novel theoretical results, which represent the main technical contribution of our work.

\begin{figure}[!ht]
\caption{\small{\bf Confidence Bands.} This figure illustrates uniform confidence bands using data from \cite{Akcigitetal2021_QJE}. The dependent variable, independent variable, and controls are the same as in Figure \ref{fig:intro-covs}. Binscatter estimates are based on weights of each state's 1940 population count using the optimal number of bins as described in Section \ref{sec:IMSE}. Shaded regions denote 95\% nominal confidence bands using a cluster-robust variance estimator with two-way clustering by year and state $\times$ five-year period.}
\label{fig:intro-band}
\begin{subfigure}[t]{0.5\columnwidth}
\centering	
\caption{\it Binscatter and Conf. Band}
\includegraphics[trim={0cm 1cm 0cm 1cm}, clip, scale=.275]{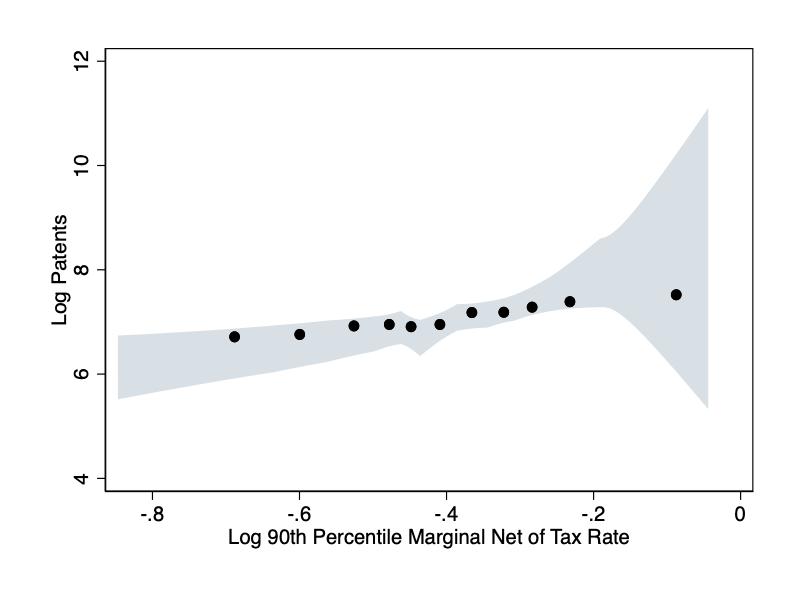}
\end{subfigure}\hfill
\begin{subfigure}[t]{0.5\columnwidth}
\centering	
\caption{\it Conf. Band}
\includegraphics[trim={0cm 1cm 0cm 1cm}, clip, scale=.275]{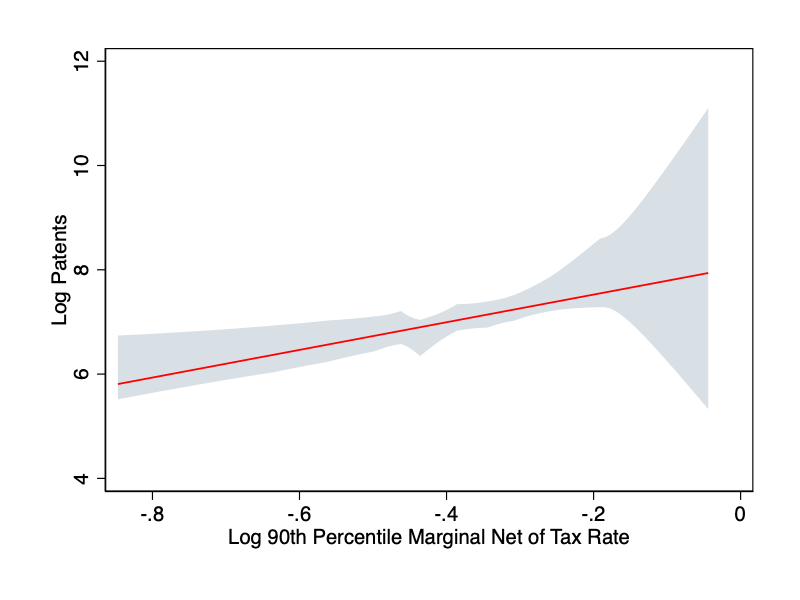}
\end{subfigure}\hfill
\end{figure}

Figure \ref{fig:intro-band}(a) shows a confidence band for AGNS, relying on our data-driven choice of $J$ and robust bias correction methods to ensure the inference is valid. The binscatter itself is quite linear in appearance, in contrast with the original Figure \ref{fig:intro-covs}(a). Moreover, Figure \ref{fig:intro-band}(b) shows that a linear function can be drawn within the confidence band (red line), so we can validly conclude that linearity is consistent with these data. In this case, our novel methods bolster the case for the paper's original linear regression analysis. (In Section \ref{sec:applications} we show an application where linearity is not supported, but our methods nonetheless reinforce an empirical conclusion and extend it in economically interesting ways.)

The paper proceeds as follows. We next briefly review the related literature and summarize our technical contributions. Section \ref{sec:setup} formalizes binned scatter plots as a nonparametric estimator, including clarifying the parameter of interest and the correct method for adding control variables. Section \ref{sec:IMSE} discusses the choice of the number of bins $J$. Section \ref{sec:Quantifying Uncertainty} studies uncertainty quantification for both visualization and testing. Throughout, we use the application of AGNS for illustration. In addition, Section \ref{sec:applications} contains a second application, where we revisit \cite{Moretti2021_AER}. Both applications highlight the usefulness of our results in empirical settings. Section \ref{sec:theory} presents our main theoretical results and further discussion of the technical contributions of the paper. Finally, Section \ref{sec:conclusion} concludes. An online Appendix provides additional discussion and detail omitted from the main text, proofs of all our results, and a thorough account of our technical contributions. All of our methodological results are available in fully-featured \texttt{Stata}, \texttt{R}, and \texttt{Python} packages (see \citet*{Cattaneo-Crump-Farrell-Feng_2023_Stata} and \url{https://nppackages.github.io/binsreg/}).

\subsection*{Related Literature}

Our paper fits into several literatures. Our work speaks most directly to the applied literature using binscatter methods, which is too large to enumerate here. \citet{Starr-Goldfarb_2020_SMJ} gives an overview and many references. Beyond binscatter itself, binning has a long history in both visualization and formal estimation. The most familiar case is the classical histogram. Applying binning to regression problems dates back at least to the regressogram of \cite{Tukey_1961_Berkeley}. The core idea has been applied in such diverse areas as climate studies, for nonlinearity detection \citep{schlenker2009nonlinear}; program evaluation, called subclassification \citep{Stuart_2010_StatSci}; empirical finance, called portfolio sorting \citep{Cattaneo-Crump-Farrell-Schaumburg_2020_RESTAT}; and applied microeconomics, for visualization in bunching \citep{Kleven_2016_ARE} and regression discontinuity designs \citep{Cattaneo-Titiunik_2022_ARE}.

In recent years, there has been related research looking at the importance and limitations of graphical analysis in different applied areas. For example, \citet{Korting-Lieberman-Matsudaira-Pei-Shen_2023_wp}
conducts a field experiment to investigate the role of visual inference and graphical representation in regression discontinuity designs via RD plots \citep{Calonico-Cattaneo-Titiunik_2015_JASA}. They conclude that unprincipled graphical methods could lead to misleading or incorrect empirical conclusions. Similar concerns regarding graphical analysis are raised by \citet{Freyaldenhoven-Hansen-Perez-Shapiro_2021_WorldCrongress} in the context of event study designs, where they proposed principled visualization methods. Graphical and visualization methods are also being actively discussed in the machine learning community \citep[see][and references therein, for an overview of the literature]{wang2021survey}, where the importance of focusing on principled methods with well-understood properties for both in-sample and out-of-sample learning has been highlighted. Our paper contributes to this literature by offering principled approaches for visualization and inference employing binscatter methodology. Furthermore, well-executed visualization techniques can help with issues of statistical nonsignificance in empirical economics employing big data \citep{abadie2020statistical}.

Finally, our technical work contributes to the literature on nonparametric regression, particularly for uniform distributional approximations. Binning as a nonparametric procedure has been studied in the past, but existing theory is insufficient for our purposes for two main reasons. First, the extant literature cannot generally accommodate data-driven bin breakpoints, such as splitting the support by empirical quantiles. Such a choice of breakpoints generates random basis functions and so are not nested in previously obtained results on nonparametric series estimators. Second, where results are available, they imply overly stringent conditions on smoothing parameters ruling out simple averaging within each bin (which amounts to local constant fitting) and are thus not applicable to binscatter. Circumventing these limitations with new theoretical results is crucial to directly study the empirical practice of binned scatter plots.

\citet{Gyorfi-etal_2002_book} gives a textbook introduction to binning in nonparametric regression, where the procedure is known as partitioning regression. Recent work on partitioning, always assuming known breakpoints, includes convergence rates and pointwise distributional approximations \citep{Ling-Hu_2008_JNS,Cattaneo-Farrell_2013_JoE}, and uniform distributional approximations and robust bias correction methods \citep{Cattaneo-Farrell-Feng_2020_AoS}. Partition regression is intimately linked to spline and wavelet methods, and the general results in our online Appendix treat these estimators as well, improving over earlier work by \cite{Shen-Wolfe-Zhou_1998_AoS}, \cite{Huang_2003_AoS}, \citet{Belloni-Chernozhukov-Chetverikov-Kato_2015_JoE}, \citet{Cattaneo-Farrell-Feng_2020_AoS}, and references therein. We discuss these technical contributions in more detail in Section \ref{sec:theory} and in the online Appendix.

\section{Canonical Binscatter and Covariate Adjustments}\label{sec:setup}

The observed data is a random sample $(y_i,x_i,\bw_i')$, $i=1,2,\dots,n$, where $y_i$ is the outcome, $x_i$ is the main regressor of interest, and $\bw_i$ is a $d$-vector of other covariates (e.g., pre-intervention characteristics or fixed effects). A binscatter has three key elements: the binning of the support of the covariate $x_i$, the estimation within each bin, and the way in which the controls $\bw_i$ are handled. We discuss each of these in turn.

The partition of the support requires a choice of the number of bins, $J$, as well as how to divide the space. The choice of $J$ is the tuning parameter of this estimator, and in current practice it is often set independently of the data and equal to $J=10$ or $J=20$. We discuss the choice of $J$ in Section \ref{sec:IMSE}, but for now we take $J < n$ as given. For the spacing of the $J$ bins, we follow standard empirical practice and use the marginal empirical quantiles of $x_i$. Let $x_{(i)}$ denote the $i$-th order statistic of the sample $(x_1,x_2,\dots,x_n)$ and $\lfloor \cdot \rfloor$ denote the floor operator. Then, the partitioning scheme is defined as $\widehat{\Delta} = \{\widehat{\mathcal{B}}_1, \widehat{\mathcal{B}}_2, \dots, \widehat{\mathcal{B}}_J\}$, where
\[\widehat{\mathcal{B}}_j = \begin{cases}
                            \Big[x_{(1)}, x_{(\lfloor n/J \rfloor)}\Big)                            & \qquad \text{if } j=1\\
                            \Big[x_{(\lfloor n(j-1)/J \rfloor)}  , x_{(\lfloor nj/J \rfloor)}\Big)  & \qquad \text{if } j=2,3,\dots,J-1\\
                            \Big[x_{(\lfloor n(J-1)/J \rfloor)}, x_{(n)}\Big]                             & \qquad \text{if } j=J
                            \end{cases}.
\]
Each estimated bin $\widehat{\mathcal{B}}_j$ contains (roughly) the same number of observations $N_j=\sum_{i=1}^n \I_{\widehat{\mathcal{B}}_j}(x_i)$, where $\I_\mathcal{A}(x)=\I(x\in \mathcal{A})$ is the indicator function. The notation $\widehat{\Delta}$ emphasizes that the partition is estimated from the data. Handling this randomness requires novel nonparametric statistical theory (Section \ref{sec:theory}). Our theory can accommodate quite general partitioning schemes, both random and nonrandom, provided high-level conditions are satisfied. In some cases the bins may be determined by the empirical application (e.g., income ranges, or schooling levels), while in others equally spaced bins may be more appropriate. However, given the ubiquity of quantile binning in economics, we focus on $\widehat{\Delta}$ as defined above.

We begin with the bivariate case, where there are no covariates $\bw_i$. Given the partition $\widehat{\Delta}$, which encompasses a choice of the number of bins $J$, a binscatter is the collection of $J$ sample averages of the response variable: for each bin $\widehat{\mathcal{B}}_j$, we obtain $\bar{y}_j = \frac{1}{N_j} \sum_{i=1}^n \I_{\widehat{\mathcal{B}}_j}(x_i) y_i$; under our assumptions $\min_{1\leq j\leq J}N_j>0$ with probability approaching one in large samples. These sample averages are plotted as a ``scatter'' of points along with another, parametric estimate of the regression function $\upsilon_0(x_i)=\E[y_i|x_i]$, often an ordinary least squares fit using the raw data. This construction is shown in Figures \ref{fig:intro}(b) and \ref{fig:intro}(c).

For fixed $J$, under regularity conditions, a binscatter can be interpreted as estimating $\xi_0(j) = \E[y_i|x_i\in\mathcal{B}_j]$, $j=1,2,\dots,J$, where $\mathcal{B}_j$ denotes the $j$th bin based on the population quantiles of $x_i$. This interpretation of the binscatter ignores the shape of the underlying conditional expectation within each bin, as it targets a likely misspecified constant model: $\xi_0(j)$ and $\upsilon_0(x)$ can be quite different for different values $x\in\mathcal{B}_j$, except in special cases. If $x_i$ was discrete with relatively few unique values, in which case binning would be unnecessary to begin with, or if the bins $\mathcal{B}_1,\mathcal{B}_2,\dots,\mathcal{B}_J$ had a natural economic interpretation (e.g., income ranges), then the $J$-dimensional parameter $\bxi_0 = (\xi_0(1),\xi_0(2),\dots,\xi_0(J))'$ could be of interest in applications. This parameter is intrinsically parametric in nature (for fixed $J$) and, as discussed below, all the results in the paper apply to $\bxi_0$ without modification.

When $x_i$ is continuously distributed or exhibits many distinct values, and the binning structure has no useful economic interpretation in and of itself, it is more natural to view the binscatter as a nonparametric approximation of $\upsilon_0(x)=\E[y_i|x_i]$ for appropriately chosen tuning parameter $J$. This approach characterizes misspecification errors (within and across bins) as well as nonparametric uncertainty in a principled way. Thus, we formalize a binscatter as a nonparametric estimator of $\upsilon_0(x)$ by recasting it as a piecewise constant fit: $\widehat{\upsilon}(x) = \bar{y}_j$ for all $x\in\widehat{\mathcal{B}}_j$, $j=1,2,\dots,J$. This is a least-squares series regression using a zero-degree piecewise polynomial. Formally, we define
\begin{equation}
    \label{eq:binscatter-canonical}
    \widehat{\upsilon}(x)=\widehat{\bb}(x)'\widehat{\bxi}, 
    \qquad 
    \widehat{\bxi} = \argmin_{\bxi\in\mathbb{R}^J} \sum_{i=1}^{n} (y_i-\widehat{\bb}(x_i)'\bxi)^2,
\end{equation}
where $\widehat{\bb}(x) = [\I_{\widehat{\mathcal{B}}_1}(x),\I_{\widehat{\mathcal{B}}_2}(x),\cdots,\I_{\widehat{\mathcal{B}}_J}(x)]'$ is the binscatter basis given by a $J$-dimensional vector of orthogonal indicator variables, that is, the $j$-th component of $\widehat{\bb}(x)$ records whether the evaluation point $x$ belongs to the $j$-th bin in the partition $\widehat{\Delta}$. This piecewise constant fit is shown in Figure \ref{fig:intro}(d), and from an econometric point of view, is identical to the dots of Figures \ref{fig:intro}(b) and \ref{fig:intro}(c). In the online Appendix, we present results for a general polynomial fit within each bin, allowing for smoothness constraints across bins, which is useful to reduce misspecification bias.

\subsection{Residualized Binscatter}

We highlight an important methodological mistake with most applications of binscatter with covariates, including the Stata packages \texttt{binscatter} and \texttt{binscatter2}. Widespread empirical practice for covariate adjustment proceeds by first regressing out the covariates $\bw_i$ from $x_i$ and $y_i$, and then applying the bivariate binscatter approach \eqref{eq:binscatter-canonical} to the residualized variables. This approach is heuristically motivated by the usual Frisch–Waugh–Lovell theorem for ``regressing/partialling out'' other covariates in linear regression settings.

From a nonparametric perspective, under regularity conditions, the residualized binscatter is consistent for
\begin{equation}
    \label{eq:wrong plim}
    \E\big[y_i-\text{L}(y_i|\bw_i) \; \big| \; x_i - \text{L}(x_i|\bw_i)\big]
\end{equation}
with $\text{L}(a_i|\bw_i)=(1,\bw_i')(\E[(1,\bw_i')'(1,\bw_i')])^{-1}\E[(1,\bw_i')'a_i]$, and thus $\text{L}(y_i|\bw_i)$ and $\text{L}(x_i|\bw_i)$ can be interpreted as the best (in mean square) linear approximations to, respectively, $\E[y_i|\bw_i]$ and $\E[x_i|\bw_i]$ \citep[see][Chapter 2]{Wooldridge_2010_Book}. $\text{L}(y_i|\bw_i)$ and $\text{L}(x_i|\bw_i)$ are, in general, misspecified approximations of the conditional expectations $\E[y_i|\bw_i]$ and $\E[x_i|\bw_i]$. Unless the true model is linear, the probability limit in \eqref{eq:wrong plim} is difficult to interpret and does not align with standard economic reasoning. Furthermore, the shape of the function in \eqref{eq:wrong plim} and even its support may be incorrect, and therefore can lead to incorrect empirical findings. The same problems arise when interpreting residualized binscatter from a fixed-$J$ perspective or when $x_i$ is discrete. 

We therefore refer to the popular residualized binscatter method for covariate adjustment as incorrect or inconsistent for two main reasons. First, even when assuming a semi-linear conditional mean function $\E[y_i|x_i,\bw_i] = \mu_0(x_i) + \bw_i'\bgamma_0$, residualized binscatter does not, in general, consistently estimate $\upsilon_0(x)$, $\mu_0(x)$, or $\E[y_i|x_i=x,\bw_i=\bw]$ for some evaluation point $\bw$, despite being motivated by standard least squares methods. Only when $\mu_0(x)$ is linear does \eqref{eq:wrong plim} reduce to $\mu_0(x)$, which need not equal $\upsilon_0(x)$ because $\E[y_i|x_i] = \E[\E[y_i|x_i,\bw_i]|x_i] = \mu_0(x_i)+\E[\bw_i|x_i]'\bgamma_0$ under the semi-linear conditional mean structure. Therefore, from a point estimation and visualization perspective, residualized binscatter is not recommended for empirical work.

Second, from the perspective of assessing linearity or other shape features of the regression functions, the residualized binscatter is also not recommended. If $\E[y_i|x_i,\bw_i] = \mu_0(x_i) + \bw_i'\bgamma_0$, with $\mu_0(x)$ a linear function of $x$, then the residualized binscatter plot will appear linear (for sufficiently large $n$ and an appropriate choice of $J$). However, linearity of the regression functions is only sufficient, not necessary: for some nonlinear $\mu_0(x_i)$ the plot will appear linear, while for other nonlinear $\mu_0(x_i)$ it will appear nonlinear. Thus, relying on residualized binscatter to assess linearity is not recommended because researchers may incorrectly conclude that $\mu_0(x_i)$ (and hence $\E[y_i|x_i,\bw_i]=\mu_0(x_i) + \bw_i'\bgamma_0$ for some value of $\bw_i$) is linear from visual inspection or informal testing, thereby rendering subsequent empirical results based on a parametric linear regression potentially misleading.

Section SA-1.1 in the online Appendix presents two simple parametric examples illustrating the potential biases introduced by residualized binscatter. The first example considers a Gaussian polynomial regression model, where $\mu_0(x)=x^m$ for some $m\in\mathbb{N}$, $d=1$, and $(y_i,x_i,w_i)'\thicksim \mathsf{Normal}$, and shows precisely how the different parameters underlying the model can change the shape of $\mu_0(x)$ as well as the concentration of 
$x_i-\text{L}(x_i|\bw_i)$, thereby affecting visually and formally the ``shape'' and ``support'' of \eqref{eq:wrong plim}. The second example considers $\mu_0(x)$ unrestricted, $d=1$, $w_i\thicksim \mathsf{Bernoulli}$, and $x_i|w_i=0\thicksim \mathsf{Uniform}$ and $x_i|w_i=1\thicksim\mathsf{Uniform}$ with disjoint supports, and shows how residualized binscatter can turn a nonlinear $\mu_0(x)$ into a linear function in \eqref{eq:wrong plim} with incorrect support. These analytical examples complement our empirical applications (see Figure \ref{fig:intro-covs} and Figure \ref{fig:M_repl}), which illustrate with real data the detrimental effects of employing residualized binscatter for understanding the true form of the regression function relating the outcome $y_i$ to $x_i$ and $\bw_i$.

\subsection{Covariate-Adjusted Binscatter}

With only bivariate data $(y_i, x_i)$, the binscatter \eqref{eq:binscatter-canonical} naturally provides (a visualization of) an estimate of the conditional mean function, $\upsilon_0(x_i)=\E[y_i|x_i]$, which has a straightforward interpretation. Controlling for additional covariates complicates interpretation: we want to visually assess how $y_i$ and $x_i$ relate while ``controlling'' for $\bw_i$ in some precise sense. There is not a universal answer to this problem, and the empirical literature employing binscatter methods is usually imprecise.

Motivated by \eqref{eq:binscatter-canonical}, a more principled way to incorporate the covariates $\bw_i$ into the binscatter is via semiparametric partially linear regression, as is commonly done in applied econometrics and program evaluation \citep{Abadie-Cattaneo_2018_ARE,Angrist-Pischke_2008_Book,Wooldridge_2010_Book}. We define the covariate-adjusted binscatter as
\begin{equation}
    \label{eq:binscatter-covadj}
    \widehat{\mu}(x) = \widehat{\bb}(x)'\widehat{\bbeta}, \qquad
  \begin{bmatrix}\;\widehat{\bbeta}\;\\\;\widehat{\bgamma}\;\end{bmatrix}
  = \argmin_{\bbeta\in\mathbb{R}^J,\bgamma\in\mathbb{R}^d} \sum_{i=1}^{n} (y_i-\widehat{\bb}(x_i)'\bbeta-\bw_i'\bgamma)^2.
\end{equation}
In this paper we take the semi-linear covariate-adjusted binscatter implementation \eqref{eq:binscatter-covadj} as the starting point of analysis, and thus view $\widehat{\Upsilon}(x_i,\bw_i) = \widehat{\mu}(x_i) + \bw_i'\widehat{\bgamma}$ as the plug-in estimator of
\begin{equation}\label{eq:partially linear model}
    \E[y_i|x_i,\bw_i] = \mu_0(x_i)+\bw_i'\bgamma_0 = \Upsilon_0(x_i,\bw_i),
\end{equation}
where we assume the usual identification restriction that $\E[\V[\bw_i|x_i]]$ is positive definite. The imposed additive separability between $x_i$ and $\bw_i$ of the conditional mean function follows standard empirical practice, but affects interpretation in certain cases. Our theoretical results continue to hold under misspecification of $\E[y_i|x_i,\bw_i]$, provided the probability limit of $\widehat{\Upsilon}(x_i,\bw_i)$ is interpreted as a best mean square approximation of $\E[y_i|x_i,\bw_i]$ using functions of the form $g(x,\bw) = \mu(x)+\bw'\bgamma$. More precisely, under regularity conditions, the best mean square approximation would be $\text{P}(y_i|x_i,\bw_i) = \mu^\star_0(x_i) + \bw_i'\bgamma^\star_0$ with
\[\mu^\star_0(x_i) = \E[y_i|x_i] - \E[\bw_i|x_i]'\bgamma^\star_0 \qquad\text{and}\qquad
  \bgamma^\star_0 = \big(\E\big[\V[\bw_i|x_i]\big]\big)^{-1}\E\big[\Cov[\bw_i,y_i|x_i]\big].\]
In particular, $\mu^\star_0(x_i)=\mu_0(x_i)$ and $\bgamma^\star_0 = \bgamma_0$ if \eqref{eq:partially linear model} holds.

We adopt the semi-linear structure \eqref{eq:partially linear model} throughout the paper because it is often invoked (explicitly or implicitly) for interpretation in empirical work. \cite{Cattaneo-Crump-Farrell-Feng_2023_Generalized} generalize binscatter methods to settings beyond the semi-linear conditional mean, including quantile regression, other nonlinear models such as logistic regression, and first-order interactions with a discrete covariate (e.g., a subgroup indicator). Those generalizations allow for a richer class of semiparametric parameters of interest and associated binscatter methods.

Given the working model \eqref{eq:partially linear model}, it remains to determine the (functional) parameter of interest. For visualization, a natural choice is a partial mean effect:
\begin{equation}\label{eq:estimand-partialmeaneffect}
    \Upsilon_0(x) = \Upsilon_0(x,\E[\bw_i]) = \mu_0(x) + \E[\bw_i]'\bgamma_0,
\end{equation}
which captures the average effect of $x_i$ on $y_i$ for units with covariates $\bw_i$ at their average value $\E[\bw_i]$, and thus gives an intuitive notion of the mean relationship of $x_i$ and $y_i$ after controlling for covariates $\bw_i$ at their average values. The plug-in estimator is
\begin{equation}\label{eq:binscatter-longreg}
    \widehat{\Upsilon}(x) = \widehat{\mu}(x) + \bar{\bw}'\widehat{\bgamma}
\end{equation}
with $\bar{\bw}=\frac{1}{n}\sum_{i=1}^n \bw_i$.

The structure imposed and the parameter considered are not innocuous, but lead to several advantages over other options. First, the target parameter in \eqref{eq:estimand-partialmeaneffect} has a natural partial mean interpretation because $\Upsilon_0(x) = \int \Upsilon_0(x,\bw) \text{d}F(\bw) = \mu_0(x_i) + \E[\bw_i]'\bgamma_0$, where $F(\bw)=\P[\bw_i\leq \bw]$ is the marginal distribution function of the covariates. In addition, if $\bw_i$ is mean independent of $x_i$, that is, if $\E[\bw_i|x_i]=\E[\bw_i]$, then $\upsilon_0(x_i)=\E[y_i|x_i] = \E[\E[y_i|x_i,\bw_i]|x_i] = \mu_0(x_i)+\E[\bw_i|x_i]'\bgamma_0 = \Upsilon_0(x_i)$. For example, if $x_i$ is a randomly assigned treatment dose and $\bw_i$ are pre-intervention covariates, then $\Upsilon_0(x)$ corresponds to the dose-response average causal effect.

Second, $\Upsilon_0(x)$ matches the goal of examining potential nonlinearities (and other features) only along the $x_i$ dimension. The goal in a binscatter analysis is to control for $\bw_i$, not to allow for (or discover) heterogeneity along these variables. This is why the covariates $\bw_i$ are typically controlled for linearly, and without interactions with $x_i$, in the post-visualization regression analysis. 

Third, $\Upsilon_0(x)$ has practical advantages.  To see why, consider the alternative of estimating the fully-flexible conditional mean, $\E[y_i|x_i,\bw_i]$, and then integrating over the marginal distribution of $\bw_i$.  Although we would avoid imposing any structure on the conditional mean function, this approach would be impractical in common empirical settings as it would require nonparametric estimation in many dimensions.  
Taking the case of our running example to illustrate, AGNS control for four continuous variables, $49$ state fixed effects, and $60$ year fixed effects, so that $\dim(\bw_i) = 113$.  Furthermore, even when the partially linear model is adopted, there may still be a curse of dimensionality when interest lies in $\mu_0(x)$ because $\upsilon_0(x_i) =\mu_0(x_i)+\E[\bw_i|x_i]'\bgamma_0$, implying that the potentially high-dimensional conditional expectation $\E[\bw_i|x_i]$ needs to be estimated. For example, in AGNS this would require fitting $113$ preliminary nonparametric regressions to estimate $\E[\bw_i|x_i]$.

Finally, because $\Upsilon_0(x)$ is a special case of the more general partial mean $x\mapsto\Upsilon_0(x,\bw) = \mu_0(x) + \bw'\bgamma_0$ for some fixed value $\bw$, it is possible to use other choices for the evaluation point $\bw$. For example, setting the discrete components of $\bw$ to a base category (such as zero) is a natural alternative. The choice of $\bw$ will affect the interpretation of the estimand and the statistical properties of the estimator: see Section SA-1.2 in the online Appendix for more discussion. In the remainder of the paper, we focus on $\Upsilon_0(x)$, but our theoretical results cover other choices of the evaluation point $\bw$ (see Section \ref{sec:theory} and the online Appendix).

Figure \ref{fig:intro-covs} in the Introduction showed how the results can change when using the correct and incorrect residualization (recall that panels (a) and (b) use the incorrect residualization). First, in Figure \ref{fig:intro-covs}(a) the shape does not appear linear. Second, Figure \ref{fig:intro-covs}(b) shows the extreme compression of the support of the estimate using the incorrect residualization by restoring the proper scale.  This generally comes about because the variability of both the dependent and independent variables of interest have been overly suppressed. Finally, Panel (c) shows our estimator $\widehat{\Upsilon}(x)$ defined in \eqref{eq:binscatter-longreg}, using the correct residualization \eqref{eq:binscatter-covadj}. We can observe a much clearer shape of the estimate of the conditional expectation. In this case our methods give stronger visual support for the linear regression used by AGNS, in contrast to the apparent nonlinearity in the original binscatter. It is important to remember that although binscatters such as Figure \ref{fig:intro-covs} visually resemble conventional scatter plots of a data set, the plotted dots are actually a point estimate of a function (though in the case of Figure \ref{fig:intro-covs}(a) and (b), not necessarily a useful function, see \eqref{eq:wrong plim}). 

We can also accommodate covariates in the fixed-$J$ case in a principled way.  In this case,  the estimator $\widehat{\Upsilon}(x)$ remains the same but the estimand, $\Upsilon_0(x)$, is replaced by its fixed-$J$ analogue: $\bXi_0= (\Xi_0(1),\Xi_0(2),\dots,\Xi_0(J))'$ with $\Xi_0(j)=\bb(x)'\bbeta_J+\E[\bw_i]'\bgamma_J$ for $x\in\mathcal{B}_j$ with
\[\begin{bmatrix}\;\bbeta_J\;\\\;\bgamma_J\;\end{bmatrix} = \argmin_{\bbeta\in\mathbb{R}^J,\bgamma\in\mathbb{R}^d} \E[(y_i-\bb(x_i)'\bbeta-\bw_i'\bgamma)^2]\]
where $\bb(x) = [\I_{\mathcal{B}_1}(x),\I_{\mathcal{B}_2}(x),\cdots,\I_{\mathcal{B}_J}(x)]'$. Under mild regularity conditions, as the number of bins increases, each bin becomes smaller and thus the fixed-$J$ parameter $\Xi_0(j)$ approximates $\Upsilon_0(x)$ for $x\in\mathcal{B}_j$ for all $j=1,2,\dots,J$ uniformly: $\max_{1\leq j\leq J} \sup_{x\in\mathcal{B}_j}|\Xi_0(j)-\Upsilon_0(x)|\to 0$ as $J\to\infty$. A small number of bins in finite samples, however, can make these parameters quite different due to misspecification errors induced by the local constant approximation within bins.

\section{Choosing the Number of Bins}
    \label{sec:IMSE}

The final element of the binscatter estimator to formalize is the choice of $J$, the number of bins. It is common to encounter applications of binscatter where $J =\mathtt{J}$ for a fixed natural number $\mathtt{J}$, regardless of the data features. For example, the default in the Stata packages \texttt{binscatter} and \texttt{binscatter2} is $\mathtt{J}=20$, while AGNS used $\mathtt{J}=100$. As already mentioned, from a fixed $J$ perspective, the canonical binscatter \eqref{eq:binscatter-canonical} estimates $\bxi_0$ and the covariate-adjusted binscatter \eqref{eq:binscatter-covadj} estimates $\bXi_0$, neither of which may be the parameter of interest in a specific application. For example, the two parameters $\xi_0(j)=\E[y_i|x_i\in\mathcal{B}_j]$ and $\upsilon_0(x)$ for $x\in\mathcal{B}_j$ can lead to substantially different interpretations from both statistical and economic perspectives within bin $\mathcal{B}_j$. Furthermore, when comparing across bins, $(\xi_0(j):j=1,2,\dots,J)$ can be substantially different from $(\upsilon_0(x):x\in\mathcal{X})$. The choice of the tuning parameter $J$ determines the interpretation of the binscatter plot and estimand. In this section, we illustrate these concepts and discuss the choice of $J$ in practice. 

We view binscatter as a sequence of approximating models indexed by $J$, where the larger $J$ (more bins) is, the less bias but more variance the estimator will exhibit. In other words, we view binscatter as most useful when the focus is on recovery of $\upsilon_0(x)$ or $\Upsilon_0(x)$, allowing us to visualize and conduct inference on those unknown functions. It is only by recovering $\upsilon_0(x)$ or $\Upsilon_0(x)$ that we can answer substantive questions regarding functional form or shape restrictions. In what is perhaps the leading case, if we wish to use a binscatter plot to precede a linear regression, then our interest is in whether $\upsilon_0(x)$ or $\Upsilon_0(x)$ is linear, so we must recover the true function. Recovering the coarsened version, as with a fixed $J=\mathtt{J}$, is not sufficient. The same reasoning applies to any statement regarding other shape constraints such as whether the relationship is monotonic or convex.

\begin{figure}[!ht]
\caption{\small{\bf Choice of $J$.} This figure illustrates the role of the choice of $J$ using data from \cite{Akcigitetal2021_QJE}.  The dependent variable, independent variable, and controls are the same as in Figure \ref{fig:intro-covs}.  The left and right plots show a binned scatter plot with $J=5$ and $J=50$, respectively.  The middle plot shows the binned scatter plot using the optimal choice of $J=11$ based on a cluster-robust variance estimator with two-way clustering by year and state $\times$ five-year period.  Binscatter estimates are based on weights of each state's 1940 population count.}
\label{fig:choice of J}
\begin{subfigure}[t]{0.333\columnwidth}
\centering	
\caption{\it Oversmoothed}
\includegraphics[trim={0cm 1cm 0cm 1cm}, clip, scale=.2]{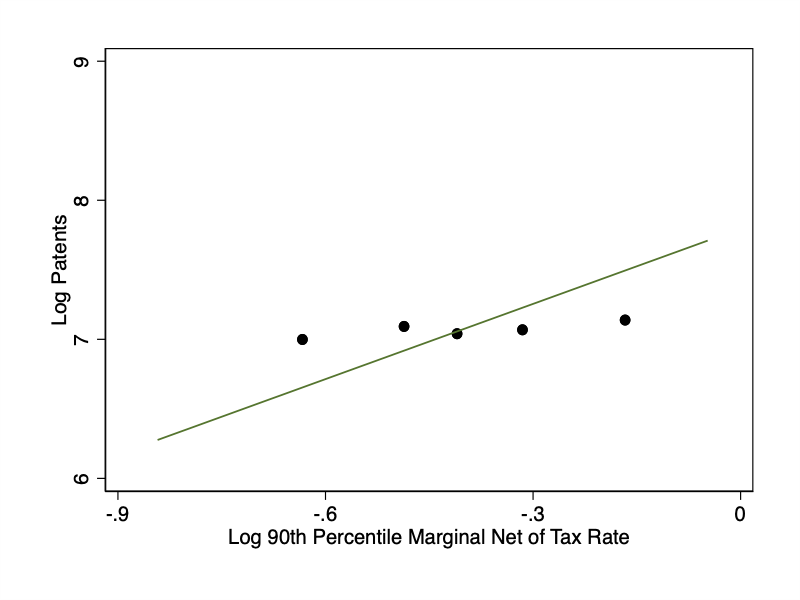}
\end{subfigure}
\begin{subfigure}[t]{0.333\columnwidth}
\centering	
\caption{\it Optimal}
\includegraphics[trim={0cm 1cm 0cm 1cm}, clip, scale=.2]{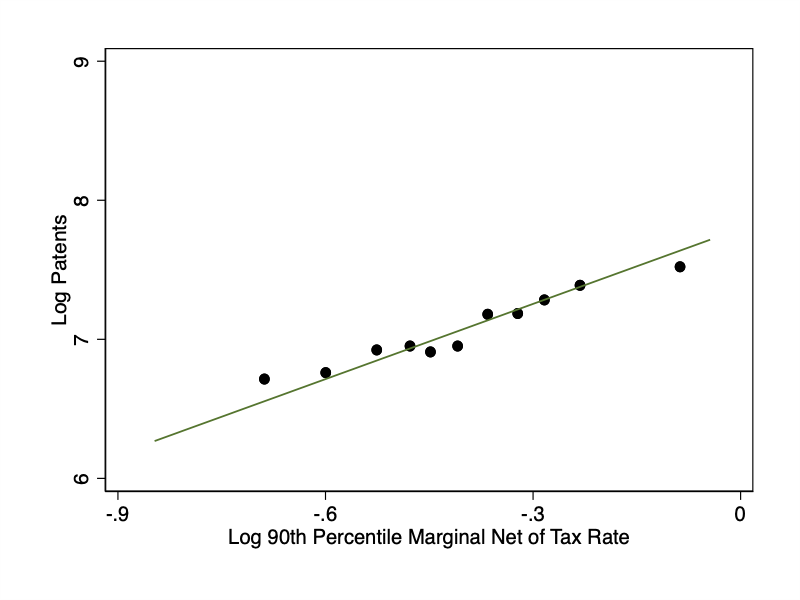}
\end{subfigure}
\begin{subfigure}[t]{0.333\columnwidth}
\centering	
\caption{\it Undersmoothed}
\includegraphics[trim={0cm 1cm 0cm 1cm}, clip, scale=.2]{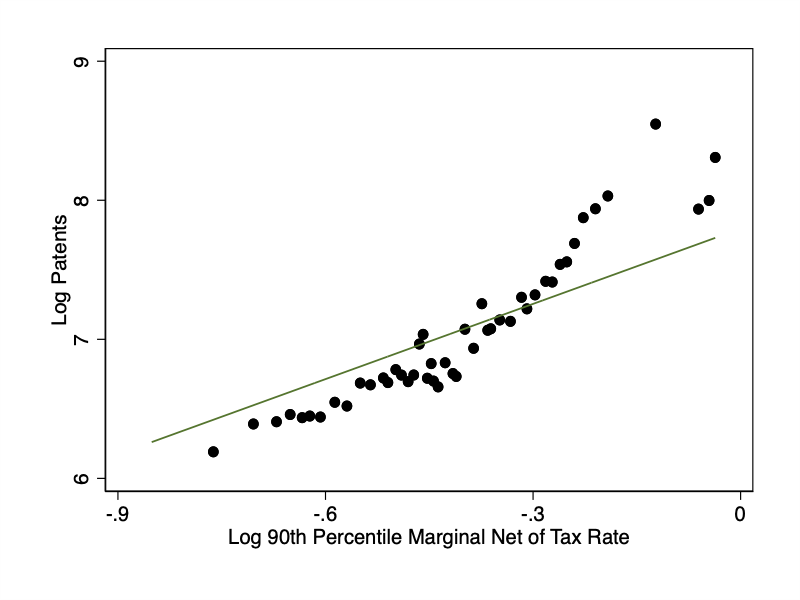}
\end{subfigure}
\end{figure}

Consistent nonparametric estimation of $\upsilon_0(x)$ or $\Upsilon_0(x)$ requires $J$ to diverge with the sample size, but neither too rapidly nor too slowly.  To remove the approximation bias, a sufficiently large $J$ is required to overcome the limited flexibility of the constant fit within bins: intuitively, as $J$ diverges, the bin width collapses, and $\xi_0(j) = \E[y_i|x_i\in\mathcal{B}_j] \approx \upsilon_0(x)$ for $x\in\mathcal{B}_j$ because the width of the bin $\mathcal{B}_j$ shrinks as $J$ increases. However, the variance of the estimator increases with $J$ because variance is controlled by the bin-specific sample sizes, which are roughly $n/J$. Thus, as is familiar in nonparametric estimation, we face a bias-variance 
trade-off when choosing $J$. Figure \ref{fig:choice of J} illustrates this bias-variance trade-off in our running application. In Panel (a) we use $\mathtt{J} = 5$. If we consider this choice as fixed, we can use these results to, for example, compare the productivity of those subject to the highest quintile of all tax rates on high earners to those in areas where taxes on high earners are in the lowest quintile. But for the purpose of nonparametric estimation and inference, the estimator is oversmoothed: the number of bins is too small to remove sufficient bias. At the other extreme, Panel (c) uses 50 bins, and the estimator is undersmoothed (too wiggly) to provide a reliable visualization of the conditional mean.

A wide range of choices for $J$ will, in large sample theory, ensure that both bias and variance are adequately controlled and thus yield a consistent estimator and valid distributional approximation. However, such rate restrictions are not informative enough to guide practice. It is therefore important to have tight guidance for empirical research. To accomplish this, we develop a selector for $J$ that is optimal in terms of integrated mean square error (IMSE). As is standard in nonparametrics, the IMSE-optimal $J$ balances variance and (squared) bias, resulting in 
\begin{equation}
    \label{eq:J IMSE}
    J_{\mathtt{IMSE}} = \left\lceil \left(\frac{2\mathscr{B}_n}{\mathscr{V}_n}
	\right)^{1/3} \; n^{1/3} \right\rceil,
\end{equation}
The terms $\mathscr{V}_n$ and $\mathscr{B}_n$ capture the asymptotic variance and (squared) bias of the binscatter, respectively. We give complete expressions in the online Appendix. All that matters at present is that (i) both are generally bounded and bounded away from zero under minimal assumptions, (ii) the variance accounts for heteroskedasticity and clustering, and (iii) both incorporate the additional covariates appropriately, so that the optimal $J$ depends on the presence of $\bw_i$. A formal IMSE expansion is discussed in Section \ref{sec:theory} and given in Theorem SA-3.4 in the online Appendix, along with a uniform consistency result in Corollary SA-3.1, which has the same rate up to a $\log(J)$ factor. A feasible version, $\widehat{J}_{\mathtt{IMSE}}$, is straightforward to implement. Details are given in Section SA-4 in the online Appendix. 

The formula for $J_{\mathtt{IMSE}}$ intuitively reflects the trade-off as depicted in Figure \ref{fig:choice of J}. If the data are highly variable $\mathscr{V}_n$ will be large, driving down $J_{\mathtt{IMSE}}$, so that each bin has a large sample size. On the other hand, if $\mu_0(x)$ is highly nonsmooth, $\mathscr{B}_n$ will be large, and more bins are required to adequately remove bias. Figure \ref{fig:choice of J}(b) shows our feasible IMSE-optimal choice in the data of AGNS, where we find $\widehat{J}_{\mathtt{IMSE}} = 11$. With this choice, we obtain a  visualization and optimal nonparametric estimation of $\mu_0(x)$ and $\Upsilon_0(x)$. We will also base our uncertainty visualization and quantification around this implementation, to ensure validity, as we detail in the next section.

Even if a fixed $J=\mathtt{J}$ is chosen for an application, the data-driven choice $\widehat{J}_{\mathtt{IMSE}}$ can provide a useful benchmark to understand better the bias-variance trade-off underlying the binscatter implementation. For example, choosing a $\mathtt{J}$ that is much larger than $\widehat{J}_{\mathtt{IMSE}}$ will yield a binscatter that is likely to exhibit considerably more variability than bias, given the data generating process. Thus, the data-driven choice $\widehat{J}_{\mathtt{IMSE}}$ can help applied researchers discipline and improve their fixed $J=\mathtt{J}$ binscatter implementations. 

In the remainder of the paper we focus on the covariate-adjusted binscatter estimate \eqref{eq:binscatter-longreg} implemented using $J_{\mathtt{IMSE}}$, or its fixed-$J$ analogue when appropriate for concreteness. Our technical results in the online Appendix accommodate other choices of $J$ as a function of the sample size with and without covariate-adjustment, thereby covering, in particular, the canonical binscatter estimate \eqref{eq:binscatter-canonical} implemented using its corresponding $J_{\mathtt{IMSE}}$. See Section \ref{sec:theory} for a brief overview.

\section{Quantifying Uncertainty}
    \label{sec:Quantifying Uncertainty}

We provide both visualization and analytical tools to capture the uncertainty underlying the mean estimate $\widehat{\Upsilon}(x) = \widehat{\mu}(x) + \bar{\bw}'\widehat{\bgamma}$, valid simultaneously for all values of $x\in\mathcal{X}$. This uniformity over $x\in\mathcal{X}$ is required both to answer the substantive questions of interest in empirical work and to provide a correct visualization of the uncertainty for the function $\Upsilon_0(x)$. Uniform inference theory is a major technical contribution of this paper (see Section \ref{sec:theory} and the online Appendix). Confidence bands directly enhance the visualization capabilities of binned scatter plots by summarizing and displaying the uncertainty around the estimate $\widehat{\Upsilon}(x)$. Loosely speaking, a confidence band is simply a confidence ``interval'' for a function, and is interpreted much like a traditional confidence interval.  

A typical confidence interval for a single parameter (such as a mean or regression coefficient) is a range between two endpoint values that, in repeated samples, covers the true parameter with a prespecified probability. The width of a confidence interval increases with the uncertainty in the data. Intuitively, the interval shows the values of the parameter that are compatible with the data. For example, if the interval contains zero, then zero is a plausible value for the true parameter. That is, the null hypothesis of zero cannot be rejected.

A confidence band is essentially the same, but as a function of $x$, and can therefore be directly plotted. It is the area between two endpoint functions that contains all the functions $\Upsilon_0(x)$ that are compatible with the data for some pre-set probability. Matching the use of a confidence interval, the band can be used to evaluate hypotheses. For example, if the band contains a linear function, then linearity is a plausible form for $\mu_0(x)$ (as in  Figure \ref{fig:intro-band}(b)). That is, the null hypothesis that $x$ enters $\Upsilon_0(x)$ linearly cannot be rejected. The same logic can be used for other shape restrictions: if the band contains monotonic functions, then monotonicity of $\Upsilon_0(x)$ is consistent with the data. This is illustrated below. Thus, adding a confidence band is an important step in any binscatter, to visually assess and communicate the uncertainty, just as the addition of standard errors is an important step and good empirical practice in any regression analysis. The reader can see not only the estimate of the relationship (the ``dots'' of the binscatter), but also the uncertainty surrounding this estimate.

The construction and theory of our confidence bands also intuitively match standard confidence intervals. First, our confidence bands reflect the underlying heteroskedastic variance in the data uniformly over the support of $x_i$. While the visualizations do reflect these quantities, they are not directly shown or formally accounted for. This is analogous to how a simple confidence interval for the mean reflects only estimation uncertainty about the parameter, even though the interval depends on the variance of the data. For visualizing the ``spread'' and detecting outliers conditional quantiles may be more useful \citep[see][]{Cattaneo-Crump-Farrell-Feng_2023_Generalized}.  Second, the upper/lower endpoint functions are given by the point estimate plus/minus a critical value times a standard error. In this way, the width of the band at any point depends on the overall uncertainty and the heteroskedasticity. 

Before presenting the confidence band formulation, we must be precise about the object we intend to cover with the confidence band. If $J$ is taken as fixed, the parameter is $\bXi_0$, and inference is parametric because there is no misspecification bias for that parameter. However, as explained before, in many applications the parameter of interest will not be $\bXi_0$ but rather $\Upsilon_0(x)$, leading to unavoidable misspecification errors introduced by the binscatter approximation to the true function. Thus, we focus on a band to cover the function $\Upsilon_0(x)$ given in \eqref{eq:estimand-partialmeaneffect}. It is only in this case that the band can be used to assess properties of the function of interest. Testing linearity (prior to a regression analysis) is the most common use case, but binned scatter plots are also utilized to assess other shape restrictions (see, for example, \cite{ShapiroWilson2021} or \cite{FeigenbergMiller2021}).  Regardless of the application, the band must be constructed from a nonparametric perspective (i.e., assuming $J$ diverging to account explicitly for misspecification error).

To ensure validity of the nonparametric confidence band, we will use $J_{\mathtt{IMSE}}$ given in \eqref{eq:J IMSE} together with debiasing to remove the first-order nonparametric misspecification bias introduced by employing the IMSE-optimal binscatter. More specifically, we employ a simple application of the standard robust bias correction method for debiasing \citep{Calonico-Cattaneo-Farrell_2018_JASA,Cattaneo-Farrell-Feng_2020_AoS,Calonico-Cattaneo-Farrell_2022_Bernoulli}. Section \ref{sec:theory} discusses the theoretical foundations, and the online Appendix provides all the details, while here we describe the key ideas heuristically. The confidence band for $\Upsilon_0(x)$ is
\begin{equation}\label{eq:CB formula}
    \widehat{I}_{\tt RBC}(x) = \Big[ \; \widehat{\Upsilon}_{\tt BC}(x) \pm \cval_{\tt RBC} \cdot \sqrt{\widehat{\Omega}_{\tt RBC}(x)/n} \; \Big],
\end{equation}
where $\widehat{\Upsilon}_{\tt BC}(x)$ denotes the covariate-adjusted debiased binscatter estimator of $\Upsilon_0(x)$, $\widehat{\Omega}_{\tt RBC}(x)/n$ is its variance estimator, and $\cval_{\tt RBC}$ is the appropriate quantile to make the confidence band uniformly valid. The exact formulas are given in Section \ref{sec:theory}. Intuitively, $\widehat{\Upsilon}_{\tt BC}(x) = \widehat{\Upsilon}(x) - \widehat{\textsf{Bias}}[\widehat{\Upsilon}(x)]$, where $\widehat{\textsf{Bias}}[\widehat{\Upsilon}(x)]$ denotes the bias correction, and $\widehat{\Omega}_{\tt RBC}(x)/n = \widehat{\textsf{Var}}[\widehat{\Upsilon}_{\tt BC}(x)]$ is an estimator of the variance of $\widehat{\Upsilon}_{\tt BC}(x)$ not just of $\widehat{\Upsilon}(x)$. The key idea underlying the robust bias correction method is that debiasing introduces additional estimation uncertainty that must be incorporated explicitly into the standard error formula. While there are many ways of debiasing the IMSE-optimal point estimator $\widehat{\Upsilon}(x)$, a simple one proceeds by fitting a constrained linear regression within each bin where the estimated coefficients are restricted to ensure that the binscatter estimator is continuous;  that is, the constraints force the piecewise linear fits within bins to be connected at the boundary of the bins. This construction ensures that the associated confidence bands are also continuous.  More details are given in Section \ref{sec:theory} and in the online Appendix.

Our results rely on standard regularity conditions for valid uniform distribution theory with robust bias correction discussed in Section \ref{sec:theory}, and the online Appendix gives results under more general, and in some cases weaker, conditions. More precisely, for $\alpha\in(0,1)$, we show that
\begin{equation}\label{eq:CB validity}
    \P\Big[ \Upsilon_0(x) \in \widehat{I}_{\tt RBC}(x), \; \text{for all } x\in\mathcal{X} \Big] \to 1 -\alpha
\end{equation}
giving formal validity, that is, in repeated samples the area covers the true function $\Upsilon_0(x)$ with a pre-specified probability $1-\alpha$. Recall that $\Upsilon_0(x)$, by definition, uses the mean of $\bw_i$; other possible choices and their impact on the confidence band are discussed in Section SA-1.2 of the online Appendix.

The result in \eqref{eq:CB validity} shows how to add valid confidence bands to any binned scatter plot. This visual assessment of uncertainty is an important step in any analysis. Our discussion focused on the nonparametric uncertainty quantification when employing the IMSE-optimal binscatter constructed using $J=J_\mathtt{IMSE}$ bins and debiasing using within-bin linear regression, but our theoretical results remain valid more generally for other choices of $J$ and debiasing approaches. Furthermore, the bands continue to be valid when $J = \mathtt{J}$ is fixed provided the estimand $\Upsilon_0(x)$ is switched to its fixed-$J$ analogue $\bXi_0$. In this latter case, robust bias correction is not technically needed because the misspecification error is removed by assumption (i.e., by redefining the parameter of interest).

If $x_i$ is discrete, or the researcher is content with the coarsened version of the parameter $\bXi_0$ under a fixed-$J$ approach, our results provide (uniformly) valid inference for the covariate-adjusted outcome mean conditional on falling in each bin. This amounts to adding pointwise confidence intervals to a plot --  which is common practice in many uses -- and making corrections for multiple testing. These can be used directly to assess uncertainty about the mean for a masspoint of $x_i$ (or within a given quantile range), but cannot be used to assess functional features of the regression function $\Upsilon_0(x)$ as a whole. 

\begin{figure}[!ht]
\caption{\small{\bf Quantifying Uncertainty: The Role of $J$.} This figure illustrates uniform confidence bands using data from \cite{Akcigitetal2021_QJE}.  The dependent variable, independent variable, and controls are the same as in Figure \ref{fig:intro-covs}.  The left plot presents a confidence band for $\bXi_0$ whereas the right plot shows the confidence band for $\Upsilon_0(x)$.  Binscatter estimates are based on weights of each state's 1940 population count.  Shaded regions denote 95\% nominal confidence bands using a cluster-robust variance estimator with two-way clustering by year and state $\times$ five-year period.}
\label{fig:band}
\begin{subfigure}[t]{0.5\columnwidth}
\centering	
\caption{\it Fixed $J$}
\includegraphics[trim={0cm 1cm 0cm 1cm}, clip, scale=.275]{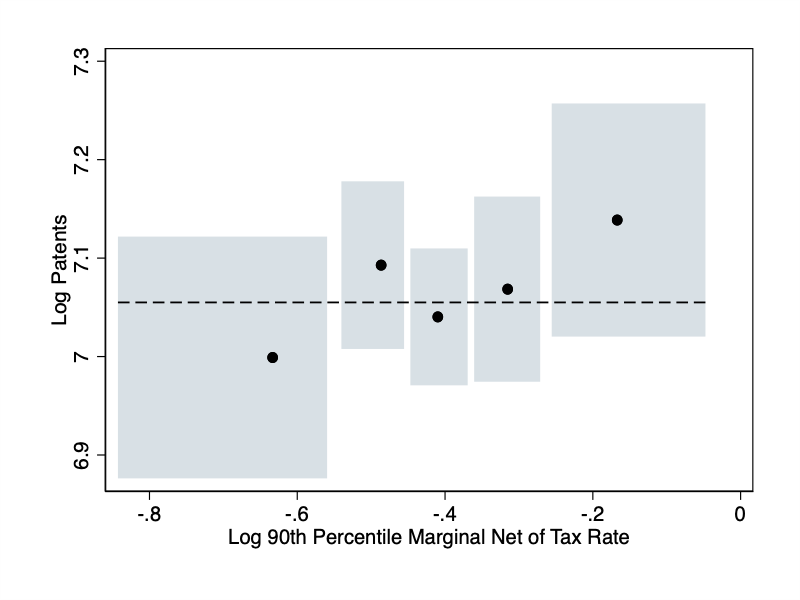}
\end{subfigure}
\begin{subfigure}[t]{0.5\columnwidth}
\centering	
\caption{\it Diverging $J$}
\includegraphics[trim={0cm 1cm 0cm 1cm}, clip, scale=.275]{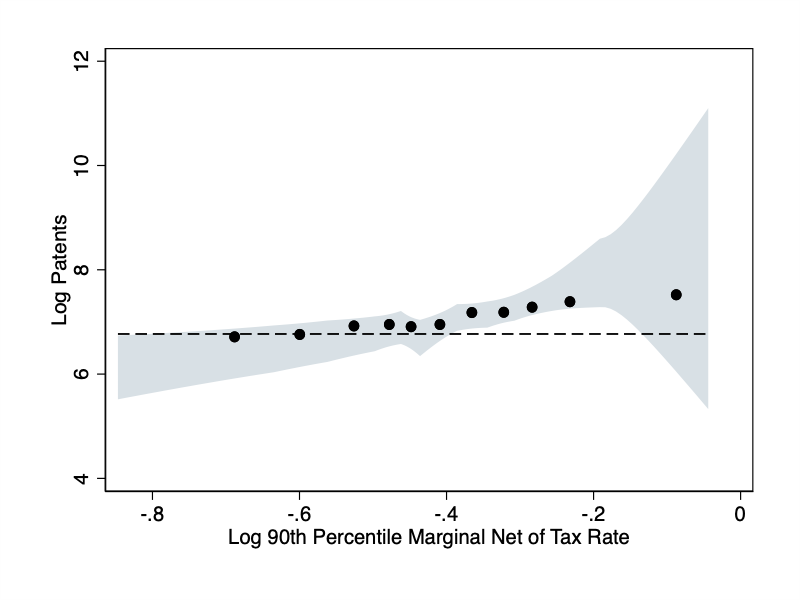}
\end{subfigure}
\end{figure}

Figure \ref{fig:band} compares these two cases, fixed-$J$ versus large-$J$, using the data of AGNS. Figure \ref{fig:band}(a) shows confidence bands for $\mathtt{J} = 5$, with the interpretation of studying the conditional expectation of log patents given marginal tax rates in a specific quintile controlling for additional covariates (i.e., $\bXi_0$).  As we saw in Figure \ref{fig:choice of J}(a), the point estimates are all relatively similar across quintiles and, in fact, we cannot rule out that all five conditional means are the same.  This can be gleaned by the dashed horizontal line in Figure \ref{fig:band}(a) which comfortably sits in all five shaded regions.  In Figure \ref{fig:band}(b) we consider inference on $\Upsilon_0(x)$ using the optimal choice of $J$ (as in Figures \ref{fig:intro-band} and \ref{fig:choice of J}(b)).  We have already discussed that the confidence band is consistent with a linear relation between the variables.  However, we can also highlight classes of functions that the confidence band excludes.  The dashed horizontal line is set to the upper bound of the confidence band at the smallest value of $x$ in the support.  We can immediately observe that the confidence band rules out any horizontal lines, i.e., we can reject that log patents have no relationship with marginal tax rates.  This horizontal line is also a useful visual cue to evaluate the class of monotonically decreasing functions.  Clearly, we can also reject a monotonically decreasing relation between the two variables.  Figure \ref{fig:band} illustrates that the use of confidence bands for investigating the attributes of the true functional form is simple and straightforward.

In addition to employing confidence bands for testing substantive hypotheses about $\Upsilon_0(x)$ such as positivity, monotonicity, or concavity, we develop formal hypothesis testing based on canonical binscatter methods in the online Appendix for completeness. These methods are also available in our companion software implementations \citep{Cattaneo-Crump-Farrell-Feng_2023_Stata}, and can be used to complement the empirical analysis based on canonical binscatter discussed previously, offering potential power improvements as well as more precise econometric conclusions (e.g., formal p-values). Since using the confidence bands for testing is already a valid, easy, and intuitive econometric methodology for empirical work employing canonical binscatter, we offer further technical discussion of the companion formal hypothesis testing methods for parametric specification and shape restrictions in \cite{Cattaneo-Crump-Farrell-Feng_2023_Generalized}, covering \textit{generalized} binscatter methods based on both least squares and other loss functions (e.g., quantile or logistic regression).

\section{Another Empirical Illustration}
    \label{sec:applications}

As an additional empirical application we revisit \cite{Moretti2021_AER}, which examined the relation between the productivity of top inventors and high-tech clusters, where clusters are defined as activity in a city of a specific research field (e.g., computer scientists in  Silicon Valley).  The paper estimates an elasticity of number of patents in a year with respect to cluster size of 0.0676.  The statistically significant positive relationship aligns with the empirical observation that increasingly large subsidies are being offered by states and localities for high-tech firms to relocate within their regions.  

\begin{figure}[!ht]
\caption{\small{\bf Relation Between Productivity of Top Inventors and High-Tech Clusters.} This figure uses the data from \cite{Moretti2021_AER}. The dependent variable is the log number of patents per inventor per year, and the independent variable is the log cluster size.  The top left plot shows a raw scatter plot of the data.  The top right plot replicates Figure 4 in \cite{Moretti2021_AER} which controls for year, research field, and city effects while the middle left plot shows the implied estimated conditional mean function \eqref{eq:wrong plim}.  The incorrect residualization versus the semi-linear specification introduced in Section \ref{sec:setup} (both for 40 bins) is shown in the middle right chart. The bottom left chart uses the optimal choice of $J$ introduced in Section \ref{sec:IMSE}.  The bottom right chart again uses the optimal choice of $J$ but for the main specification of \citet[Table 3, Columnn (8)]{Moretti2021_AER}.  Shaded regions denote 95\% confidence bands using a cluster-robust variance estimator with clustering by city $\times$ field.}
\label{fig:M_repl}
\begin{subfigure}[t]{0.5\columnwidth}
\centering	
\caption{\it Raw Scatter plot}
\includegraphics[trim={0cm 1cm 0cm 1cm}, clip, scale=.275]{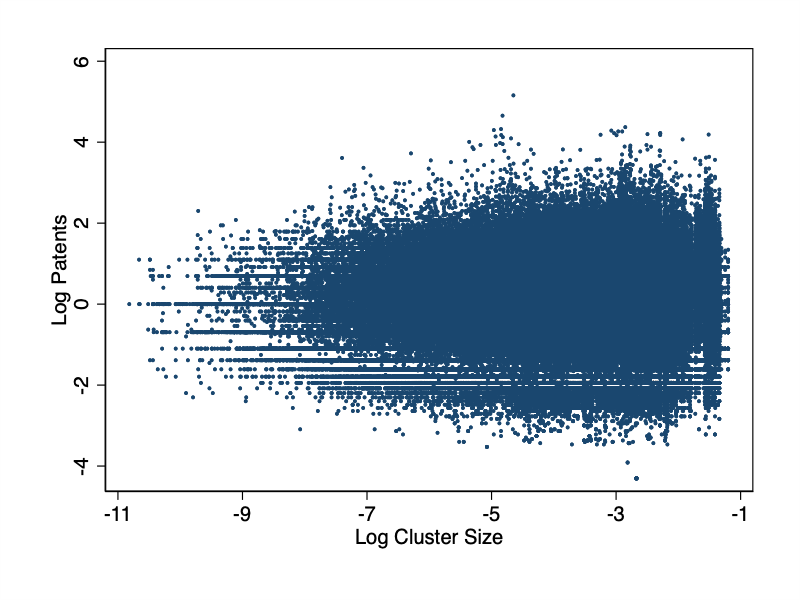}
\end{subfigure}\hfill
\begin{subfigure}[t]{0.5\columnwidth}
\centering	
\caption{\it Fig. 4 of Moretti (2021)}
\includegraphics[trim={0cm 1cm 0cm 1cm}, clip, scale=.275]{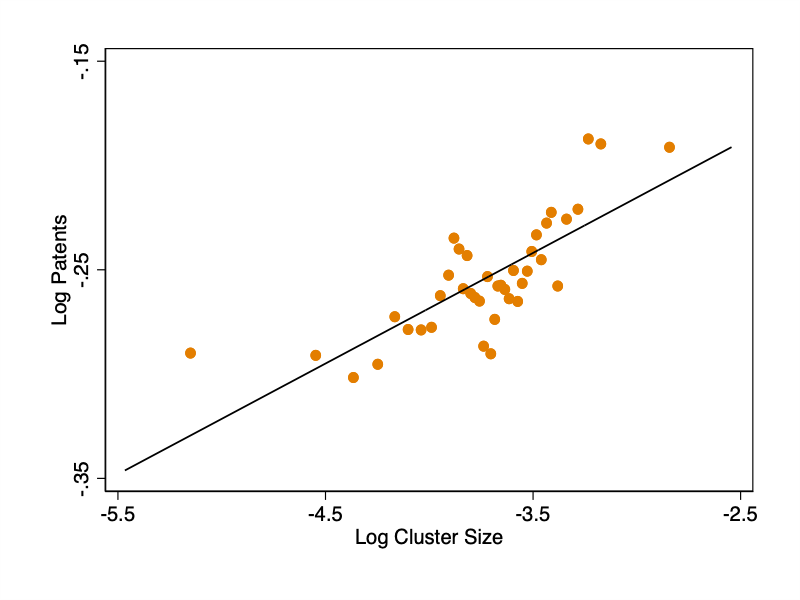}
\end{subfigure}\\[2 ex]
\begin{subfigure}[t]{0.5\columnwidth}
\centering	
\caption{\it Incorrect Residualization}
\includegraphics[trim={0cm 1cm 0cm 1cm}, clip, scale=.275]{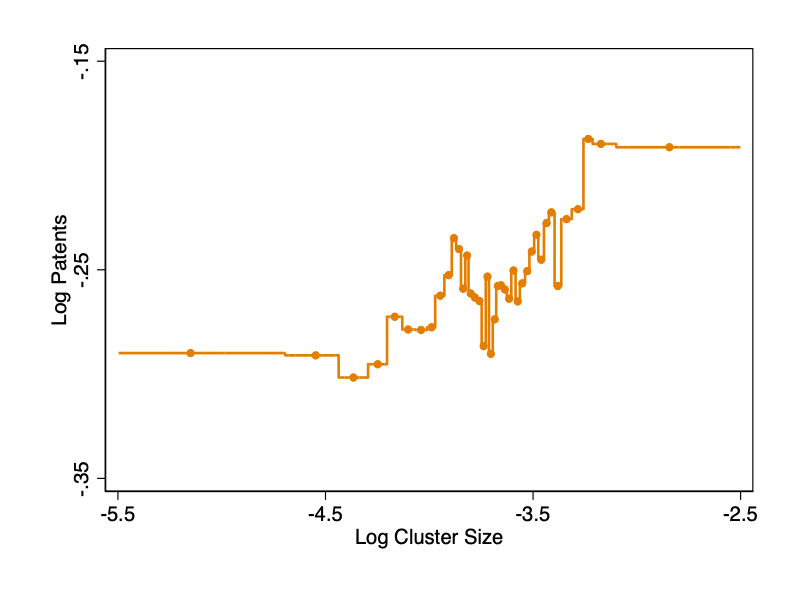}
\end{subfigure}\hfill
\begin{subfigure}[t]{0.5\columnwidth}
\centering	
\caption{\it Covariate Adjustment}
\includegraphics[trim={0cm 1cm 0cm 1cm}, clip, scale=.275]{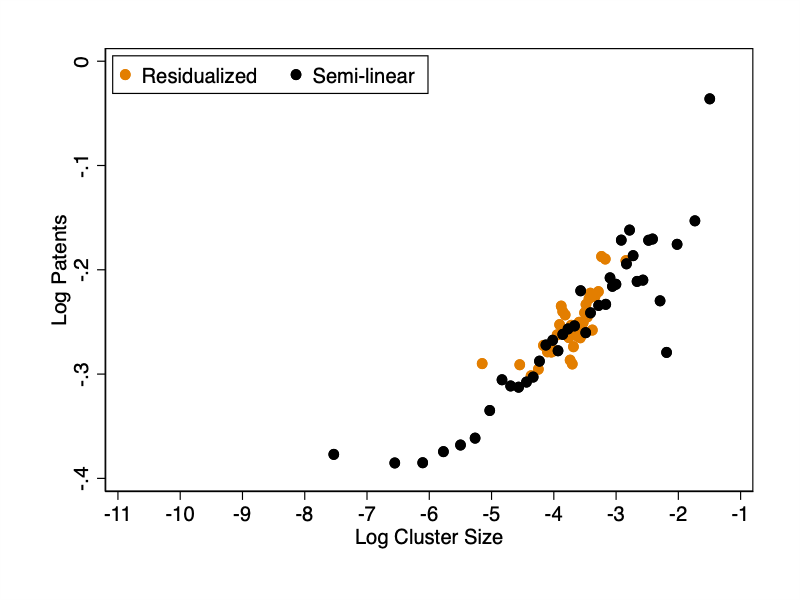}
\end{subfigure}\\[2 ex]
\begin{subfigure}[t]{0.5\columnwidth}
\centering	
\caption{\it Confidence Band}
\includegraphics[trim={0cm 1cm 0cm 1cm}, clip, scale=.275]{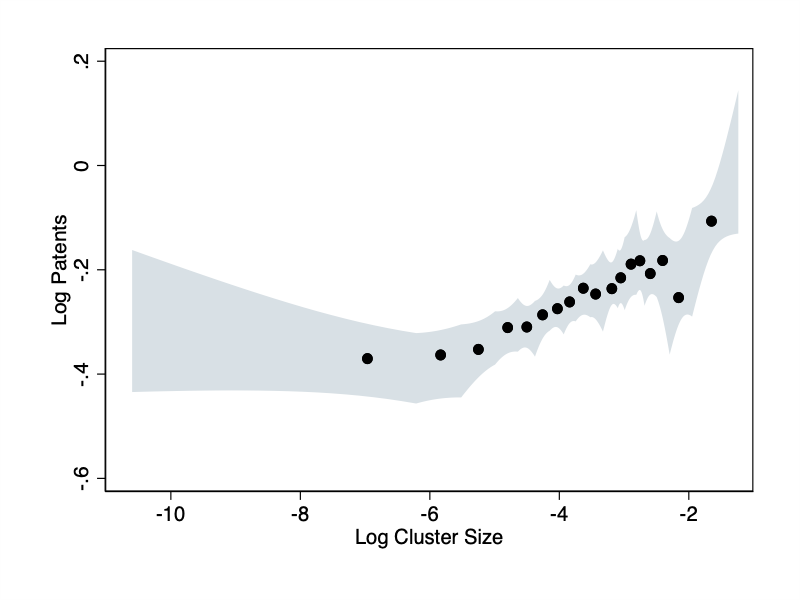}
\end{subfigure}\hfill
\begin{subfigure}[t]{0.5\columnwidth}
\centering	
\caption{\it Confidence Band (Full Specification)}
\includegraphics[trim={0cm 1cm 0cm 1cm}, clip, scale=.275]{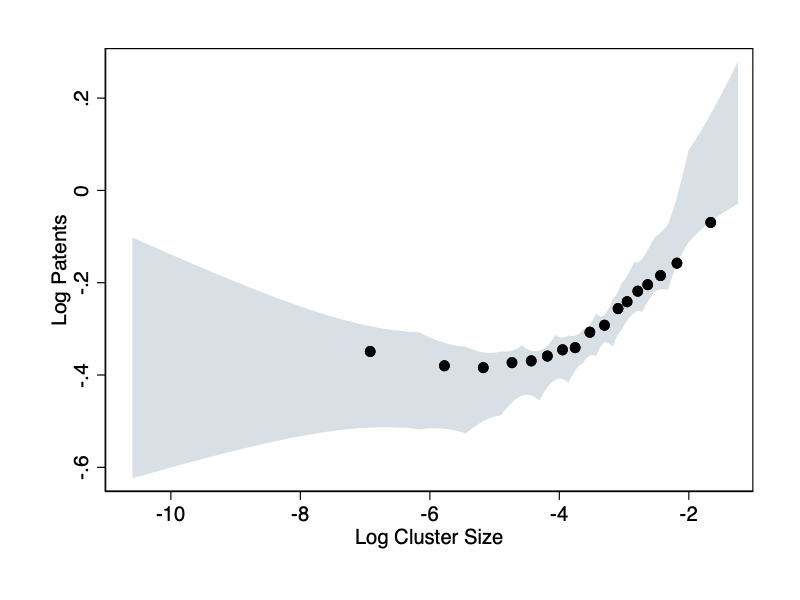}
\end{subfigure}\hfill
\end{figure}

We begin our analysis with a raw scatter plot of the data (top left of Figure \ref{fig:M_repl}).  With close to one million observations, the scatter plot is both dense and uninformative.  In the top right plot we replicate Figure 4 in \cite{Moretti2021_AER} which is a binned scatter plot controlling for year, research field, and city effects. It is intuitive to view and interpret this figure as one would a conventional scatter plot -- a cloud of points with a regression line fit to the ``data'' -- and we would conclude that there may be a positive but noisy relationship between these two variables. This interpretation is tempting, and indeed the very name ``binscatter'' invites this, but as previously discussed it is incorrect: the dots here are not data points but estimates of the conditional mean function. 

This is emphasized in Figure \ref{fig:M_repl}(c) which is the implied estimate of the conditional mean function.  This plot is \emph{formally identical} to the figure in the original paper (Figure \ref{fig:M_repl}(b)), but visually very different; moreover, assuming that the wiggly step function is well-approximated by a line seems inappropriate. However, there are two issues here: the incorrect residualization has been performed and the number of bins is too large, leading to substantial undersmoothing. Figure \ref{fig:M_repl}(d) addresses the former issue, applying our corrected approach to covariates overlaying the incorrectly residualized version now at the correct scale, making the difference starker. Correctly adjusting for covariates presents a much clearer picture of the empirical conclusions to be drawn from the data than do Figures \ref{fig:M_repl}(b) and \ref{fig:M_repl}(c).

This visual pattern is even more apparent in the bottom left plot where we utilize the IMSE-optimal choice of $J$ ($J_{\mathtt{IMSE}} = 18$).   Now, the point estimate of the conditional expectation function is thrown into sharper relief.  For smaller cluster sizes, the conditional expectation appears roughly flat whereas for larger cluster sizes, the estimate rises sharply.  This gives the appearance of a nonlinear relation between productivity and high-tech clusters.  We can formalize this conclusion by utilizing the associated confidence band also shown in Figure \ref{fig:M_repl}(e).  We clearly reject the null of no relationship between the variables as the confidence band does not contain a horizontal line.  Furthermore, we can also clearly reject linearity as no linear function can be wholly enveloped by the confidence band.  However, we  fail to reject convexity given the shape of the confidence band.  Taken in sum, these  results suggest a nonlinear relation between the number of patents and cluster size.  Figure \ref{fig:M_repl}(f) replicates this analysis for the main specification in \citet[Table 3, Columnn (8)]{Moretti2021_AER} which includes 11 different fixed effects.  We draw the same conclusions, with strong evidence against a linear functional form.  This added nuance to the results of \cite{Moretti2021_AER} obtained through our new tools is not inconsequential.  Taken at face value, it would imply that states and localities which have only small clusters of inventors might have to offer very generous incentives in order to grow their cluster size sufficiently large to generate the positive agglomeration effects presented in \cite{Moretti2021_AER}.

\section{Theoretical Foundations}
    \label{sec:theory}

The online Appendix reports our novel theoretical results for partitioning-based estimators with semi-linear covariate-adjustment and random binning based on empirical quantiles, which provide all the necessary econometric tools to formally study canonical and covariate-adjusted binscatter least squares methods. This section overviews those results, and discusses them in connection with the previous sections.

We study a covariate-adjusted estimator with more flexible basis functions allowing for polynomial fitting within bins and smoothness constraints across bins. The $p$-th order polynomial, $(s-1)$-times continuously differentiable, covariate-adjusted \textit{extended} binscatter estimator is
\begin{equation}
    \label{eq:binscatter-extended}
    \widehat{\mu}^{(v)}(x) = \widehat{\bb}_{p,s}^{(v)}(x)'\widehat{\bbeta}, \qquad
  \begin{bmatrix}\;\widehat{\bbeta}\;\\\;\widehat{\bgamma}\;\end{bmatrix}
  = \argmin_{\bbeta,\bgamma} \sum_{i=1}^{n} (y_i-\widehat{\bb}_{p,s}(x_i)'\bbeta-\bw_i'\bgamma)^2, \qquad 0 \leq v,s \leq p.
\end{equation}
where $\widehat{\bb}_{p,s}(x)=\widehat{\mathbf{T}}_s[\widehat{\bb}(x)\otimes(1,x,\dots,x^p)']$, $\widehat{\mathbf{T}}_s$ is a $[(p+1)J-(J-1)s]\times (p+1)J$ matrix of linear restrictions ensuring that the $(s-1)$-th derivative of the estimate is continuous, $\otimes$ denotes the Kronecker product, and $g^{(v)}(x)=\frac{\text{d}^v}{\text{d}x^v}g(x)$. (See Section SA-2 for further details.) For example, $s=1$ returns a continuous but nondifferentiable function ($\widehat{\mathbf{T}}_1$ constrains the polynomial fits within bins to be connected at the boundary of the bins), while $s=0$ gives a discontinuous function ($\widehat{\mathbf{T}}_0$ is the identity matrix). The form of $\widehat{\mathbf{T}}_s$ is given in the online Appendix, and it depends on the estimated quantiles. If $p=0$ (forcing $s=v=0$), then \eqref{eq:binscatter-extended} reduces to \eqref{eq:binscatter-covadj} because $\widehat{\bb}_{0,0}(x)=\widehat{\bb}(x)$ which is equivalent to  the Haar basis or a zero-degree spline. The additional generality of allowing for polynomial basis functions, beyond piecewise constant functions, is useful for estimating derivatives of the function of interest ($v>0$), as well as for reducing the smoothing bias of the estimator. The online Appendix treats the general case $0\leq v,s \leq p$, but in the paper we only consider $s=p$, with $p=0$ for binscatter estimation and $p\geq1$ for inference, and thus we set $\widehat{\bb}_{p}(x)=\widehat{\bb}_{p,p}(x)$ to simplify notation (and note that $\widehat{\bb}(x)=\widehat{\bb}_{0}(x)=\widehat{\bb}_{0,0}(x)$). More specifically, the implementations of robust bias correction discussed in Section \ref{sec:Quantifying Uncertainty} sets $(p,s,v)=(1,1,0)$. 

The following assumption gives a simplified version of the conditions imposed in the online Appendix.
\begin{assumption}
    \label{ass:main}
    The sample $(y_{i}, x_{i}, \bw_{i}')$, $i=1,2,\dots,n$, is i.i.d. and satisfies \eqref{eq:partially linear model}. The functions $\mu_0(x)$ and $\E[\bw_i|x_i=x]$ are $(p+2)$-times continuously differentiable. The covariate $x_{i}$ has a Lipschitz continuous density function $f_X(x)$ bounded away from zero on the compact support $\mathcal{X}$. The minimum eigenvalue of $\V[\bw_i|x_i=x]$ is uniformly bounded away from zero. For $\epsilon_{i}=y_i - \mu_0(x_i) - \bw_i'\bgamma_0$, $\sigma^2(x)=\E[\epsilon_{i}^2|x_{i}=x]$ is Lipschitz continuous and bounded away from zero, and $\E[\|\bw_i\|^4|x_i=x]$, $\E[\epsilon_{i}^4|x_{i}=x]$, and $\E[\epsilon_{i}^2|x_{i}=x,\bw_i=\bw]$ are uniformly bounded, where $\|\cdot\|$ is the Euclidean norm.
\end{assumption}

Section SA-3.1 presents new technical lemmas for random partitions based on empirical quantiles. Those results include general characterizations of the ``regularity'' of the random partitioning scheme (Lemmas SA-3.1 and SA-3.2) and of the associated random basis functions (Lemmas SA-3.3 and SA-3.4). These results give sharp control on the underlying random binning scheme of binscatter methods. 

Sections SA-3.2--SA-3.7 study large sample point estimation and distributional properties of the extended covariate-adjusted binscatter estimator. Preliminary technical results include: (i) technical lemmas for the Gram matrix (Lemma SA-3.5), asymptotic variance (Lemmas SA-3.6 and SA-3.7), approximation error (Lemma SA-3.8), and covariate adjustments (Lemma SA-3.9); (ii) stochastic linearization and uniform convergence rates (Theorem SA-3.1 and Corollary SA-3.1) and variance estimation (Theorem SA-3.2); and (iii) pointwise distributional approximation (Theorem SA-3.3). All these results explicitly account for the random binning scheme.

Using our new technical results, Section SA-3.5 also establishes a density-weighted IMSE expansion of the binscatter estimator (Theorem SA-3.4). Letting $\mathsf{IMSE}[\widehat{\Upsilon}^{(v)}] = \int \E[(\widehat{\Upsilon}^{(v)}(x) - \Upsilon^{(v)}_0(x))^2|x_1, \dots, x_n, \bw_1, \dots, \bw_n] f_X(x) dx$, a simplified version of our general result follows. 

\begin{thm}[IMSE]
    \label{thm:IMSE}
	Let Assumption \ref{ass:main} hold, $0\leq v \leq p$, $J\log(J)/n\to0$, and $nJ^{-4p-5}\to0$.
 Then, $\mathsf{IMSE}[\widehat{\Upsilon}^{(v)}] = \frac{J^{1+2v}}{n} \mathscr{V}_n(p,s,v) + J^{-2(p+1-v)} \mathscr{B}_n(p,s,v) + o_\P( \frac{J^{1+2v}}{n} + J^{-2(p+1-v)})$, where $\mathscr{V}_n(p,s,v)$ and $\mathscr{B}_n(p,s,v)$ are non-random, $n$-varying bounded sequences (see Section SA-3.5).
\end{thm}

Optimizing the leading terms over $J$ gives the optimal choice $J_{\mathtt{IMSE}}(p,s,v)$, and specializing it to $p=s=v=0$ gives \eqref{eq:J IMSE}. Feasible IMSE-optimal tuning parameter selection is discussed in Section SA-4. All these results explicitly account for the random binning scheme and the covariate adjustment.

Section SA-3.6 reports our most noteworthy novel technical result: a conditional strong approximation for the extended binscatter estimator, which circumvents a fundamental lack of uniformity of the random binning basis $\widehat{\bb}_{p}(x)$, while still delivering a sufficiently fast uniform coupling,  requiring only $J^2/n \to 0$ (up to $\log(n)$ terms). In fact, if a subexponential moment restriction holds for $\epsilon_{i}$, it suffices that $J/n\rightarrow 0$ (up to $\log(n)$ terms). Our rate conditions not only improve on previous results in the literature, but also allow for canonical binscatter (i.e., there exists a sequence $J\to\infty$ such that bias and variance are simultaneously controlled even when $p=s=0$).
    
The starting point is the Studentized $t$-statistic that centers and scales the extended binscatter estimator $\widehat{\Upsilon}^{(v)}(x)=\widehat{\mu}^{(v)}(x)+\I(v=0)\bar{\bw}'\widehat{\bgamma}$ of the extended parameter of interest $\Upsilon_0^{(v)}(x)=\mu_0^{(v)}(x)+\I(v=0)\E[\bw_i]'\bgamma_0$. We index important objects with $p$ (recall that $s=p$ in the paper, but the online Appendix treats the general case). We study the $t$-statistic 
\begin{equation*}
    \label{eq:T_p(x)}
    T_p(x) = \frac{\widehat{\Upsilon}^{(v)}(x) - \Upsilon^{(v)}_0(x)}{\sqrt{\widehat{\Omega}(x)/n}},
\end{equation*}
where $\widehat{\Omega}(x) = \widehat{\bb}_p^{(v)}(x)'\widehat{\bQ}^{-1}\widehat{\bSigma}\widehat{\bQ}^{-1}\widehat{\bb}^{(v)}_p(x)$, $\widehat{\bQ} = \frac{1}{n}\sum_{i=1}^n \widehat{\bb}_p(x_i)\widehat{\bb}_p(x_i)'$, and $\widehat{\bSigma} = \frac{1}{n}\sum_{i=1}^n \widehat{\bb}_p(x_i)\widehat{\bb}_p(x_i)' (y_{i} -\widehat{\bb}_p(x_i)'\widehat{\bbeta} - \bw_i'\widehat{\bgamma})^2$. We seek a distributional approximation for the entire stochastic process $(T_p(x):x\in\mathcal{X})$ because this allows us to study the visualization and econometric properties of the entire binscatter fit $(\widehat{\Upsilon}^{(v)}(x):x\in\mathcal{X})$ simultaneously. Using this strong approximation we can compute the critical values for valid confidence bands and hypothesis testing. Our approach gives a simple, tractable method for computing critical values based on random draws from the Gaussian distribution.

The randomness of the partition $\widehat{\Delta}$ (which is inherited by the basis functions themselves) is not just ruled out by the assumptions of prior work, but rather it is not even possible to obtain a valid strong approximation for the entire stochastic process $(T_p(x):x\in\mathcal{X})$ exactly because this  randomness causes uniformity to fail. As an alternative, we establish a conditional Gaussian strong approximation as the key building block for uniform inference. Heuristically, our strong approximation begins by establishing the following two approximations uniformly over $x\in\mathcal{X}$:
\begin{align*}
    \sqrt{n}\big(\widehat{\Upsilon}^{(v)}(x) - \Upsilon^{(v)}_0(x)\big)
    & \; \approx_\P \; \widehat{\bb}_p^{(v)}(x)'\widehat{\bQ}^{-1}\frac{1}{\sqrt{n}}\sum_{i=1}^n \widehat{\bb}_p(x_i) \epsilon_i\\
    & \; \approx_{\rm{d}} \; \widehat{\bb}^{(v)}_p(x)'\widehat{\bQ}^{-1}\widehat{\bSigma}^{1/2}\bN_{p+J}^\star,
\end{align*}
where $\bN_{p+J}^\star$ denotes a $(p+J)$-dimensional standard Gaussian random vector, independent of the data. The first approximation is a stochastic linearization (Theorem SA-3.1) and directly implies the variance formula $\widehat{\Omega}(x)$. This step is reminiscent of standard least squares algebra. The second approximation corresponds to a conditional coupling (Theorems SA-3.5 and SA-3.6). It is not difficult to show that $\widehat{\bQ}$ and $\widehat{\bSigma}$ are sufficiently close in probability to well-defined non-random matrices in the necessary norm (Lemma SA-3.5 and Theorem SA-3.2). However, $\widehat{\bb}_p^{(v)}(x)$ fails to be close in probability to its non-random counterpart \textit{uniformly} in $x\in\mathcal{X}$ due to the sharp discontinuity introduced by the indicator functions entering the binning procedure. Nevertheless, inspired by the work in \cite{Chernozhukov-Chetverikov-Kato_2014a_AoS,Chernozhukov-Chetverikov-Kato_2014b_AoS}, our approach circumvents that technical hurdle by first developing a strong approximation that is conditionally Gaussian, retaining some of the randomness introduced by $\widehat{\Delta}$, and then using such coupling to deduce a distributional approximation for  specific functionals of interest (e.g., suprema); see Section SA-3.6 for details.

We state the formal results in two steps: the first derives an infeasible strong approximation and the second shows that, given the data, a feasible version can be constructed.

\begin{thm}[Feasible Strong Approximation]
    \label{thm:strong approximation}
    Let Assumption \ref{ass:main} hold and let $\{a_n: n\geq 1\}$ be a sequence of non-vanishing constants such that $n^{-1/2}J(\log J)^2+J^{-1}+nJ^{-2p-3}=o(a_n^{-2})$.
	Then, on a properly enriched probability space, there exists a standard Gaussian random vector $\bN_{p+J}$, of length $p+J$, such that for any $\xi>0$,	
	\[
    	\P\Big(\sup_{x\in\mathcal{X}}|T_p(x)-Z_p(x)|>\xi a_n^{-1}\Big)=o(1), \qquad
    	Z_p(x)=\frac{\widehat{\bb}_p^{(v)}(x)'\bQ_0^{-1}\bSigma_0^{1/2}}{\sqrt{\Omega(x)}}\bN_{p+J}.
	\]
    Also, there exists a standard Gaussian random vector $\bN_{p+J}^\star$, of length $p+J$, independent of the data $\bD=\{(y_{i}, x_{i}, \bw_{i}'): i=1,2,\dots,n\}$, such that for any $\xi>0$,
	\[
    	\P\Big(\sup_{x\in\mathcal{X}}|\widehat{Z}_p(x)-Z_p(x)|>\xi a_n^{-1}\Big|\bD\Big)=o_\P(1),
    	\qquad
    	\widehat{Z}_p(x)=\frac{\widehat{\bb}_p^{(v)}(x)'\widehat{\bQ}^{-1}\widehat{\bSigma}^{1/2}}{\sqrt{\widehat{\Omega}(x)}}\bN_{p+J}^\star.
	\]
\end{thm}

This result forms the basis of the inference tools proposed in our paper. In principle, we can now approximate the distribution of any functional of the $t$-statistic process $T_p(x)$ using a plug-in approach based on $\widehat{Z}_p(x)$. This prescription is easy to put into practice, because it depends only on Gaussian draws and the already-computed elements $\widehat{\bb}_p(x)$, $\widehat{\bQ}$, $\widehat{\bSigma}$, and $\widehat{\Omega}(x)$, and therefore the process $\widehat{Z}_p(x)$ is simple to simulate. For example, the distribution of $\sup_{x\in\mathcal{X}} \big|T_{p}(x)\big|$ is well approximated by that of $\sup_{x\in\mathcal{X}} \big|\widehat{Z}_p(x)\big|$, conditional on the data, and we can use this to obtain critical values for testing or forming confidence bands. 

However, and crucially for applied practice, one must choose $J$ such that the approximation is valid. In addition, ideally, the choice of $J$ would be optimal in some way and the resulting inference would be robust to small fluctuations in $J$. The IMSE-optimal choice $J_{\mathtt{IMSE}}(p,s,v)$ cannot be directly used, as it is too ``small'' to remove enough bias for the $t$-statistic $T_p(x)$ to be correctly centered. Feasible implementation of $J_{\mathtt{IMSE}}(p,s,v)$ would also require additional smoothness assumptions, rendering the resulting point estimator $\widehat{\Upsilon}^{(v)}(x)$ suboptimal from a point estimation minimax perspective \citep{Tsybakov_2009_Book}. Different approaches for tuning parameter selection are available in the literature, including undersmoothing or ignoring the bias \citep{hall2001bootstrapping}, bias correction \citep{hall1992effect}, robust bias correction \citep{Calonico-Cattaneo-Farrell_2018_JASA,Calonico-Cattaneo-Farrell_2022_Bernoulli}, and Lepskii's method \citep{Lepski-Spokoiny_1997_AOS,birge2001alternative}. In this paper, we employ robust bias correction based on an IMSE-optimal binscatter, that is, without altering the partitioning scheme $\widehat{\Delta}$ used. This inference approach is easy to implement and more robust to the choice of $J$: for a choice of $p$, we construct the binscatter (point) estimate $\widehat{\Upsilon}^{(v)}(x)$ based on the random binning $\widehat{\Delta}$ using the (feasible) method of Section \ref{sec:IMSE}, and then for inference we employ $T_{p+1}(x)$. Thus, in Section \ref{sec:Quantifying Uncertainty}, we set $J=J_{\mathtt{IMSE}}(0,0,0)$, $p=s=1$, $v=0$, $\widehat{\Upsilon}_{\tt BC}(x) = \widehat{\Upsilon}(x)$, $\widehat{\Omega}_{\tt RBC}(x) = \widehat{\Omega}(x)$, and $\cval_{\tt RBC} = \inf\big\{c\in\mathbb{R}_+:\P\big[ \sup_{x\in\mathcal{X}} |\widehat{Z}_{1}(x)| \leq c \;\big| \;\bD \big]\geq 1-\alpha \big\}$.

All our results explicitly account for the random binning scheme and the semi-linear covariate-adjustment with random evaluation point. Another noteworthy novel result in Section SA-3.6 is the proof technique to transform our strong approximation results (Theorem SA-3.5), and their feasible versions (Theorem SA-3.6), into statements about the Kolmogorov distance for the suprema and related functionals of the $t$-statistic processes of interest (Theorem SA-3.7). Our technical approach again circumvents a fundamental lack of uniformity of the random binning basis $\widehat{\bb}_{p}^{(v)}(x)$, while still delivering a sufficiently fast uniform coupling, requiring only $J^2/n \to 0$ (up to $\log(n)$ terms). Our proof technique can also be used to analyze other functionals such as the $L_p$ distance, Kullback–Leibler divergence, and $\argmax$ statistic.

Finally, from a theoretical point of view, the rate conditions of Theorem \ref{thm:strong approximation} are seemingly minimal and improve on prior results. In fact, it can be shown that when $a_n=\sqrt{\log n}$ and a subexponential moment restriction holds for the error term, it suffices that  $J/n=o(1)$, up to $\log n$ terms. In contrast, a strong approximation of the $t$-statistic process for general series estimators was obtained based on Yurinskii coupling in \cite{Belloni-Chernozhukov-Chetverikov-Kato_2015_JoE}, which requires $J^5/n=o(1)$, up to $\log n$ terms. Alternatively, a strong approximation of the \emph{supremum} of the $t$-statistic process can be obtained under weaker rate restrictions, such as the requirement of $J/n^{1-2/\nu}=o(1)$ used by \cite{Chernozhukov-Chetverikov-Kato_2014a_AoS}, up to $\log n$ terms, where $\nu$ is related to the moment assumptions imposed in the online Appendix, but their result applies exclusively to the suprema of the stochastic process. Our theoretical improvements have direct practical consequences as the rate conditions are weak enough to accommodate the canonical binscatter (i.e., the piecewise constant $p=0$ estimator), which would otherwise not be possible. See the online Appendix for more details.

\section{Conclusion}
    \label{sec:conclusion}

Data visualization is a powerful device for effectively conveying empirical results in a simple and intuitive form.  Binned scatter plots have become a popular tool to present a flexible, yet cleanly interpretable, estimate of the relationship between an outcome and a covariate of interest. However, despite their visual simplicity and conceptual appeal, there has been no work to establish that they provide a high-quality, or even accurate, visualization of the data. This hampers their reliability and usability in applications.

We introduce a suite of formal and visual tools based on binned scatter plots to improve, and in some cases correct, empirical practice. Our methods offer novel visualization tools, principled covariate adjustment, estimation of conditional mean functions, visualization of variance and precise uncertainty quantification, and tests of hypotheses such as linearity or monotonicity. We illustrate our methods with two substantive empirical applications, revisiting recently published papers \citep{Akcigitetal2021_QJE,Moretti2021_AER} in economics, and show, in particular, the pitfalls of employing binned scatter methods incorrectly in practice. Further, our empirical reanalysis showcases how applying binned scatter plots correctly can strengthen the empirical findings in those papers. All of our results are fully implemented in publicly available software \citep{Cattaneo-Crump-Farrell-Feng_2023_Stata}.

In this paper our focus is on binned scatter plots, and hence the case of a scalar variable $x_i$. However, all of our results (including covariate adjustment) extend immediately to cover the case where  $\dim(x_i)>1$. One important application is a heat map, which is used in applied work to show some feature of the conditional distribution of $y_i$ (the ``heat'') given positioning in two-dimensional space (the ``map'').  For recent examples, see \cite{Crawford-Shcherbakov-Shum2019_AER} and \cite{GHSS2022}. Finally, the results herein cover conditional means only, while \cite{Cattaneo-Crump-Farrell-Feng_2023_Generalized} treats nonlinear settings such as conditional quantiles and other nonlinear features.

\clearpage
\section*{References} 
\begingroup
\renewcommand{\section}[2]{}	
\bibliography{CCFF_2024_AER--bib}
\bibliographystyle{aer}
\endgroup

\end{document}


\title{\vspace{-0.0in}Online Appendix\\
	On Binscatter\thanks{Cattaneo gratefully acknowledges financial support from the National Science Foundation through grants SES-1947805, SES-2019432, and SES-2241575. Feng gratefully acknowledges financial support from the National Natural Science Foundation of China (NSFC) through grants 72203122, 72133002, and 72250064. The views expressed in this supplemental appendix are those of the authors and do not necessarily reflect the position of the Federal Reserve Bank of New York or the Federal Reserve System.}
\bigskip }
\author{Matias D. Cattaneo\thanks{Department of Operations Research and Financial Engineering, Princeton University.} \and
	    Richard K. Crump\thanks{Macrofinance Studies, Federal Reserve Bank of New York.} \and
	    Max H. Farrell\thanks{Department of Economics, UC Santa Barbara.} \and 
	    Yingjie Feng\thanks{School of Economics and Management, Tsinghua University.}}
\date{March 15, 2024}
\maketitle

\begin{abstract}
	This supplement presents additional methodological results, general theoretical results encompassing those reported in the paper, and all technical proofs. Our new theoretical results for least squares partitioning-based semi-linear series estimation and inference are of independent interest. Companion general-purpose software and replication files are available at \url{https://nppackages.github.io/binsreg/}.
\end{abstract}
\thispagestyle{empty}

\clearpage

\singlespacing
\addtocontents{toc}{\protect\thispagestyle{empty}}
\setcounter{tocdepth}{2}
\tableofcontents
\thispagestyle{empty}

\clearpage
\doublespacing
\setcounter{page}{1}

\section{Additional Methodological Results}

We discuss two important issues related to the results in the main text. First, building on Section I.A, we provide two simple and stylized analytical examples which explicitly characterize the effect of using the incorrect covariate adjustment for binscatter. Second, building on Section I.B, we discuss the role of the choice of the evaluation point $\bw$ for visualization, estimation, and inference for $\Upsilon_0(x,\bw)=\E[y_i|x_i=x,\bw_i=\bw]$.

\subsection{Bias of Residualized Binscatter}\label{SA section: residualized binscatter}

We present two examples to showcase the potential problems resulting from the incorrect residualization method. In the following we use $m{!}{!}$ to denote the double factorial of a number $m$, $\mathsf{U}(a,b)$ to denote the uniform distribution on $[a, b]$ and $\mathsf{Bernoulli}(p)$ to denote the Bernoulli distribution with mean equal to $p$.

\subsubsection{Example 1: Gaussian Polynomial Regression Model}

Suppose that for some integer $m>1$,
\[y_i=x_i^m+w_i\gamma_0+\epsilon_i, \qquad \gamma_0=0, \qquad
  \begin{bmatrix}
    x_i\\
    w_i\\
    \epsilon_i
\end{bmatrix}
\thicksim \mathsf{Normal}\left(
\begin{bmatrix}
    0\\0\\0
\end{bmatrix},
\begin{bmatrix}
    \sigma_x^2&\rho\sigma_x&0\\
    \rho\sigma_x&1&0\\
    0&0&\sigma^2
\end{bmatrix}
\right).
\]
Thus, using the notation in the paper, $\mu_0(x_i)=x_i^m$ and $\bw_i$ is scalar ($d=1$).

Residualizing the covariate $x_i$ with respect to the control $w_i$ in this Gaussian model yields
\[x_i-\text{L}(x_i|w_i)=x_i-\rho\sigma_xw_i,\]
The residualized covariate $x_i-\rho\sigma_xw_i$ is still supported on the whole real line, but its variance shrinks to $(1-\rho^2)\sigma_x^2$. In addition, residualizing the outcome $y_i$ with respect to $w_i$ yields 
\[y_i-\text{L}(y_i|w_i) = y_i-(1\quad w)(\E[x_i^m] \quad \E[x_i^m w_i])' = y_i-\alpha_0-\alpha_1 w_i \]
where
\[
\alpha_0=\begin{cases}
    0 & \text{if $m$ is odd}\\
    \sigma_x^m(m-1){!}{!} & \text{if $m$ is even}
\end{cases} \quad\text{and}\quad 
\alpha_1=\begin{cases}
    m\rho\sigma_x^{m}(m-2){!}{!} & \text{ if $m$ is odd}\\
    0 & \text{ if $m$ is even}
\end{cases}.
\]
Then, letting $z_i=x_i-\rho\sigma_xw_i$, we have
\[\E[y_i-\text{L}(y_i|w_i)|x_i-\text{L}(x_i|w_i)] = \E[x_i^m-\alpha_0-\alpha_1w_i|z_i] = \E[x_i^m|z_i]-\alpha_0.\]
Note that $x_i|z_i\thicksim\mathsf{N}(z_i, \rho^2\sigma_x^2)$. Then, we can concisely write
\[
\E[x_i^m|z_i]=\sum_{0\leq l\leq m \atop m-l \text{ is even}}\binom{m}{l}z_i^{l}|\rho\sigma_x|^{m-l}(m-l-1){!}{!}.
\]
For instance, if the true underlying model is a quadratic regression model ($m=2$) we obtain
\[\E[y_i-\text{L}(y_i|w_i)|z_i]=(\rho^2-1)\sigma_x^2+z_i^2 \qquad\qquad (\text{for } m=2 ),\]
while for a cubic regression model ($m=3$) we obtain
\[\E[y_i-\text{L}(y_i|w_i)|z_i]=3\rho^2\sigma_x^2z_i+z_i^3 \qquad\qquad (\text{for } m=3).\]
Clearly, for $m=2$, the residualization leads to a vertical shift of the true function (quadratic monomial).
For $m=3$, however, the problem is more severe: residualization adds a linear function of the covariate to the true function (cubic monomial), and when $|\rho\sigma_x|$ is large, the linear component $3\rho^2\sigma_x^2z_i$ will visually dominate in a binscatter plot, leading to an incorrect ``linear'' specification of the model. Moreover, in any sample, this effect is likely to be amplified because $z_i$ is more concentrated around its mean than $x_i$ is.

Using the above results, we can even obtain the functional form of the residualized binscatter when $\mu_0$ is any polynomial function and all variables are multivariate normal. Generally, the residualized binscatter yields a polynomial relationship between the residualized $y_i$ and the residualized $x_i$ that may be different from the original polynomial $\mu_0$.

\subsubsection{Example 2: Semiparametric Bernoulli Model}

Suppose that
\[y_i = \mu_0(x_i)+w_i\gamma_0+\epsilon_i, \qquad \gamma_0=0,\]
where 
\[w_i\thicksim \textsf{Bernoulli}(0.5),\qquad 
    x_i|w_i=0\thicksim \textsf{U}(0,1), \qquad 
    x_i|w_i=1\thicksim\textsf{U}(1,2),\qquad
    \epsilon_i \ind  (x_i, w_i).\]

It follows that $x_i\thicksim\mathsf{U}(0,2)$. Residualizing the covariate $x_i$ with respect to $w_i$ yields
\[
x_i-\text{L}(x_i|w_i)=x_i-0.5-w_i\in
\left[-0.5,0.5\right].
\]
The support of this residualized covariate is different from that of the original one, not only in the location but also in the length.

In addition, residualizing the outcome $y_i$ with respect to $w_i$ yields 
\[
y_i-\text{L}(y_i|w_i)=y_i-\alpha_0-\delta_0 w_i
\]
where $\alpha_0=\E[\mu_0(x_i)|w_i=0]$, and $\delta_0=\E[\mu_0(x_i)|w_i=1]-\E[\mu_0(x_i)|w_i=0]$.
Then, letting $z_i=x_i-0.5-w_i$, we have 
\begin{align*}
    &\E[y_i-\text{L}(y_i|w_i)|x_i-\text{L}(x_i|w_i)]\\
    &\qquad = \E[y_i-\alpha_0-\delta_0w_i|z_i]\\
    &\qquad = (\mu_0(z_i+0.5)-\alpha_0)\times\P(w_i=0|z_i)+
    (\mu_0(z_i+1.5)-\alpha_0-\delta_0)\times\P(w_i=1|z_i)\\
    &\qquad = \frac{1}{2} \mu_0(z_i+0.5)+\frac{1}{2}\mu_0(z_i+1.5)-\alpha_0-\frac{1}{2}\delta_0.
\end{align*}
Ignoring the constants, the residualized binscatter in this example characterizes a linear combination of two ``horizontally shifted'' versions of the true function $\mu_0(\cdot)$, which in general can be very different from the original $\mu_0(\cdot)$. For instance, consider 
\[
\mu_0(x)=x^2\I(x\in[0,1))+(2-(x-2)^2)\I(x\in[1,2]),
\]
which is continuously differentiable. This specification actually implies that $y_i$ and $x_i$ have a quadratic relationship which is heterogeneous across the two groups with $w_i=0$ and $w_i=1$. However, the residualized binscatter yields
\[\E[y_i-\text{L}(y_i|w_i)|x_i-\text{L}(x_i|w_i)] = z_i+1-\alpha_0-\frac{1}{2}\delta_0
\]
which becomes a linear function in $z_i$, thereby giving a (visually and theoretically) wrong functional form for the true underlying conditional expectation.

\subsection{Impact of Evaluation Point \texorpdfstring{$\bw$}{\textbf{w}} }\label{SA section: role of w point}

This supplemental appendix will focus on estimation and inference for the conditional expectation $\Upsilon_0(x,\bw)=\E[y_i|x_i=x,\bw_i=\bw]$ and its derivatives with respect to $x$, where $\bw$ is a user-specified value of control variables at which $\Upsilon_0(x,\cdot)$ is evaluated, such as $\bw=\mathbf{0}$, $\E[\bw_i]$, or $\text{median}(\bw_i)$, where $\mathbf{0}$ denotes a vector of zeros and $\text{median}(\bw_i)$ denotes the population median of each component in $\bw_i$. In the paper attention is restricted to $\Upsilon_0(x)=\Upsilon_0(x,\E[\bw_i])$. In this section we provide a detailed discussion regarding the role of the evaluation point $\bw$, which may be important for interpretation and for numerical results, and even for the visualization itself.

One might expect that since the additional controls are modeled as additively linear, the evaluation point $\bw$ (and the coefficient $\bgamma_0$) should not impact conclusions about the nonparametric relationship between $y$ and $x$. But this intuition overlooks the fact that the function $\mu_0(x)$ is only defined relative to how $\bw_i$ is coded.
We will show that the results of parametric specification tests and confidence bands for the mean function $\Upsilon_0(x,\bw)$ might be sensitive to the choice of $\bw$, and how this issue may be circumvented by focusing instead on the derivative of the mean function, highlighting the importance of our theoretical contributions which can accommodate the estimation of derivatives.

Let us first consider the hypothesis testing procedure behind the informal practice of checking if the ``dots'' are roughly linear, and then running ordinary least squares regression of $y_i$ on $x_i$ and  $\bw_i$. This idea motivates the standard practice of plotting the fitted regression line along with the binned scatter plot, as in Figures 1 and 2 in the paper. 
In this case, the null hypothesis is \emph{not} merely that $\mu_0(x) = \theta_0 + \theta_1 x$, i.e., a linear function, but rather that the full model is linear, so that $\Upsilon_0(x,\bw)=\theta_0 + x\theta_1 + \bw'\bgamma_0$.
Under the partially linear assumption of the model \eqref{SA eq:model-withcovs} below, these would seem identical, because in either case $\bw$ enters linearly. But this is not so in practice for two reasons: the estimates of the coefficients $\bgamma_0$ will differ in general, as will the implied intercepts, and the chosen $\bw$ will impact the uncertainty about the estimate of $\theta_0$. 

\begin{figure}[h!]
\caption{\footnotesize{\bf Role of the Evaluation Point.} This figure demonstrates that the choice of $\widehat{\bw}$ shifts both the absolute position (note the $y$ axis) of the visualization and estimator, but also affects the comparison to parametric fits. The data is the same as in Figure 2 in the paper except that state and year fixed effects are omitted for simplicity.}
\label{fig:line away from dots}
\begin{subfigure}[t]{0.33\columnwidth}
\centering	
\caption{\it $\widehat{\bw} = \bw_{min}$}
\includegraphics[trim={0cm 1cm 0cm 1cm}, clip, scale=.2]{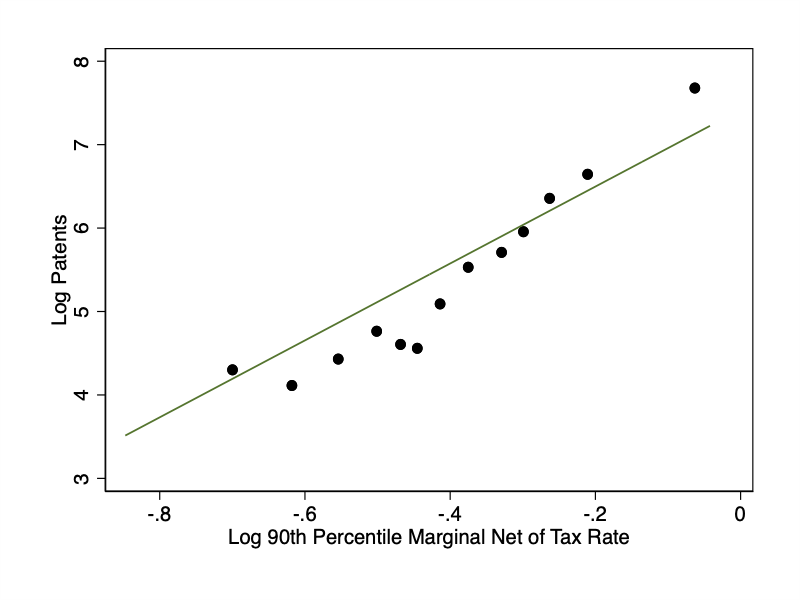}
\end{subfigure}\hfill
\begin{subfigure}[t]{0.34\columnwidth}
\centering	
\caption{\it $\widehat{\bw} = \bar{\bw}$}
\includegraphics[trim={0cm 1cm 0cm 1cm}, clip, scale=.2]{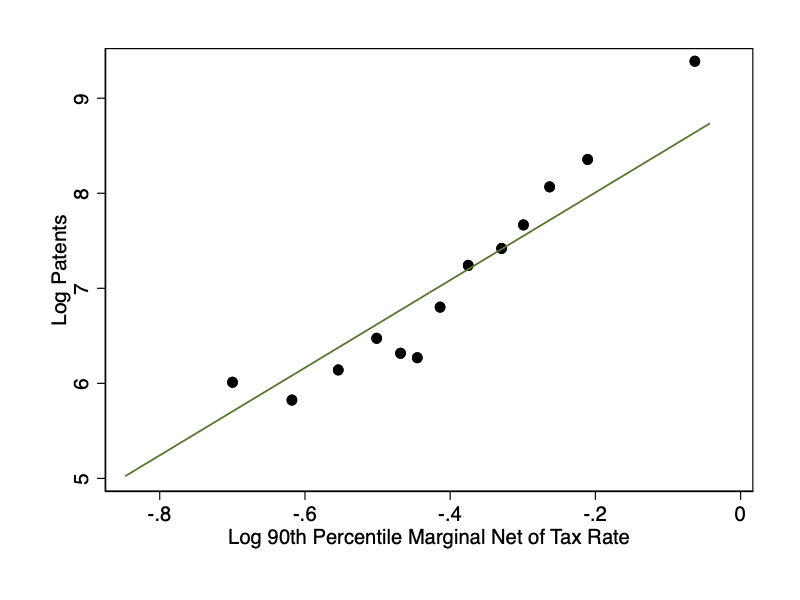}
\end{subfigure}\hfill
\begin{subfigure}[t]{0.33\columnwidth}
\centering	
\caption{\it $\widehat{\bw} = \bw_{max}$}
\includegraphics[trim={0cm 1cm 0cm 1cm}, clip, scale=.2]{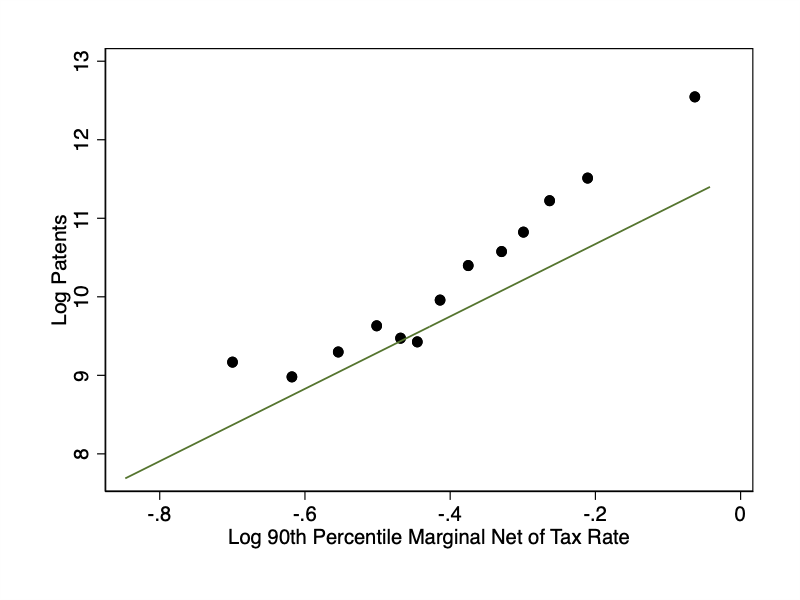}
\end{subfigure}\hfill
\end{figure}

In a standard binscatter plot such as Figure 1 in the paper, the ``dots'' show the semiparametric estimate $\widehat{\Upsilon}(x,\widehat\bw) = \widehat{\mu}(x) + \widehat{\bw}'\widehat{\bgamma}$, defined in \eqref{SA eq:binscatter-QMLE} below, while the plotted line is the parametric fit $\widetilde{\theta}_0 + x\widetilde{\theta}_1 + \widehat{\bw}'\widetilde{\bgamma}$, obtained from least squares regression. Thus, while we are only interested in assessing the linearity of $\mu_0(x)$, we are \emph{actually} testing these two functional forms for $\Upsilon_0(x,\bw)$, and the fact that $\widehat{\bgamma} \neq \widetilde{\bgamma}$ becomes important. Moreover, because $\widetilde{\theta}_0 + x\widetilde{\theta}_1 + \widehat{\bw}'\widetilde{\bgamma}$ is a global parametric fit while $\widehat{\Upsilon}(x,\widehat{\bw}) = \widehat{\mu}(x) + \widehat{\bw}'\widehat{\bgamma}$ is local and nonparametric, the implied intercept when plotted depends on the chosen $\widehat{\bw}$, and this can shift the line away from the dots. Figure \ref{fig:line away from dots} demonstrates this by example: everything is identical between the three plots except for choice of $\widehat{\bw}$. Notice the shift in absolute position (note the $y$ axis) and the change in the relative position of the line and the binscatter. This phenomenon is unavoidable in this setting, and the user must select $\widehat{\bw}$ appropriately. (Note that this does not occur when using the incorrect residualization because the covariates are mishandled.)

Beyond the visual inspection of a plot like Figure \ref{fig:line away from dots}, 
we can also consider a formal test for the hypothesis $\Upsilon_0(x,\bw)=M(x,\bw;\btheta,\bgamma_0)=m(x;\btheta)+\bw'\bgamma_0$. (In the case of linearity, $\btheta=(\theta_0,\theta_1)'$ and $m(x;\btheta)=\theta_0 + x\theta_1$.) 
This is a special case of the specification tests discussed in Section \ref{SA sec: uniform inference}:
\begin{align*}
    \dot{\mathsf{H}}_0&:\quad \sup_{x\in\mathcal{X}} \Big|\Upsilon_0(x,\bw) - M(x,\bw;\btheta,\bgamma_0)\Big|=0, \quad \text{ for some } \btheta, \qquad vs.\\
    \dot{\mathsf{H}}_\text{A}&:\quad \sup_{x\in\mathcal{X}} \Big|\Upsilon_0(x,\bw) - M(x,\bw;\btheta,\bgamma_0)\Big|>0, \quad \text{ for all } \btheta.
\end{align*}
One rejects $\dot{\mathsf{H}}_0$ if and only if $\sup_{x\in\mathcal{X}} |\dot{T}_p(x)|\geq \cval$ for some critical value $\cval$ where $\dot{T}_p(x) = \frac{\widehat{\Upsilon}(x,\widehat{\bw}) - M(x,\widehat{\bw};\widetilde{\btheta},\widetilde{\bgamma})}{\sqrt{\widehat{\Omega}(x)/n}}$.

This testing procedure formalizes the idea of visually examining a binned scatter plot compared to a parametric specification; a common step before regression analysis. But it also formalizes the problematic dependency on the evaluation point $\bw$ and the difference between $\widehat{\bgamma}$ and $\widetilde{\bgamma}$. Despite the fact that $\bw'\bgamma_0$ cancels out in both the null and alternative statements, the numerator of the $t$-statistic depends on $\widehat{\bw}'(\widehat{\bgamma} - \widetilde{\bgamma})$, because in finite samples $\bgamma_0$ is unknown. Therefore our uncertainty about how $x$ enters the model depends on the controls $\bw_i$. As mentioned above, this comes about because $\mu_0(x)$ is only defined relative to $\bw_i$. 

Consider the case where $\bw_i$ is an indicator (or fixed effect). Then setting $\widehat{\bw} = \mathbf{0}$ would seem to remove the problem, because the numerator of $\dot{T}_p(x)$ depends only on $\widehat{\mu}(x)$ and $m(x;\widetilde{\btheta})$, while setting $\widehat{\bw} = \mathbf{1}$ maximizes it. This is correct, but is then sensitive to how the researcher has coded $\bw_i$, i.e., which category is considered the baseline. Thus we can get a different answer to the test depending on which category of $\bw$ we consider, even though the hypothesis applies to both. This is intuitively the same as the fact that in a linear model with dummy variables the standard error of the intercept changes depending on how $\bw$ is coded. The case of a continuous $\bw_i$ (especially with large support, such as annual income) is perhaps worse: if $\widehat{\bgamma} \neq \widetilde{\bgamma}$, then there is \emph{always} some value $\widehat{\bw}$ for which we reject the null. Thus, using the procedure described above to test parametric specifications is potentially confusing at best, and at worst is vulnerable to $p$-hacking. It is worth noting that in most papers studying the partially linear model, the parameter of interest is $\bgamma_0$, and so these concerns have gone largely unnoticed. (And are masked by construction when using the incorrect residualization approach.)

To avoid these issues, and motivated by the fact that the central point of binscatter is to study how $y_i$ relates to $x_i$, controlling for $\bw_i$, we advocate reformulating the hypothesis as pertaining to the \emph{derivative} of $\mu_0(x)$, instead of the level. Under the partially linear model maintained throughout, any derivative of $\E[y_i|x_i=x,\bw_i=\bw]$ is exactly $\mu_0^{(v)}(x)$, and is by definition $\Upsilon^{(v)}_0(x,\bw)$. Therefore, instead of testing the null $\Upsilon_0(x,\bw)=m(x;\btheta)+\bw'\bgamma_0$, we test the equivalent hypothesis that $\Upsilon^{(v)}_0(x,\bw)=m^{(v)}(x;\btheta)$ for some $v \geq 1$. For example, instead of testing that $\mu_0(x)$ is linear, we test that it has constant first derivative. To test if $\mu_0(x)$ itself is constant, the null would be that $\mu_0^{(1)}(x) = m^{(1)}(x;\btheta) = 0$.

Such (more robust) tests are still special cases of the specification tests discussed in Section \ref{SA sec: uniform inference}:  for some $v\geq 1$,
\begin{align*}
    \dot{\mathsf{H}}_0&:\quad \sup_{x\in\mathcal{X}} \Big|\Upsilon_0^{(v)}(x,\bw) - m^{(v)}(x;\btheta)\Big|=0, \quad \text{ for some } \btheta, \qquad vs.\\
    \dot{\mathsf{H}}_\text{A}&:\quad \sup_{x\in\mathcal{X}} \Big|\Upsilon^{(v)}_0(x,\bw) - m^{(v)}(x;\btheta)\Big|>0, \quad \text{ for all } \btheta.
\end{align*}
One rejects $\dot{\mathsf{H}}_0$ if and only if $\sup_{x\in\mathcal{X}} |\dot{T}_p(x)|\geq \cval$ for some critical value $\cval$ where $\dot{T}_p(x)=\frac{\widehat{\mu}^{(v)}(x) - m^{(v)}(x;\widetilde{\btheta})}{\sqrt{\widehat{\Omega}(x)/n}}$.

Finally, notice that the visual appearance of the confidence band for the mean function $\Upsilon_0(x,\bw)=\E[y_i|x_i=x,\bw_i=\bw]$ will also be impacted by the evaluation point $\bw$ (or its feasible version $\widehat{\bw}$). This is important to keep in mind when evaluating binscatter plots. By definition, each binscatter plot shows only one choice of $\bw$, and therefore while the shape of $\widehat{\Upsilon}(x,\widehat{\bw})$ is unchanged, a level shift will occur and the size of the band can change. For an intuitive example, again consider the case where $\bw$ is categorical, and some categories have much larger or smaller sample sizes. These different sample sizes will naturally be reflected in the uncertainty for $\Upsilon_0(x,\bw)$.

For this reason, we must be careful when using confidence bands as visual aids in parametric specification testing. If we plot $\widehat{\Upsilon}(x,\widehat{\bw})$ and its associated confidence band, it is tempting to say that if this band does not contain a line (or quadratic function), then we say that at level $\alpha$ we reject the null hypothesis that $\mu_0(x)$ is linear (or quadratic). Although this is formally justified, we must interpret such analyses with caution because of the role of the evaluation point.

\section{General Setup and Notation}\label{SA section: Setup}

To present all our complete theoretical results we first review and generalize the notation introduced in the main text. Suppose that $(y_i,x_i, \bw_i')$, $1\leq i\leq n$, is a random sample where $y_i\in\mathcal{Y}$ is a scalar response variable, $x_i\in\mathcal{X}$ is a scalar covariate, and $\bw_i\in\mathcal{W}$ is a vector of additional control variables of dimension $d$.  Define the following least squares estimand: 
\begin{equation}\label{SA eq: parameter}
	(\mu_0(\cdot), \bgamma_0)=\argmin_{\mu\in\mathcal{M}, \bgamma\in\mathbb{R}^d}\; \E\big[\big(y_i-\mu(x_i)-\bw_i'\bgamma\big)^2\big],
\end{equation}
where $\mathcal{M}$ is a space of functions satisfying certain smoothness conditions to be specified later.

We study binscatter estimators in the partially linear regression model:
\begin{equation}\label{SA eq:model-withcovs}
	y_i = \mu_0(x_i) + \bw_i'\bgamma_0 + \epsilon_i,  \qquad \E[\epsilon_i|x_i,\bw_i]=0.
\end{equation}
The parameter of interest is
\[
\Upsilon^{(v)}_0(x, \bw) = \frac{\partial^v}{\partial x^v} \E[y_i|x_i=x,\bw_i=\bw], \qquad v\in\mathbb{N}_0,
\]
for some evaluation points $x$ and $\bw$. Given the assumption $\E[\epsilon_i|x_i,\bw_i]=0$ in \eqref{SA eq:model-withcovs}:
\[\Upsilon_0(x, \bw)=\Upsilon^{(0)}_0(x, \bw)=\mu_0(x)+\bw'\bgamma_0 \qquad\text{ and }\qquad \Upsilon^{(v)}_0(x, \bw)=\mu_0^{(v)}(x) \text{ for } v>0.\] 
In the paper, we focused on $\Upsilon_0^{(v)}(x)=\Upsilon_0^{(v)}(x,\E[\bw_i])$, one special case of $\Upsilon_0^{(v)}(x,\bw)$ defined above for some evaluation point $\bw$.

The following basic conditions on the data generating process are imposed throughout.

\begin{assumDGP}[Data Generating Process]\label{SA Assumption DGP}
	$\{(y_{i}, x_{i}, \bw_i'): 1\leq i\leq n\}$ is i.i.d. satisfying \eqref{SA eq: parameter} with $\mathcal{X}$ a compact interval;
    $x_{i}$ has a distribution function $F_X(x)$ with a Lipschitz continuous (Lebesgue) density $f_X(x)$ bounded away from zero on $\mathcal{X}$; and
    $\mu_0(x)$ is $\varsigma_\mu$-times continuously differentiable for some $\varsigma_\mu\geq p+1$.
\end{assumDGP}

We next impose a condition that is specific to the least squares binscatter. Binscatters in more general models are studied in \cite{Cattaneo-Crump-Farrell-Feng_2023_Generalized}.
Section \ref{SA section: Notation} defines standard notation.

\begin{assumLS}[Least Squares]\label{SA Assumption LS} \leavevmode
	\begin{enumerate}[label=\bfseries(\roman*)]
		\item $\E[\epsilon_{i}|x_i,\bw_i]=0$;
              $\sigma^2(x):=\E[\epsilon_i^2|x_i=x]$ is Lipschitz continuous and bounded away from zero on $\mathcal{X}$; and
              $\sup_{x\in\mathcal{X}}\E[|\epsilon_{i}|^\nu|x_{i}=x]\lesssim 1$ for some $\nu>2$.
              
		\item
		$\max_{1\leq i\leq n}\E[\epsilon_i^2|\bw_i, x_i]\lesssim_\P 1$; 
		$\E[\bw_i|x_i=x]$ is $\varsigma_w$-times continuously differentiable for some $\varsigma_w\geq 1$;  $\sup_{x\in\mathcal{X}}\E[\|\bw_i\|^{\nu}|x_i=x]\lesssim 1$;  
		$\max_{1\leq i\leq n}\E[\|\bw_i-\E[\bw_i|x_i]\|^4|x_i]\lesssim_\P 1$; and
		$\min_{1\leq i\leq n}\lambda_{\min}(\E[(\bw_i-\E[\bw_i|x_i])(\bw_i-\E[\bw_i|x_i])'|x_i])\gtrsim_\P 1$.
	\end{enumerate}
\end{assumLS}

Part (i) imposes some moment conditions on the error term which are commonly used in the nonparametric series estimation literature. Part (ii) includes a set of conditions similar to those used in \cite*{Cattaneo-Jansson-Newey_2018_ET,Cattaneo-Jansson-Newey_2018_JASA} to  analyze the semiparametric partially linear regression model. They ensure the negligibility of the estimation error of $\widehat{\bgamma}$. To reduce notation, we use the same constant $\nu>2$ in the conditional moment bounds for $\epsilon_i$ and $\bw_i$.

Binscatter estimators are typically constructed based on quantile-spaced partitions, and a major innovation herein is accounting for this additional randomness. Our results allow for other options as well, including evenly spaced partitioning. Specifically, the relevant support of $x_i$ is partitioned into $J$ disjoint intervals employing the empirical quantiles, leading to the partitioning scheme $\widehat{\Delta} = \{\widehat{\mathcal{B}}_1, \widehat{\mathcal{B}}_2, \dots, \widehat{\mathcal{B}}_J\}$, where

\[\widehat{\mathcal{B}}_j = \begin{cases}
\big[x_{(1)}, x_{(\lfloor n/J \rfloor)}\big)                               & \qquad \text{if } j=1\\
\big[x_{(\lfloor (j-1)n/J \rfloor)}  , x_{(\lfloor jn/J \rfloor)}\big) & \qquad \text{if } j=2,3,\dots,J-1\\
\big[x_{(\lfloor (J-1)n/J \rfloor)}, x_{(n)}\big]                             & \qquad \text{if } j=J
\end{cases},
\]
$x_{(i)}$ denotes the $i$-th order statistic of the sample $\{x_1,x_2,\dots,x_n\}$, and $\lfloor \cdot \rfloor$ is the floor operator. The number of bins $J$ plays the role of tuning parameter for the binscatter method, and is assumed to diverge: $J\to\infty$ as $n\to\infty$ throughout the supplement, unless explicitly stated otherwise.

The piecewise polynomial basis of degree $p$, for some choice of $p=0,1,2,\dots$, is defined as
\[\Big[\begin{array}{cccc}
\I_{\widehat{\mathcal{B}}_1}(x)&
\I_{\widehat{\mathcal{B}}_2}(x)&
\cdots&
\I_{\widehat{\mathcal{B}}_J}(x)
\end{array} \Big]' \otimes
\Big[\begin{array}{cccc}
1& x& \cdots& x^p
\end{array} \Big]',
\]
where $\I_\mathcal{A}(x)=\I(x\in \mathcal{A})$ and $\otimes$ is the Kronecker product operator. For convenience of later analysis, we use $\widehat{\bb}_{p,0}(x)$ to denote a \textit{standardized rotated} basis, the $j$th element of which is given by
\[
\sqrt{J}\times\I_{\widehat{\mathcal{B}}_{\bar{j}}}(x)\times\Big(\frac{x-x_{(\lfloor(\bar{j}-1)n/J\rfloor)}}{\hat{h}_{\bar{j}}} \Big)^{j-1-(\bar{j}-1)(p+1)},
\quad j=1, \cdots, (p+1)J,
	\]
where $\bar{j}=\lceil j/(p+1)\rceil$, $\lceil\cdot\rceil$ is the ceiling operator, 
and $\hat{h}_{\bar{j}}=x_{(\lfloor \bar{j}n/J\rfloor)}-x_{(\lfloor(\bar{j}-1)n/J\rfloor)}$.
Thus, each local polynomial is centered at the start of each bin and scaled by the length of the bin. $\sqrt{J}$ is an additional scaling factor which  helps simplify some expressions of our results. The standardized rotated basis $\widehat{\bb}_{p,0}(x)$  is equivalent to the original piecewise polynomial basis in the sense that they represent the same (linear) function space.

To impose the restriction that the estimated function is $(s-1)$-times continuously differentiable for $1\leq s \leq p$, we introduce a new basis 
\[
\widehat{\bb}_{p,s}(x)=\Big(\widehat{b}_{p,s,1}(x),\ldots, \widehat{b}_{p,s,K_{p,s}}(x)\Big)'
=\widehat{\mathbf{T}}_s\widehat{\bb}_{p,0}(x),
\qquad K_{p,s}=(p+1)J-s(J-1),
\] 
where
$\widehat{\mathbf{T}}_s:=\widehat{\bT}_s(\widehat{\Delta})$ is a $K_{p,s}\times(p+1)J$ matrix depending on $\widehat{\Delta}$, which transforms a piecewise polynomial basis to a smoothed binscatter basis. When $s=0$, we let $\widehat{\mathbf{T}}_0=\mathbf{I}_{(p+1)J}$, the identity matrix of dimension $(p+1)J$. Thus $\widehat{\bb}_{p,0}(x)$ is the discontinuous basis without any constraints defined previously. When $s=p$, $\widehat{\bb}_{p,s}(x)$ is the well-known $B$-spline basis of order $p+1$ with simple knots, 
which is $(p-1)$-times continuously differentiable. When $0<s<p$, they can be defined similarly as $B$-splines with knots of certain multiplicities. See Definition 4.1 in Section 4 of \cite{Schumaker_2007_book} for more details. We require $s\leq p$, since if $s=p+1$, $\widehat{\bb}_{p,s}(x)$ reduces to a global polynomial basis of degree $p$.  

A key feature of the transformation matrix $\widehat{\mathbf{T}}_s$ is that on every row it has \textit{at most} $(p+1)^2$ nonzeros, and on every column it has \textit{at most} $p+1$ nonzeros. The expression of these elements is cumbersome. The proof of Lemma \ref{SA lem: spline transform} describes the structure of $\widehat{\bT}_s$ in more detail and provides an explicit representation for $\widehat{\bT}_s$.

Given a choice of basis, we consider the following least squares binscatter estimator:
\begin{equation}\label{SA eq:binscatter-QMLE}
	\widehat{\mu}^{(v)}(x) = \widehat{\bb}_{p,s}^{(v)}(x)'\widehat{\bbeta}, \qquad
	\begin{bmatrix}\;\widehat{\bbeta}\;\\\;\widehat{\bgamma}\;\end{bmatrix}
	= \argmin_{\bbeta,\bgamma} \sum_{i=1}^{n} \Big(y_i - \widehat{\bb}_{p,s}(x_i)'\bbeta - \bw_{i}'\bgamma\Big)^2,
\end{equation}
where $\widehat{\bb}_{p,s}^{(v)}(x)=\frac{d^v}{dx^v}\widehat{\bb}_{p,s}(x)$ for some $v\in\mathbb{Z}_+$ such that $v\leq p$. 
It is well known that this estimator admits the following ``backfitting'' expression, which will be convenient for later theoretical analysis:
\[ 
\widehat{\bbeta}=(\mathbf{B}'\mathbf{B})^{-1}
\mathbf{B}'(\mathbf{Y}-\mathbf{W}\widehat{\bgamma}),\qquad
\widehat{\bgamma}=(\mathbf{W}'\bM_{\mathbf{B}}\mathbf{W})^{-1}
(\mathbf{W}'\bM_{\mathbf{B}}\mathbf{Y}),
\]
where $\mathbf{Y}=(y_1, \ldots, y_n)'$,
$\mathbf{B}=(\widehat{\bb}_{p,s}(x_{1}), \ldots, \widehat{\bb}_{p,s}(x_{n}))'$,  
$\bW=(\bw_1,\cdots, \bw_n)'$ and 
 $\bM_{\mathbf{B}}=\bI_{n}-\mathbf{B}(\mathbf{B}'\mathbf{B})^{-1}\mathbf{B}'$ with $\bI_n$ denoting the identify matrix of size $n$.

Given an estimator $\widehat{\bw}$ of the evaluation point $\bw$, we have the following estimator of $\Upsilon_0^{(v)}(x, \bw)$:
\[
\widehat{\Upsilon}^{(v)}(x, \widehat\bw)=
\begin{cases}
\widehat{\mu}(x)+\widehat{\bw}'\widehat{\bgamma} \quad &\text{if } v=0 \\
\widehat{\mu}^{(v)}(x)\quad &\text{if } v\geq 1
\end{cases}.
\]
Throughout the supplement (and the paper), we always assume that the estimator $\widehat{\bw}$ is either nonrandom (e.g., a fixed value) or generated based on $\bW$.

\begin{remark}[Smoothness and Bias Correction]
	We remind readers that this supplemental appendix presents \textit{all} results under general choices of the number of bins $J$, the degree of the basis $p$, and the smoothness of the basis $s$. By contrast, for simplicity, the paper only uses the binscatter basis with $s=p$, where $p=0$ for binscatter estimation and $p = 1$ for inference. In addition, in the paper we let $J$ be the IMSE-optimal choice corresponding to  $p=\mathtt{p}$ for a fixed number $\mathtt{p}$ (see Theorem \ref{SA thm: IMSE, LS}), and inference is conducted based on the binscatter basis of degree $p=\mathtt{p}+1$. In particular, we set $\mathtt{p}=0$ to construct confidence bands in Section III. This can be viewed as a bias correction strategy \citep*{Calonico-Cattaneo-Farrell_2018_JASA,Calonico-Cattaneo-Farrell_2022_Bernoulli} which guarantees the smoothing bias of the binscatter estimator is negligible in inference under mild conditions.
\end{remark}

\subsection{Notation}\label{SA section: Notation}

For background definitions, see \citet{vandevarrt-Wellner_1996_book}, \citet{Bhatia_2013_book}, \citet{Gine-Nickl_2016_book}, and references therein.

\textit{\textbf{Matrices and Norms}}. For (column) vectors, $\|\cdot\|$ denotes the Euclidean norm, $\|\cdot\|_1$ denotes the $L_1$ norm, $\|\cdot\|_\infty$ denotes the sup-norm, and $\|\cdot\|_0$ denotes the number of nonzeros. For matrices, $\|\cdot\|$ is the operator matrix norm induced by the $L_2$ norm, and $\|\cdot\|_\infty$ is the matrix norm induced by the supremum norm, i.e., the maximum absolute row sum of a matrix. For a square matrix $\mathbf{A}$, $\lambda_{\max}(\mathbf{A})$ and $\lambda_{\min}(\mathbf{A})$ are the maximum and minimum eigenvalues of $\mathbf{A}$, respectively. $[\mathbf{A}]_{ij}$ denotes the $(i,j)$th entry of a generic matrix $\mathbf{A}$. We will use $\mathcal{S}^{L}$ to denote the unit circle in $\mathbb{R}^L$, i.e., $\|\mathbf{a}\|=1$ for any $\mathbf{a}\in \mathcal{S}^L$. For a real-valued function $g(\cdot)$ defined on a measure space $\mathcal{Z}$, let $\|g\|_{\mathbb{Q}, 2}:=(\int_\mathcal{Z} |g|^2d\mathbb{Q})^{1/2}$ be its $L_2$-norm with respect to the measure $\mathbb{Q}$. In addition, let  $\|g\|_\infty=\sup_{z\in\mathcal{\mathcal{Z}}}|g(z)|$ be $L_\infty$-norm of $g(\cdot)$, and $g^{(v)}(z)=d^vg(z)/dz^v$ be the $v$th derivative for $v\geq 0$.

\textit{\textbf{Asymptotics}}. For sequences of numbers or random variables, we use $l_n\lesssim m_n$ to denote that $\limsup_n|l_n/m_n|$ is finite, $l_n\lesssim_\P m_n$ or $l_n=O_\P(m_n)$ to denote $\limsup_{\varepsilon\rightarrow\infty}\limsup_n\P[|l_n/m_n|\geq\varepsilon]=0$, $l_n=o(m_n)$ implies $l_n/m_n\rightarrow 0$, and $l_n=o_\P(m_n)$ implies that $l_n/m_n\rightarrow_\P 0$, where $\rightarrow_\P$ denotes convergence in probability. $l_n\asymp m_n$ implies that $l_n\lesssim m_n$ and $m_n\lesssim l_n$. 

\textit{\textbf{Empirical Process}}. We employ standard empirical process notation: $\E_n[g(\bv_i)]=\frac{1}{n}\sum_{i=1}^ng(\bv_i)$, and $\G_n[g(\bv_i)]=\frac{1}{\sqrt{n}}\sum_{i=1}^n(g(\bv_i)-\E[g(\bv_i)])$ for a sequence of random variables $\{\bv_i\}_{i=1}^n$. In addition, we employ the notion of covering number extensively in the proofs. Specifically, given a measurable space $(A, \mathcal{A})$ and a suitably measurable class of functions $\mathcal{G}$ mapping $A$ to $\mathbb{R}$ equipped with a measurable envelop function $\bar{G}(z)\geq \sup_{g\in\mathcal{G}}|g(z)|$, 
the \textit{covering number} of $N(\mathcal{G}, L_2(\mathbb{Q}), \varepsilon)$ is the minimal number of $L_2(\mathbb{Q})$-balls of radius $\varepsilon$ needed to cover $\mathcal{G}$ for a measure $\mathbb{Q}$. The covering number of $\mathcal{G}$ relative to the envelope is denoted as $N(\mathcal{G}, L_2(\mathbb{Q}), \varepsilon\|\bar{G}\|_{\mathbb{Q},2})$. 

\textit{\textbf{Partitions}}. Given the random partition $\widehat{\Delta}$, we use the notation $\E_{\widehat{\Delta}}[\cdot]$ to denote that the expectation is taken with the partition $\widehat{\Delta}$ understood as fixed. To further simplify notation, we let $\{\hat{\tau}_0\leq \hat{\tau}_1\leq \cdots\leq \hat{\tau}_J\}$ denote the empirical quantile sequence employed by $\widehat{\Delta}$ and $\hat{h}_j=\hat{\tau}_j-\hat{\tau}_{j-1}$ be the width of the $j$-th bin $\widehat{\mathcal{B}}_j$. Accordingly, 
let $\{\tau_0\leq \cdots\leq \tau_J\}$ be the population quantile sequence, i.e., $\tau_j=F_X^{-1}(j/J)$ for $0\leq j\leq J$.
Then $\Delta_0=\{\mathcal{B}_1, \ldots, \mathcal{B}_J\}$ denotes the partition based on population quantiles, i.e.,
\[\mathcal{B}_j = \begin{cases}
\big[\tau_0, \tau_1\big)                               & \qquad \text{if } j=1\\
\big[\tau_{j-1}  , \tau_j\big) & \qquad \text{if } j=2,3,\dots,J-1\\
\big[\tau_{J-1}, \tau_{J}\big]                             & \qquad \text{if } j=J
\end{cases}.
\]
Let $h_j=F_X^{-1}(j/J)-F_X^{-1}((j-1)/J)$ be the width of $\mathcal{B}_j$.
Analogously to $\widehat{\bb}_{p,s}(x)$, $\bb_{p,s}(x)$ denotes the binscatter basis of degree $p$ that is  $(s-1)$-times continuously differentiable and is constructed based on the \textit{nonrandom} partition $\Delta_0$. We sometimes write $\bb_{p,s}(x;\Delta)=(b_{p,s,1}(x;\Delta), \ldots, b_{p,s,K_{p,s}}(x;\Delta))'$ to emphasize a binscatter basis is constructed based on a particular partition $\Delta$. Therefore, $\widehat{\bb}_{p,s}(x)=\bb_{p,s}(x;\widehat{\Delta})$ and $\bb_{p,s}(x)=\bb_{p,s}(x;\Delta_0)$.

For any given partition $\Delta$, the \textit{population} least squares projection of $\mu_0(\cdot)$ is given by $\bb_{p,s}(\cdot;\Delta)'\bbeta_0(\Delta)$ with 
\begin{equation}\label{SA eq: L2 projection}
	\bbeta_0(\Delta):=\argmin_{\bbeta\in\mathbb{R}^{K_{p,s}}}\;
	\E[(\mu_0(x_i)-\bb_{p,s}(x_i;\Delta)'\bbeta)^2].
\end{equation}
Accordingly, given the random partition $\widehat{\Delta}$ and the nonrandom partition $\Delta_0$, we have
\[
\begin{split}
&\widehat{\bbeta}_0:=\bbeta_0(\widehat{\Delta}):=\argmin_{\bbeta\in\mathbb{R}^{K_{p,s}}}\;
\E_{\widehat{\Delta}}[(\mu_0(x_i)-\bb_{p,s}(x_i;\widehat{\Delta})'\bbeta)^2],\quad\text{and}\\
&\bbeta_0:=\bbeta_0(\Delta_0):=\argmin_{\bbeta\in\mathbb{R}^{K_{p,s}}}\;
\E[(\mu_0(x_i)-\bb_{p,s}(x_i;\Delta_0)'\bbeta)^2].
\end{split}\]
The corresponding $L_2$ projection error is $r_{0,v}(x;\Delta)=\mu_0^{(v)}(x)-\bb_{p,s}^{(v)}(x;\Delta)'\bbeta_0(\Delta)$.
We therefore define the approximation errors
\[\widehat{r}_{0,v}(x):=r_{0,v}(x;\widehat{\Delta}), \qquad\text{ and }\qquad r_{0,v}(x):=r_{0,v}(x;\Delta_0).\]
For $v=0$, we write $\widehat{r}_{0}(x):=\widehat{r}_{0,0}(x)$ and $r_{0}(x):=r_{0,0}(x)$

\textit{\textbf{Other}}. Let
$\bX = [x_1, \ldots, x_n]'$, $\bW=[\bw_1,\cdots, \bw_n]'$, and $\bD= [(y_i, x_i, \bw_i')' : i=1,2,\dots,n]$. 
$\lceil z \rceil$ outputs the smallest integer no less than $z$ and $a\wedge b=\min\{a,b\}$.
``w.p.a. $1$'' means ``with probability approaching one''.

\section{Theoretical Results}\label{SA section: LS}

Our main theoretical results are presented in this section. We will focus on the estimator $\widehat{\Upsilon}^{(v)}(x,\widehat\bw)$ of $\Upsilon_0^{(v)}(x,\bw)$. The estimator $\widehat\Upsilon(x)$ of $\Upsilon_0^{(v)}(x)=\Upsilon_0^{(v)}(x,\E[\bw_i])$ discussed in the paper is covered as a special case.

\subsection{Properties of Quantile-Based Partition and Binscatter Basis}\label{SA section: prelim lemmas}

In this section we first give some preliminary lemmas concerning the basic properties of the quantile-based partition and the binscatter basis, which are necessary for our main analysis and may be of independent interest.

The asymptotic properties of partitioning-based estimators require a partition that is not too ``irregular''. In the binscatter setting, we let 
$\bar{f}_X=\sup_{x\in\mathcal{X}}f_X(x)$ and $\underline{f}_X=\inf_{x\in\mathcal{X}} f_X(x)$, and for any partition $\Delta$ with $J$ bins, we let $h_j(\Delta)$ denote the length of the $j$th bin in $\Delta$. Therefore, $\hat{h}_j=h_j(\widehat{\Delta})$ and $h_j=h_j(\Delta_0)$. Then, we introduce the family of partitions:
\begin{equation} \label{SA eq: quasi-uniform}
	\Pi = \Big\{ \Delta:
	\frac{\max_{1\leq j \leq J} h_j(\Delta)}
	{\min_{1\leq j \leq J} h_j(\Delta)}\leq \frac{3\bar{f}_X}{\underline{f}_X} \Big\}.
\end{equation}
Intuitively, if a partition belongs to $\Pi$, then the lengths of its bins do not differ ``too'' much, a property usually referred to as ``quasi-uniformity'' in approximation theory. Our first lemma shows that a quantile-spaced partition possesses this property with probability approaching one.

\begin{lem}[Quasi-Uniformity of Quantile-Spaced Partitions]
	\label{SA lem: quantile partition}
	Suppose that Assumption \ref{SA Assumption DGP} holds. If
	$\frac{J\log J}{n}=o(1)$ and $\frac{\log n}{J}=o(1)$, then
	(i) $\max_{1\leq j \leq J}|\hat{h}_j-h_j|\lesssim_\P J^{-1}\Big(\frac{J\log J}{n}\Big)^{1/2}$, and (ii) $\widehat{\Delta}\in\Pi$ w.p.a. $1$.
\end{lem}

As discussed previously, $\widehat{\bT}_s$ links the more complex spline basis with a simple piecewise polynomial basis. Recall that $\widehat{\bT}_s=\widehat{\bT}_s(\widehat{\Delta})$ depends on the empirical-quantile-based partition $\widehat{\Delta}$. The next lemma describes its key features. We let  $\bT_s:=\bT_s(\Delta_0)$ be the transformation matrix corresponding to the nonrandom basis $\bb_{p,s}(x)$, i.e., $\bb_{p,s}(x)=\bT_s\bb_{p,0}(x)$.

\begin{lem}[Transformation Matrix]\label{SA lem: spline transform}
	Suppose that Assumption \ref{SA Assumption DGP} holds. 
	If $\frac{J\log J}{n}=o(1)$ and $\frac{\log n}{J}=o(1)$,
	then
	$\widehat{\bb}_{p,s}(x)=\widehat{\mathbf{T}}_s\widehat{\bb}_{p,0}(x)$ with
	$\|\widehat{\mathbf{T}}_s\|_\infty\lesssim_\P 1$, $\|\widehat{\mathbf{T}}_s\|\lesssim_\P 1$,  
	$\|\widehat{\mathbf{T}}_s-\mathbf{T}_s\|_\infty\lesssim_\P\Big(\frac{J\log J}{n}\Big)^{1/2}$, and $\|\widehat{\mathbf{T}}_s-\mathbf{T}_s\|\lesssim_\P\Big(\frac{J\log J}{n}\Big)^{1/2}$.
\end{lem}

The following lemma provides some simple bounds on the basis.
\begin{lem}[Local Basis] \label{SA lem: local basis}
Suppose that Assumption \ref{SA Assumption DGP} holds. Then,
$\sup_{x\in\mathcal{X}}\|\widehat{\bb}_{p,s}^{(v)}(x)\|_0\leq (p+1)^2$. If, in addition, $\frac{J\log J}{n}=o(1)$ and $\frac{\log n}{J}=o(1)$, then
$\sup_{x\in\mathcal{X}}\|\widehat{\bb}_{p,s}^{(v)}(x)\|\lesssim_\P J^{\frac{1}{2}+v}$.	
\end{lem}

The following lemma characterizes the approximation error $\widehat{r}_{0,v}(x)$ in terms of the sup norm.
\begin{lem}[Approximation Error] \label{SA lem: uniform approx rate}
	Suppose that Assumption \ref{SA Assumption DGP} holds. If $\frac{J\log J}{n}=o(1)$ and $\frac{\log n}{J}=o(1)$, then 
     \begin{align*}
         \sup_{\Delta\in\Pi}\sup_{x\in\mathcal{X}}|\bb_{p,s}^{(v)}(x;\Delta)'\bbeta_0(\Delta)-\mu_0^{(v)}(x)|\lesssim J^{-p-1+v}\quad \text{and}
         \quad \sup_{x\in\mathcal{X}}|\widehat{\bb}_{p,s}^{(v)}(x)'\widehat{\bbeta}_0-\mu_0^{(v)}(x)|\lesssim_\P J^{-p-1+v}.
     \end{align*}
\end{lem}

\begin{remark}[Improvements over literature]
	Lemmas \ref{SA lem: quantile partition}--\ref{SA lem: uniform approx rate} show some basic characteristics of the binscatter basis, which are used in the subsequent main analysis. Compared with other studies of splines (see, e.g., \citealp*{Shen-Wolfe-Zhou_1998_AoS,Huang_2003_AoS,Schumaker_2007_book}), we formally take into account the randomness of the partition formed by empirical quantiles.
\end{remark}

\subsection{Preliminary Technical Lemmas} \label{SA section: tech lemmas}
This section collects a set of technical lemmas, which are key ingredients of our main theorems. 

We first introduce the following quantities that will be frequently used:
\begin{align*}
	&\widehat{\bQ}:=\widehat{\bQ}(\widehat{\Delta}):=
	\E_n[\widehat{\bb}_{p,s}(x_i)\widehat{\bb}_{p,s}(x_i)'],
	\quad
	\bQ_0:=\bQ(\Delta_0):=\E[\bb_{p,s}(x_i)\bb_{p,s}(x_i)'], \\
	&\widehat{\bSigma}:=\widehat{\bSigma}(\widehat{\Delta}):=\E_n[\widehat{\bb}_{p,s}(x_i)\widehat{\bb}_{p,s}(x_i)'\widehat{\epsilon}_i^2], \quad
	\bar{\bSigma}:=\bar{\bSigma}(\widehat{\Delta}):=
	\E_n\Big[\E[\widehat{\bb}_{p,s}(x_i)\widehat{\bb}_{p,s}(x_i)'\epsilon_i^2|\bX]\Big], \quad\\
	&\bSigma_0:=\bSigma(\Delta_0):=\E[\bb_{p,s}(x_i)\bb_{p,s}(x_i)'\epsilon_i^2], \\
	&\widehat{\Omega}(x):=\widehat{\Omega}(x;\widehat{\Delta}):=\widehat{\bb}_{p,s}^{(v)}(x)'\widehat{\bQ}^{-1}\widehat{\bSigma}\widehat{\bQ}^{-1}\widehat{\bb}_{p,s}^{(v)}(x),\\
	&\bar{\Omega}(x):=\bar{\Omega}(x;\widehat{\Delta}):=
	\widehat{\bb}_{p,s}^{(v)}(x)'\widehat{\bQ}^{-1}\bar{\bSigma}\widehat{\bQ}^{-1}\widehat{\bb}_{p,s}^{(v)}(x), \; \text{ and } \\ &\Omega(x):=\Omega(x;\widehat{\Delta}):=\widehat{\bb}_{p,s}^{(v)}(x)'\bQ_0^{-1}\bSigma_0\bQ_0^{-1}\widehat{\bb}_{p,s}^{(v)}(x),
\end{align*}
where $\widehat{\epsilon}_i=y_i-\widehat{\bb}_{p,s}(x_i)'\widehat{\bbeta}-\bw_i'\widehat{\bgamma}$. 
All quantities with $\;\widehat{}\;$ or $\;\bar{}\;$ depend on the random partition $\widehat{\Delta}$, and those without any accents are nonrandom with the only exception of $\Omega(x)$, where the basis $\widehat{\bb}_{p,s}^{(v)}(x)$ still depends on $\widehat{\Delta}$. The dependence on $p$, $s$ and $v$ is often omitted for simplicity.

The following lemma characterizes the properties of the Gram matrix of the binscatter basis.

\begin{lem}[Gram] \label{SA lem: Gram, LS}
	Suppose that Assumption \ref{SA Assumption DGP} holds. Then, 
	$1\lesssim \lambda_{\min}(\bQ_0)\leq \lambda_{\max}(\bQ_0)\lesssim 1$.
	If, in addition, $\frac{J\log J}{n}=o(1)$ and $\frac{\log n}{J}=o(1)$, then
	\[
		\|\widehat{\bQ}-\bQ_0\|
		\lesssim_\P \Big(\frac{J\log J}{n}\Big)^{1/2},\quad
		\|\widehat{\bQ}^{-1}\|_\infty\lesssim_\P 1,\quad\text{and}\quad
		\|\widehat{\bQ}^{-1}-\bQ_0^{-1}\|_\infty\lesssim_\P \Big(\frac{J\log J}{n}\Big)^{1/2}.
	\]
\end{lem}

The next lemma shows that the limiting variance of $\widehat{\mu}^{(v)}(x)$ is bounded from above and below if properly scaled. Recall that $\bar{\Omega}(x)=\bar{\Omega}(x;\widehat{\Delta})$ and $\Omega(x)=\Omega(x;\widehat{\Delta})$.

\begin{lem}[Asymptotic Variance] \label{SA lem: asymp variance, LS}
	Suppose that Assumptions \ref{SA Assumption DGP} and \ref{SA Assumption LS}(i) hold. If $\frac{J\log J}{n}=o(1)$ and $\frac{\log n}{J}=o(1)$,
	then w.p.a. $1$,
	\[
		J^{1+2v}\lesssim\inf_{x\in\mathcal{X}}\bar{\Omega}(x)
		\leq\sup_{x\in\mathcal{X}}\bar{\Omega}(x)
		\lesssim J^{1+2v}\quad \text{and}\quad
		J^{1+2v}\lesssim\inf_{x\in\mathcal{X}}\Omega(x)
		\leq\sup_{x\in\mathcal{X}}\Omega(x)
		\lesssim J^{1+2v}.
	\]
\end{lem}

The next lemma gives a bound on the variance component of the binscatter estimator, which is the main building block of uniform convergence.

\begin{lem}[Uniform Convergence: Variance] \label{SA lem: uniform converge var part, LS}
	Suppose that Assumptions \ref{SA Assumption DGP} and \ref{SA Assumption LS}(i) hold. If $\frac{J^{\frac{\nu}{\nu-2}}\log J}{n}=o(1)$ and $\frac{\log n}{J}=o(1)$, then
	\[
	\sup_{x\in\mathcal{X}}\Big|\widehat{\bb}_{p,s}^{(v)}(x)'\widehat{\bQ}^{-1}\E_n[\bb_{p,s}(x_i)\epsilon_i]\Big|\lesssim_\P J^v\Big(\frac{J\log J}{n}\Big)^{1/2}.
	\]
\end{lem}

As explained before, $\widehat{r}_0(x)$ is understood as the $L_2$ approximation error of least squares estimators for $\mu_0(x)$. The next lemma establishes the bound on  the projection of $\widehat{r}_0(x)$ onto the space spanned by $\widehat{\bb}_{p,s}(x)$ in terms of sup-norm.

\begin{lem}[Projection of Approximation Error] \label{SA lem: proj approx error, LS}
	Under Assumption \ref{SA Assumption DGP}, if $\frac{J\log J}{n}=o(1)$ and $\frac{\log n}{J}=o(1)$, then $$\sup_{x\in\mathcal{X}}\Big|\widehat{\bb}_{p,s}^{(v)}(x)'\widehat{\bQ}^{-1}\E_n[\widehat{\bb}_{p,s}(x_{i})\widehat{r}_0(x_{i})]\Big|\lesssim_\P J^{-p-1+v}\Big(\frac{J\log J}{n}\Big)^{1/2}.$$
\end{lem}

The last lemma in this subsection characterizes the convergence of the parametric component in the expression of $\widehat{\bbeta}$. 
\begin{lem}[Covariate Adjustment] \label{SA lem: parametric part}
	Suppose that Assumptions \ref{SA Assumption DGP} and \ref{SA Assumption LS} hold. If $\frac{J\log J}{n}=o(1)$
	and $\frac{\log n}{J}=o(1)$,
	then
	\begin{equation*}
		\|\widehat{\bgamma}-\bgamma_0\|\lesssim_\P \frac{1}{\sqrt{n}}+J^{-p-1-(\varsigma_w\wedge(p+1))}\quad
		\text{and}\quad
		\|\widehat{\bb}_{p,s}^{(v)}(x)'\widehat{\bQ}^{-1}\E_n[\widehat{\bb}_{p,s}(x_i)\bw_i']\|_\infty\lesssim_\P J^v \;\text{ for each } x\in\mathcal{X}.
	\end{equation*}
	If, in addition, 
	$\frac{J^{\frac{\nu}{\nu-2}}\log J}{n}\lesssim 1$, then
	$\sup_{x\in\mathcal{X}}\|\widehat{\bb}_{p,s}^{(v)}(x)'\widehat{\bQ}^{-1}
	\E_n[\widehat{\bb}_{p,s}(x_i)\bw_i']\|_\infty\lesssim_\P J^v.$
\end{lem}
Let $(a_n: n\geq 1)$ be a sequence of non-vanishing constants, which will be used later to characterize the strong approximation rate. Lemma \ref{SA lem: parametric part} implies that if  $\frac{a_n}{\sqrt{J}}=o(1)$ and 
$a_n\sqrt{n}J^{-p-(\varsigma_w\wedge (p+1))-\frac{3}{2}}=o(1)$, then we have
$$\|\widehat{\bgamma}-\bgamma_0\|=o_\P(a_n^{-1}\sqrt{J/n}).$$
This result suffices to make the estimation error of $\widehat{\bgamma}$  negligible in the large sample inference on $\mu_0^{(v)}(\cdot)$ or $\Upsilon_0(\cdot, \bw)$.

\begin{remark}[Improvements over literature]
	The results in this subsection give novel rates of approximations for semi-linear partitioning-based estimators with random partitions. Compared to standard semi-linear regression results, our results provide sharper approximation rates due to the specific binscatter basis, and also formally take into account the randomness of the partition formed by empirical quantiles. See \citet*{Cattaneo-Jansson-Newey_2018_ET,Cattaneo-Jansson-Newey_2018_JASA}, and reference therein, for related literature.
\end{remark}

\subsection{Stochastic Linear Approximation and Point Estimation}

\begin{thm}[Stochastic Linear Approximation]\label{SA thm: Stochastic Linear Approximation, LS}
	Suppose that Assumptions \ref{SA Assumption DGP} and \ref{SA Assumption LS} hold. If $\frac{J^{\frac{\nu}{\nu-2}}\log J}{n}\lesssim 1$ and $\frac{\log n}{J}=o(1)$, then
	\begin{align*}
 	&\sup_{x\in\mathcal{X}}
	\Big|\widehat{\Upsilon}^{(v)}(x, \widehat\bw)-\Upsilon_0^{(v)}(x, \bw)-
	\widehat{\bb}_{p,s}^{(v)}(x)'\widehat{\bQ}^{-1}\E_n[\widehat{\bb}_{p,s}(x_i)\epsilon_i]\Big|\\
	\lesssim_\P
	&\; J^v\Big(\frac{1}{\sqrt{n}}+J^{-p-1-(\varsigma_w\wedge(p+1))}+
	J^{-p-1}\Big)+\|\widehat{\bw}-\bw\|\I(v=0).
	\end{align*}
    
\end{thm}

An immediate corollary of Theorem \ref{SA thm: Stochastic Linear Approximation, LS} is the uniform convergence of $\widehat{\Upsilon}^{(v)}(\cdot, \widehat\bw)$.

\begin{coro}[Uniform Convergence] \label{SA coro: uniform convergence, LS}
	Suppose that Assumptions \ref{SA Assumption DGP} and \ref{SA Assumption LS} hold. If 
	$\sqrt{n}J^{-p-(\varsigma_w\wedge (p+1))-\frac{3}{2}}=o(1)$ and  $\frac{J^{\frac{\nu}{\nu-2}}\log J}{n}\lesssim 1$, then
	\[
	\sup_{x\in\mathcal{X}}\Big|\widehat{\mu}^{(v)}(x)-\mu_0^{(v)}(x)\Big|\lesssim_\P
	J^v\Big(\frac{J\log J}{n}\Big)^{1/2} + J^{-p-1+v}.
	\]
	If, in addition, $\|\widehat{\bw}-\bw\|\lesssim_\P \sqrt{\frac{J\log J}{n}}+J^{-p-1}$, then
	\[
	\sup_{x\in\mathcal{X}}
	\Big|\widehat{\Upsilon}^{(0)}(x,\widehat{\bw})-
	\Upsilon^{(0)}(x, \bw)\Big|\lesssim_\P
	\Big(\frac{J\log J}{n}\Big)^{1/2} + J^{-p-1}.
	\]
\end{coro}

Based on the above facts, we can also show that the proposed variance estimator is consistent.

\begin{thm}[Variance Estimate] \label{SA thm: meat matrix, LS}
	Suppose that Assumptions \ref{SA Assumption DGP} and \ref{SA Assumption LS} hold.
	If $\frac{J^{\frac{\nu}{\nu-2}}(\log J)^{\frac{\nu}{\nu-2}}}{n}=o(1)$ and
	$\sqrt{n}J^{-p-(\varsigma_w\wedge (p+1))-\frac{3}{2}}=o(1)$, then
	\[\Big\|\widehat{\bSigma}-\bSigma_0\Big\|
	\lesssim_\P J^{-p-1}+\Big(\frac{J\log J}{n^{1-\frac{2}{\nu}}}\Big)^{1/2},\;
	\text{ and }\;
	\sup_{x\in\mathcal{X}}\Big|\widehat{\Omega}(x)-\Omega(x)\Big|\lesssim_\P J^{1+2v}\Big(J^{-p-1}+\Big(\frac{J\log J}{n^{1-\frac{2}{\nu}}}\Big)^{1/2}\Big).
	\]
\end{thm}

\begin{remark}[Improvements over literature]
	The results in this subsection improve on the linear series estimation literature \citep*{Belloni-Chernozhukov-Chetverikov-Kato_2015_JoE,Cattaneo-Farrell-Feng_2020_AoS} by formally taking into account the randomness of the partition formed by empirical quantiles, and by accounting for the semi-linear regression estimation structure. The final approximation rate in the Bahadur-type (linear) approximation is sharp for the binscatter basis (with or without random binning). 
\end{remark}

\subsection{Pointwise Distributional Approximation and Inference}

In this subsection we focus on the pointwise inference on the unknown parameter
$\Upsilon^{(v)}_0(x, \bw)=\frac{\partial^v}{\partial x^v}\E[y_i|x_i=x,\bw_i=\bw]$ and construct the $t$-statistic based on $\widehat{\Upsilon}^{(v)}(x,\widehat\bw)$:
\[
T_p(x)=\frac{\widehat{\Upsilon}^{(v)}(x,\widehat{\bw})-\Upsilon_0^{(v)}(x,\bw)}{\sqrt{\widehat{\Omega}(x)/n}}.
\]
Recall in our semi-linear model $\widehat{\Upsilon}^{(v)}(x,\widehat\bw)$ differs from $\widehat{\mu}^{(v)}(x)$ only when $v=0$ and $\widehat\bw\neq 0$. 
Therefore, the condition that $\widehat{\bw}$ converges to $\bw$ at a fast rate imposed below is needed only when $v=0$.

Let $\Phi(\cdot)$ be the cumulative distribution function of a standard normal random variable. The following theorem constructs the pointwise inference for $\Upsilon_0^{(v)}(x,\bw)$.

\begin{thm}[Pointwise Asymptotic Distribution] \label{SA thm: pointwise normality, LS}
Suppose that Assumptions \ref{SA Assumption DGP} and \ref{SA Assumption LS} hold. 
If $\sup_{x\in\mathcal{X}}\E[|\epsilon_i|^\nu|x_i=x]\lesssim 1$ for some $\nu\geq 3$, $\frac{J^{\frac{\nu}{\nu-2}}(\log J)^{\frac{\nu}{\nu-2}}}{n}=o(1)$, $nJ^{-2p-3}=o(1)$
	 and $\|\widehat{\bw}-\bw\|=o(\sqrt{J/n})$, then
		\[
	\sup_{u\in\mathbb{R}}\Big|\P(T_p(x)\leq u)-\Phi(u)\Big|=o(1),\quad
	\text{for each } x\in\mathcal{X},
	\]
 and accordingly,
 	\[
	\P\Big[\Upsilon_0^{(v)}(x,\bw)\in\widehat{I}_{p}(x)\Big]=1-\alpha+o(1), \quad
	\text{for each } x\in\mathcal{X},
	\]
 where $\widehat{I}_{p}(x)=[\widehat{\Upsilon}^{(v)}(x,\widehat{\bw})\pm \cval\sqrt{\widehat{\Omega}(x)/n}]$
 and $\cval=\Phi^{-1}(1-\alpha/2)$.
\end{thm}
\begin{remark}[Robust Bias Correction]
    In practice, we suggest employing the robust bias correction 
    method \citep*{Calonico-Cattaneo-Farrell_2018_JASA,Calonico-Cattaneo-Farrell_2022_Bernoulli} to construct valid confidence intervals. Specifically, for a given $p$, let $J$ be the corresponding IMSE-optimal choice $J_{\mathtt{IMSE}}$ (see Section \ref{SA section: implementation} for implementation details). By Theorem \ref{SA thm: IMSE, LS} and Remark \ref{SA remark: bias lower bound} below, $J_{\mathtt{IMSE}}\asymp n^{\frac{1}{2p+3}}$ in general. Then, construct the confidence intervals $\widehat{I}_{p+q}(x)$ (i.e., use $(p+q)$th-order binscatter estimator). This particular choice of $J=J_{\mathtt{IMSE}}$ satisfies $nJ^{-2p-2q-3}=o(1)$ and $\frac{J^2\log^2 J}{n}=o(1)$.
	Then, the conclusion of Theorem \ref{SA thm: pointwise normality, LS} 
 immediately applies to $\widehat{I}_{p+q}(x)$ if $\nu=4$ and $\varsigma_\mu=\varsigma_w=p+q+1$.
\end{remark}

\begin{remark}[Improvements over literature]
	The results in this subsection improve upon \citet*[Section 5]{Cattaneo-Farrell-Feng_2020_AoS}, the best results available for partitioning-based estimation, by formally taking into account the randomness of the partition formed by empirical quantiles, and by accounting for the semi-linear regression estimation structure.
\end{remark}

\subsection{Integrated Mean Squared Error}\label{SA section: IMSE, LS}

\begin{thm}[IMSE] \label{SA thm: IMSE, LS}
	Suppose that Assumptions \ref{SA Assumption DGP} and \ref{SA Assumption LS} hold. Let $\omega(x)$ be a continuous weighting function over $\mathcal{X}$ bounded away from zero. If $\sqrt{n}J^{-p-(\varsigma_w\wedge (p+1))-\frac{3}{2}}=o(1)$, $\frac{J\log J}{n}=o(1)$ and 
 $\|\widehat{\bw}-\bw\|=o_\P(\sqrt{J/n}+J^{-p-1})$, then
    \[\begin{split}
	&\;\int_{\mathcal{X}}\E\Big[\Big(\widehat{\Upsilon}^{(v)}(x,\widehat{\bw})-\Upsilon_0^{(v)}(x, \bw)\Big)^2\Big|\bX, \bW\Big]\omega(x)dx\\
	=&\;\frac{J^{1+2v}}{n}\mathscr{V}_n(p,s,v) +
	J^{-2(p+1-v)}\mathscr{B}_n(p,s,v)+o_\P\Big(\frac{J^{1+2v}}{n}+J^{-2(p+1-v)}\Big),
	\end{split}\]
 
	where
	\[
	\begin{split}
		&\mathscr{V}_n(p,s,v):=J^{-(1+2v)}\tr\Big(
		\bQ_0^{-1}\bSigma_0\bQ_0^{-1}
		\int_{\mathcal{X}}\bb_{p,s}^{(v)}(x)\bb_{p,s}^{(v)}(x)'\omega(x)dx\Big)
		\asymp 1, \\
		&\mathscr{B}_n(p,s,v):=J^{2p+2-2v}\int_{\mathcal{X}}\Big(\bb_{p,s}^{(v)}(x)'\bbeta_0-\mu_0^{(v)}(x)\Big)^2\omega(x)dx\lesssim 1.
	\end{split}
	\]
\end{thm}

\begin{remark}[Proof of Theorem 1]
	Theorem 1 stated in the paper is a special case of Theorem \ref{SA thm: IMSE, LS}. In Theorem 1  
	we let $s=p$ and $\widehat\bw=\bar\bw$ and take $\omega(x)$ in Theorem \ref{SA thm: IMSE, LS} to be $f_X(x)$; 
	Assumption 1 implies that Assumption \ref{SA Assumption DGP} holds with $\varsigma_\mu=p+2$, and 
	Assumption \ref{SA Assumption LS} holds with $\nu=4$ and $\varsigma_w=p+2$; 
	and the rate condition 
	$\sqrt{n}J^{-p-(\varsigma_w\wedge (p+1))-\frac{3}{2}}=o(1)$ in Theorem \ref{SA thm: IMSE, LS} is equivalent to $nJ^{-4p-5}=o(1)$.
\end{remark}

As a consequence, the IMSE-optimal choice of $J$ is $J_{\mathtt{IMSE}}=J_{\mathtt{IMSE}}(p,s,v) \asymp n^{\frac{1}{2p+3}}$ whenever $\mathscr{B}_n(p,s,v)\gtrsim 1$. See Remark \ref{SA remark: bias lower bound} below for discussion of the lower bound on $\mathscr{B}_n(p,s,v)$. More precisely, if $\mathscr{B}_n(p,s,v)=\mathscr{B}(p,s,v)+o(1)$ and $\mathscr{V}_n(p,s,v)=\mathscr{V}(p,s,v)+o(1)$ for some constants $\mathscr{B}(p,s,v)$ and $\mathscr{V}(p,s,v)$, then we can take
\[
J_{\mathtt{IMSE}}=J_{\mathtt{IMSE}}(p,s,v) = \bigg\lceil 
\bigg(\frac{2(p-v+1)\mathscr{B}(p,s,v)}
{(1+2v)\mathscr{V}(p,s,v)}\bigg)^{\frac{1}{2p+3}} 
n^{\frac{1}{2p+3}}
\bigg\rceil.
\]

Regarding the bias component $\mathscr{B}_n(p, s, v)$, a more explicit but more cumbersome expression is available in the proof, which forms the foundation of our bin selection procedure discussed in Section \ref{SA section: implementation}. However, for $s=0$, both variance and bias terms admit concise explicit formulas, as shown in the following corollary. To state the results, we introduce a polynomial function $\mathscr{B}_p(x)=(-1)^p\sum_{k=0}^p\binom{p}{k}\binom{p+k}{k}(-x)^k/\binom{2p}{p}$ for $p\in\mathbb{Z}_+$. $\binom{2p}{p}\mathscr{B}_p(x)$ are usually termed the \textit{shifted} Legendre polynomials on $[0,1]$, which are orthogonal on $[0,1]$ with respect to the Lebesgue measure. Also, 
let $\bm{\varphi}(z)=(1, z, \ldots, z^p)'$.

\begin{coro} \label{SA coro: IMSE, LS}
	Under the assumptions in Theorem \ref{SA thm: IMSE, LS},
	$\mathscr{V}_n(p,0,v)=\mathscr{V}(p,0,v)+o(1)$ and
	$\mathscr{B}_n(p,0,v)=\mathscr{B}(p,0,v)+o(1)$ where
	\[
	\begin{split}
		&\mathscr{V}(p,0,v):=\tr\Big\{\Big(\int_{0}^{1}\bm{\varphi}(z)\bm{\varphi}(z)'dz\Big)^{-1}\int_0^1\bm{\varphi}^{(v)}(z)\bm{\varphi}^{(v)}(z)'dz\Big\}
		\int_{\mathcal{X}}\sigma^2(x)f_X(x)^{2v}\omega(x)dx, \\
		&\mathscr{B}(p,0,v):=\frac{\int_{0}^{1}[\mathscr{B}_{p+1-v}(z)]^2 dz}
		{((p+1-v)!)^2}\int_{\mathcal{X}}\frac{[\mu_0^{(p+1)}(x)]^2}{f_X(x)^{2p+2-2v}}\omega(x)dx.
	\end{split}
	\]
\end{coro}

\begin{remark}\label{SA remark: bias lower bound}
	The above corollary implies that the bias constant $\mathscr{B}(p,0,v)$ is nonzero unless $\mu_0^{(p+1)}(x)$ is zero almost everywhere on $\mathcal{X}$. For other $s>0$, notice that $\bb_{p,s}^{(v)}(x)'\bbeta_0$ can be viewed as an approximation of $\mu_0^{(v)}(x)$ in the space spanned by piecewise polynomials of order $(p-v)$. The best $L_2(x)$ approximation error in this space, according to the above corollary, is bounded away from zero if rescaled by $J^{p+1-v}$. $\bb_{p,s}^{(v)}(x)'\bbeta_0$, as a non-optimal $L_2$ approximation in such a space, must have a larger $L_2$ error than the best one (in terms of $L_2$-norm). Since $\omega(x)$ and $f_X(x)$ are both bounded and bounded away from zero, the above fact implies that except for the quite special case mentioned previously, $\mathscr{B}(p,s,v)\asymp 1$, a slightly stronger result than that in Theorem \ref{SA thm: IMSE, LS}. We exclude this special case by assuming that the leading bias is non-degenerate, and thus $J_{\mathtt{IMSE}}\asymp n^{\frac{1}{2p+3}}$.
\end{remark}

\begin{remark}[Improvements over literature]
	The results in this subsection improve upon \citet*[Section 4]{Cattaneo-Farrell-Feng_2020_AoS}, the best results available for partitioning-based estimation, by formally taking into account the randomness of the partition formed by empirical quantiles, and by accounting for the semi-linear regression estimation structure.
\end{remark}

\subsection{Uniform Distributional Approximation}

Recall that $(a_n: n\geq 1)$ is a sequence of non-vanishing constants. 
We will first show that the (feasible) Studentized $t$-statistic process $T_p(\cdot)$ can be approximated by a Gaussian process in a proper sense at certain rate.

\begin{thm}[Strong Approximation] \label{SA thm: strong approximation, LS}
	Suppose that Assumptions \ref{SA Assumption DGP} and \ref{SA Assumption LS} hold and $\|\widehat{\bw}-\bw\|=o_\P (a_n^{-1}\sqrt{J/n})$. If 
	\[ \frac{J(\log J)^2}{n^{1-\frac{2}{\nu}}}+J^{-1}+nJ^{-2p-3}=o(a_n^{-2}), 
	\]
	then, on a properly enriched probability space, there exists some $K_{p,s}$-dimensional standard normal random vector $\bN_{K_{p,s}}$ such that for any $\xi>0$,
	\[
	\P\Big(\sup_{x\in\mathcal{X}}
	|T_p(x)-Z_p(x)|>\xi a_n^{-1}\Big)=o(1), \quad
	Z_p(x)=\frac{\widehat{\bb}_{p,0}^{(v)}(x)'\bT_s'\bQ_0^{-1}\bSigma_0^{1/2}}{\sqrt{\Omega(x)}}\bN_{K_{p,s}}.
	\]
\end{thm}

The approximating process $(Z_p(x): x\in\mathcal{X})$ is a Gaussian process conditional on $\bX$ by construction. In practice, one can replace all unknowns in $Z_p(x)$ by their sample analogues, and then construct the following feasible (conditional) Gaussian process:
\[
\widehat{Z}_p(x)=\frac{\widehat{\bb}_{p,0}^{(v)}(x)'\widehat{\bT}_s'\widehat{\bQ}^{-1}\widehat{\bSigma}^{1/2}}{\sqrt{\widehat{\Omega}(x)}}\bN_{K_{p,s}}^\star=
\frac{\widehat{\bb}_{p,s}^{(v)}(x)'\widehat{\bQ}^{-1}\widehat{\bSigma}^{1/2}}{\sqrt{\widehat{\Omega}(x)}}\bN_{K_{p,s}}^\star,
\]
where $\bN_{K_{p,s}}^\star$ denotes a $K_{p,s}$-dimensional standard normal vector independent of the data $\bD$.

\begin{thm}[Plug-in Approximation] \label{SA thm: plug-in approx, LS}
	Suppose that the conditions in Theorem \ref{SA thm: strong approximation, LS} hold. Then, on a properly enriched probability space there exists a $K_{p,s}$-dimensional standard normal random vector $\bN_{K_{p,s}}^\star$ independent of $\bD$ such that for any $\xi>0$,
	\[
	\P\Big(\sup_{x\in\mathcal{X}}|\widehat{Z}_p(x)-Z_p(x)|>\xi a_n^{-1}\Big|\bD\Big)=o_\P(1).
	\]
\end{thm}

\begin{remark}[Proof of Theorem 2]
	Theorem 2 in the paper is a special case of Theorems \ref{SA thm: strong approximation, LS} and \ref{SA thm: plug-in approx, LS}. In Theorem 2 we let $s=p$ and $\widehat\bw=\bar\bw$; 
    Assumption 1 imposed in the paper implies that Assumption \ref{SA Assumption DGP} holds with $\varsigma_\mu=p+2$ and Assumption \ref{SA Assumption LS} holds with $\varsigma_w=p+2$ and $\nu=4$. Therefore, the desired strong approximation for $\widehat{\Upsilon}^{(v)}(x,\widehat{\bw})$ follows from Theorem \ref{SA thm: strong approximation, LS} and Theorem \ref{SA thm: plug-in approx, LS}. For ease of presentation, Theorem 2 in the paper defines 
	$$Z_p(x)=\frac{\widehat{\bb}_{p,s}^{(v)}(x)'\bQ_0^{-1}\bSigma_0^{1/2}}{\sqrt{\Omega(x)}}\bN_{K_{p,s}}
	=\frac{\widehat{\bb}_{p,0}^{(v)}(x)'\widehat{\bT}_s\bQ_0^{-1}\bSigma_0^{1/2}}{\sqrt{\Omega(x)}}\bN_{K_{p,s}}.
	$$
	That is, we replace $\bT_s$ in Theorem \ref{SA thm: strong approximation, LS} with $\widehat{\bT}_s$.  
	As shown in the proof of Theorem \ref{SA thm: strong approximation, LS} (see Step 3 therein), this does not affect the strong approximation result. 
\end{remark}

\begin{remark}[Improvements over literature]
	Theorems \ref{SA thm: strong approximation, LS} and \ref{SA thm: plug-in approx, LS} offer a new easy-to-implement approach to conduct binscatter-based uniform distributional approximation and inference. We formally take into account the randomness of the empirical-quantile-based partition and approximate the \textit{whole} $t$-statistic process by a (conditional) Gaussian process under seemingly minimal rate conditions. In fact, it can be shown that when $a_n=\sqrt{\log n}$ and a subexponential moment restriction holds for the error term, it suffices that  $J/n=o(1)$, up to $\log n$ terms. In contrast, a strong approximation of the $t$-statistic process for general series estimators was obtained based on Yurinskii coupling in \cite*{Belloni-Chernozhukov-Chetverikov-Kato_2015_JoE}, which requires $J^5/n=o(1)$, up to $\log n$ terms.  Alternatively, a strong approximation of the \textit{supremum} of the $t$-statistic process can be obtained under weaker rate restrictions. For instance, \cite*{Chernozhukov-Chetverikov-Kato_2014a_AoS} requires $J/n^{1-2/\nu}=o(1)$, up to $\log n$ terms, a result that applies exclusively to the suprema of the stochastic process.
\end{remark}

Theorems \ref{SA thm: strong approximation, LS} and \ref{SA thm: plug-in approx, LS} offer a way to approximate the distribution of the \textit{whole} $t$-statistic process based on $\widehat{\Upsilon}^{(v)}(\cdot,\widehat{\bw})$. One direct application of these results is to approximate the supremum of the $t$-statistic process. The following theorem shows that our strong approximation results can be used to obtain the convergence of the Kolmogorov distance between the distributions of $\sup_{x\in\mathcal{X}}|T_p(x)|$ and its (conditionally) Gaussian analogue $\sup_{x\in\mathcal{X}}|\widehat{Z}_p(x)|$.

\begin{thm}[Supremum Approximation] \label{SA thm: sup approx}
	Let $a_n=\sqrt{\log J}$. Suppose that the conditions of Theorem \ref{SA thm: strong approximation, LS} hold. Then,
	\[\sup_{u\in\mathbb{R}}\Big|\P\Big(\sup_{x\in\mathcal{X}}
	|T_p(x)|\leq u\Big)-
	\P\Big(\sup_{x\in\mathcal{X}}|\widehat{Z}_p(x)|\leq u\Big|\bD\Big)\Big|=o_\P(1).
	\]
\end{thm}


\subsection{Uniform Inference}\label{SA sec: uniform inference}

One important application of the strong approximation results in Theorems \ref{SA thm: strong approximation, LS} and \ref{SA thm: plug-in approx, LS} is to construct uniform confidence bands. 
Let $\widehat{I}_{p}(x)=[\widehat{\Upsilon}^{(v)}(x,\widehat{\bw})\pm \cval\sqrt{\widehat{\Omega}(x)/n}]$ for some critical value $\cval$ to be specified, which is constructed based on a certain choice of $J$ and the $p$th-order binscatter basis.

\begin{thm}\label{SA thm: robust CB} 
     Let $a_n=\sqrt{\log J}$. Suppose that the conditions in Theorem \ref{SA thm: strong approximation, LS} hold.  
	If $\cval=\inf\Big\{c\in\mathbb{R}_+:\P[\sup_{x\in\mathcal{X}}|\widehat{Z}_{p}(x)|\leq c \;|\bD]\geq 1-\alpha\Big\}$, then
	\[
	\P\Big[\Upsilon^{(v)}_0(x,\bw)\in\widehat{I}_{p}(x),\text{ for all }x\in\mathcal{X}\Big]=1-\alpha+o(1).
	\]
\end{thm}
\begin{remark}[Robust Bias Correction]
    In practice, we suggest employing the robust bias correction 
    method to construct valid confidence bands. Specifically,
    for a given $p$, let $J$ be the corresponding IMSE-optimal choice $J_{\mathtt{IMSE}}$ (see Section \ref{SA section: implementation} for implementation details). By Theorem \ref{SA thm: IMSE, LS} and Remark \ref{SA remark: bias lower bound}, $J_{\mathtt{IMSE}}\asymp n^{\frac{1}{2p+3}}$ in general. Then, construct the confidence band $\widehat{I}_{p+q}(x)$ (i.e., use $(p+q)$th-order binscatter estimator). This particular choice of $J=J_{\mathtt{IMSE}}$ satisfies 
    \[
	\frac{J(\log n)^2}{\sqrt{n}}+J^{-1}+
	nJ^{-2(p+1)-3}=o(\log n^{-1}).
    \]
    Then, the conclusion of Theorem \ref{SA thm: robust CB} immediately applies to $\widehat{I}_{p+q}(x)$ if $\nu=4$ and $\varsigma_\mu=\varsigma_w=p+q+1$.

    In the paper we considered one special case of such robust bias-corrected confidence band: let $J$ be the IMSE-optimal choice corresponding to $p=s=v=0$, and construct the confidence band $\widehat{I}_1(x)$ (i.e., let $q=1$ in the above construction). 
\end{remark}

\begin{remark}
	The above results construct valid uniform confidence bands for least squares binscatter estimators under mild rate restrictions. Specifically, when $\nu\geq 4$, we require $J^{2}/n=o(1)$, up to $\log n$ terms. By contrast, \cite*{Belloni-Chernozhukov-Chetverikov-Kato_2015_JoE} considers general series-based least squares estimators, and Theorem 5.6 therein can construct confidence bands under similar rate restrictions, which relies on the strong approximation technique for the suprema of the stochastic process developed in \cite*{Chernozhukov-Chetverikov-Kato_2014a_AoS}.
\end{remark}

Using our main theoretical results, we can also test parametric specifications of the unknown function $\Upsilon_0^{(v)}(x,\bw)$. Consider the following testing problem:
\begin{align*}
\dot{\mathsf{H}}_0&:\quad \sup_{x\in\mathcal{X}} \Big|\Upsilon^{(v)}_0(x,\bw) - M^{(v)}(x,\bw;\btheta,\bgamma_0)\Big|=0, \quad \text{ for some } \btheta, \qquad vs.\\
\dot{\mathsf{H}}_\text{A}&:\quad \sup_{x\in\mathcal{X}} \Big|\Upsilon^{(v)}_0(x,\bw) - M^{(v)}(x,\bw;\btheta,\bgamma_0)\Big|>0, \quad \text{ for all } \btheta.
\end{align*}
where $M(x,\bw;\btheta, \bgamma_0)=m(x;\btheta)+\bw'\bgamma_0$.
This testing problem can be viewed as a two-sided test where the equality between two functions holds \textit{uniformly} over $x\in\mathcal{X}$. We introduce $\widetilde{\btheta}$ and $\widetilde{\bgamma}$ as consistent estimators of $\btheta$ and and $\bgamma_0$ under $\dot{\mathsf{H}}_0$, and then consider the following test statistic:
\[
\dot{T}_p(x)
:=\frac{\widehat{\Upsilon}^{(v)}(x,\widehat{\bw})-M^{(v)}(x,\widehat{\bw};\widetilde{\btheta},\widetilde{\bgamma})}
{\sqrt{\widehat{\Omega}(x)/n}}.
\]
The null hypothesis is rejected if $\sup_{x\in\mathcal{X}}|\dot{T}_p(x)|>\cval$ for some critical value $\cval$.

\begin{thm}[Parametric Specification Tests] \label{SA thm: testing specification}
	Let $a_n=\sqrt{\log J}$. Suppose that the conditions in Theorem \ref{SA thm: strong approximation, LS} hold. 
		Let $\cval=\inf\{c\in\mathbb{R}_+: \P[\sup_{x\in\mathcal{X}}|\widehat{Z}_p(x)|\leq c |\bD]\geq 1-\alpha \}$. 
		
		Under $\dot{\mathsf{H}}_0$, if 
		$\sup_{x\in\mathcal{X}} |\Upsilon^{(v)}(x,\bw)-M^{(v)}
		(x,\widehat{\bw};\widetilde{\btheta},\widetilde{\bgamma})|=o_\P\Big(\sqrt{\frac{J^{1+2v}}{n\log J}}\Big)$, then 
		$$
		\lim_{n\rightarrow\infty}\P\Big[\sup_{x\in\mathcal{X}}|\dot{T}_p(x)|>
		\cval\Big]=\alpha.
		$$
		
		Under $\dot{\mathsf{H}}_{\text{A}}$, if there exist some fixed $\bar{\btheta}$ and $\bar{\bgamma}$ such that $\sup_{x\in\mathcal{X}} |M^{(v)}(x,\widehat{\bw};\widetilde{\btheta},\widetilde{\bgamma})-M^{(v)}(x,\bw;\bar{\btheta}, \bar{\bgamma})|=o_\P(1)$, and $J^v\Big(\frac{J\log J}{n}\Big)^{1/2}=o(1)$, then 
		\[
		\lim_{n\rightarrow\infty}\P\Big[\sup_{x\in\mathcal{X}}
		|\dot{T}_p(x)|>\cval\Big]=1.
		\]
\end{thm}

\begin{remark}[Robust Bias Correction]
    In practice, we suggest employing the robust bias correction 
    method to conduct specification tests. Specifically,
    for a given $p$, let $J$ be the corresponding IMSE-optimal choice $J_{\mathtt{IMSE}}$ (see Section \ref{SA section: implementation} for implementation details). By Theorem \ref{SA thm: IMSE, LS} and Remark \ref{SA remark: bias lower bound}, $J_{\mathtt{IMSE}}\asymp n^{\frac{1}{2p+3}}$ in general. Then, construct the $t$-statistic $\dot{T}_{p+q}(x)$, (i.e., use $(p+q)$th-order binscatter estimator). This particular choice of $J=J_{\mathtt{IMSE}}$ satisfies 
    \[
	\frac{J(\log n)^2}{\sqrt{n}}+J^{-1}+
	nJ^{-2(p+1)-3}=o(\log n^{-1}).
    \]
    Also, $\frac{J^{1+2v}
	(\log J)}{n}\asymp  
n^{-\frac{2p-2v+2}{2p+3}}\log n=o(1)$ since we always require $p\geq v$.
    Then, the conclusion of Theorem \ref{SA thm: testing specification} 
 immediately applies to the test based on $\dot{T}_{p+q}(x)$ if $\nu=4$ and $\varsigma_\mu=\varsigma_w=p+q+1$.
\end{remark}

Another application of our theoretical results is to test certain shape restrictions on the unknown $\Upsilon_0^{(v)}(x,\bw)$. 
To be specific, consider the following testing problem:
\[\begin{split}
	&\ddot{\mathsf{H}}_0:\; 
	\sup_{x\in\mathcal{X}}\, (\Upsilon_0^{(v)}(x,\bw)-M^{(v)}(x,\bw; \bar{\btheta},\bar{\bgamma}))\leq 0 \text{ for certain } \bar\btheta \text{ and } \bar{\bgamma} \quad \text{v.s.}\\
	&\ddot{\mathsf{H}}_\text{A}:\;
	\sup_{x\in\mathcal{X}}\, (\Upsilon_0^{(v)}(x,\bw) -
	M^{(v)}(x,\bw;\bar{\btheta},\bar{\bgamma}))>0 \text{ for } \bar\btheta \text{ and } \bar{\bgamma},
\end{split}\]
which can be viewed as a one-sided test where the inequality holds \textit{uniformly} over $x\in\mathcal{X}$. Importantly, it should be noted that under both $\ddot{\mathsf{H}}_0$ and  $\ddot{\mathsf{H}}_\text{A}$, we fix $\bar{\btheta}$  and $\bar{\bgamma}$ to be the same values in the parameter space. We introduce 
$\widetilde{\btheta}$ and $\widetilde{\bgamma}$ as consistent estimators of $\bar{\btheta}$ and $\bar{\bgamma}$ under both $\ddot{\mathsf{H}}_0$ and $\ddot{\mathsf{H}}_\text{A}$, and then rely on the following test statistic:
\[
\ddot{T}_p(x):=
\frac{\widehat{\Upsilon}^{(v)}(x,\widehat{\bw})-M^{(v)}(x,\widehat{\bw};\widetilde{\btheta},\widetilde{\bgamma})}
{\sqrt{\widehat{\Omega}(x)/n}}.
\]
The null hypothesis is rejected if $\sup_{x\in\mathcal{X}} \ddot{T}_p(x)>\cval$ for some critical value $\cval$.

The following theorem characterizes the size and power of such tests.

\begin{thm}[Shape Restriction Tests] \label{SA thm: testing shape restriction}
	Let $a_n=\sqrt{\log J}$. Suppose that the conditions in Theorem \ref{SA thm: strong approximation, LS} hold.
	In addition, 
	$\sup_{x\in\mathcal{X}} |M^{(v)}(x,\widehat{\bw};\widetilde{\btheta},\widetilde{\bgamma})-
	M^{(v)}(x,\bw;\bar{\btheta},\bar{\bgamma})|=
	o_\P\Big(\sqrt{\frac{J^{1+2v}}{n\log J}}\Big)$. 
		Let $\cval=\inf\{c\in\mathbb{R}_+: \P[\sup_{x\in\mathcal{X}}\widehat{Z}_p(x)\leq c
		|\bD]\geq 1-\alpha \}$. 
		
		Under $\ddot{\mathsf{H}}_0$, 
		\[
		\lim_{n\rightarrow\infty}\P\Big[\sup_{x\in\mathcal{X}}
		\ddot{T}_p(x)>\cval\Big]\leq \alpha.
		\]
		
		Under $\ddot{\mathsf{H}}_\text{A}$, if $J^v\Big(\frac{J\log J}{n}\Big)^{1/2}=o(1)$, 
		\[
		\lim_{n\rightarrow\infty}\P\Big[\sup_{x\in\mathcal{X}}
		\ddot{T}_p(x)>\cval\Big]=1.
		\]
\end{thm}

\begin{remark}[Robust Bias Correction]
    In practice, we suggest employing the robust bias correction 
    method to conduct shape restriction tests. Specifically,
    for a given $p$, let $J$ be the corresponding IMSE-optimal choice $J_{\mathtt{IMSE}}$ (see Section \ref{SA section: implementation} for implementation details). By Theorem \ref{SA thm: IMSE, LS} and Remark \ref{SA remark: bias lower bound}, $J_{\mathtt{IMSE}}\asymp n^{\frac{1}{2p+3}}$ in general. Then, construct the $t$-statistic $\ddot{T}_{p+q}(x)$, (i.e., use $(p+q)$th-order binscatter estimator). This particular choice of $J=J_{\mathtt{IMSE}}$ satisfies 
    \[
	\frac{J(\log n)^2}{\sqrt{n}}+J^{-1}+
	nJ^{-2(p+1)-3}=o(\log n^{-1}).
    \]
    Also, $\frac{J^{1+2v}
	(\log J)}{n}\asymp  
n^{-\frac{2p-2v+2}{2p+3}}\log n=o(1)$ since we always require $p\geq v$.
    Then, the conclusion of Theorem \ref{SA thm: testing shape restriction} immediately applies to the test based on $\ddot{T}_{p+q}(x)$ if $\nu=4$ and $\varsigma_\mu=\varsigma_w=p+q+1$.
\end{remark}

\begin{remark}[Improvements over literature]
	The results presented in this section improve on the literature, even in the case of non-random partitioning and without covariate-adjustments, because they take advantage of the specific binscatter structure (i.e., locally bounded series basis), thereby offering faster approximation rates under weaker side restrictions \citep*[c.f.,][]{Belloni-Chernozhukov-Chetverikov-Kato_2015_JoE,Cattaneo-Farrell-Feng_2020_AoS}. Furthermore, relative to prior work, our results formally take into account the randomness of the partition formed by empirical quantiles, account for the semi-linear regression estimation structure, and consider an array of inference problems. In particular, the underlying approach to establish strong approximation and related distributional approximations for binscatter statistics may be of independent interest.
\end{remark}

\section{Feasible Number of Bins Selector} 
\label{SA section: implementation}

We discuss the implementation details for data-driven selection of the number of bins, based on the integrated mean squared error expansion for least squares binscatter estimators (see Theorem \ref{SA thm: IMSE, LS} and Corollary \ref{SA coro: IMSE, LS}). 
Thus, the selectors given below can provide a choice of $J$ that is optimal in the IMSE sense.

We offer two procedures for estimating the bias and variance constants, and once these estimates ($\widehat{\mathscr{B}}_n(p,s,v)$ and $\widehat{\mathscr{V}}_n(p,s,v)$) are available, the estimated optimal $J$ is
\[
\widehat{J}_{\mathtt{IMSE}}=\widehat{J}_{\mathtt{IMSE}}(p,s,v)
=\bigg\lceil 
\bigg(\frac{2(p-v+1)\widehat{\mathscr{B}}_n(p,s,v)}
{(1+2v)\widehat{\mathscr{V}}_n(p,s,v)}\bigg)^{\frac{1}{2p+3}} 
n^{\frac{1}{2p+3}}
\bigg\rceil.
\]
We always let $\omega(x)=f_X(x)$ as weighting function for concreteness.

\subsection{Rule-of-thumb Selector}

A rule-of-thumb choice of $J$ is obtained based on Corollary \ref{SA coro: IMSE, LS}, in which case $s=0$.

Regarding the variance constants $\mathscr{V}(p,0,v)$, the unknowns 
are the density function $f_X(x)$ and the conditional variance $\sigma^2(x)$. A Gaussian reference model is employed to get the estimate $\widehat{f}_X$ of $f_X(x)$. For the conditional variance, recall $\sigma^2(x_i,\bw_i)=\E[y_i^2|x_i, \bw_i]-(\E[y_i|x_i,\bw_i])^2$, where 
the two conditional expectations can be approximated by global polynomial regressions of degree $p+1$. Let $\widehat\sigma^2(x_i,\bw_i)$ denote the resulting estimate. 
Then, the variance constant is estimated by
\[
\widehat{\mathscr{V}}(p,0,v)=\tr\Big\{\Big(\int_{0}^{1}\bm{\varphi}(z)\bm{\varphi}(z)'dz\Big)^{-1}\int_0^1\bm{\varphi}^{(v)}(z)\bm{\varphi}^{(v)}(z)'dz\Big\}\times
\frac{1}{n}\sum_{i=1}^{n}\widehat{\sigma}^2(x_i,\bw_i)\widehat{f}_X(x_i)^{2v}.
\]

Regarding the bias constant, the unknowns are $f_X(x)$, which is estimated using the Gaussian reference model, and $\mu_0^{(p+1)}(x)$, which can be estimated based on the global polynomial regression that approximates $\E[y_i|x_i, \bw_i]$.
Then, the bias constant is estimated by
\[
\widehat{\mathscr{B}}(p,0,v)=\frac{\int_{0}^{1}[\mathscr{B}_{p+1-v}(z)]^2 dz}{((p+1-v)!)^2}\times
\frac{1}{n}\sum_{i=1}^{n}\frac{[\widehat{\mu}^{(p+1)}(x_i)]^2}{\widehat{f}_X(x_i)^{2p+2-2v}}.
\]

The resulting $J$ selector employs the correct rate but an inconsistent constant approximation. Recall that $s$ does not change the rate of $J_{\mathtt{IMSE}}$. Thus, even for other $s>0$, this selector still gives a correct rate.

\subsection{Direct-plug-in Selector}

The direct-plug-in selector is implemented based on binscatter estimators, which applies to any user-specified $p$, $s$ and $v$. It requires a preliminary choice of $J$, for which the rule-of-thumb selector previously described can be used.

More generally, suppose that a preliminary choice $J_{\mathtt{pre}}$ is given, and then a binscatter basis $\widehat{\bb}_{p,s}(x)$ (of order $p$) can be constructed immediately on the preliminary partition. Implementing a binscatter regression using this basis and partitioning, we can obtain the variance constant estimate using a standard variance estimator, such as the one in Theorem \ref{SA thm: meat matrix, LS}. 

Regarding the bias constant, we employ the uniform approximation \eqref{SA eq: uniform approx} in the proof of Theorem \ref{SA thm: IMSE, LS}. The key idea of the bias representation is to ``orthogonalize" the leading error of the uniform approximation based on splines with simple knots (i.e., $p$ smoothness constraints are imposed)  with respect to the preliminary binscatter basis $\widehat{\bb}_{p,s}(x)$. Specifically, the key unknown in the expression of the leading error is  $\mu_0^{(p+1)}(x)$, which can be estimated by implementing a binscatter regression of order $p+1$ (with the preliminary partition unchanged).
Plug it in \eqref{SA eq: orth approx error}, and all other quantities in that equation can be replaced by their sample analogues. Then, a bias constant estimate is available.

By this construction, the direct-plug-in selector employs the correct rate and a consistent constant approximation for any $p$, $s$ and $v$.

\section{Proofs} \label{SA section: proof}

\subsection{Proof of Lemma \ref{SA lem: quantile partition}}

\begin{proof}
	The first result follows by Lemma SA2 of \citet*{Calonico-Cattaneo-Titiunik_2015_JASA}. To show the second result, first consider the deterministic partition sequence $\Delta_0$ based on the population quantiles. By the mean value theorem,
	\[ h_{j}=F_X^{-1}\Big(\frac{j}{J}\Big)-F_X^{-1}\Big(\frac{j-1}{J}\Big)
	=\frac{1}{f_X(F_X^{-1}(\xi))}\cdot \frac{1}{J},
	\]
	where $\xi$ is some point between $(j-1)/J$ and $j/J$.
	Since $f_X$ is bounded and bounded away from zero, 
	$\max_{1\leq j\leq J}h_j/\min_{1\leq j\leq J}h_j\leq \bar{f}_X/\underline{f}_X$. Using the first result, we have with probability approaching one, 	
	\[
	\max_{1\leq j\leq J}|\hat{h}_j-h_{j}|\leq J^{-1}\bar{f}_X^{-1}/2.
	\]
	Then, 
	\[
	\frac{\max_{1\leq j\leq J}\hat{h}_j}{\min_{1\leq j\leq J}\hat{h}_j}=
	\frac{\max_{1\leq j\leq J}h_{j}+\max_{1\leq j \leq J}|\hat{h}_j-h_j|}
	{\min_{1\leq j\leq J}h_{j}-\max_{1\leq j \leq J}|\hat{h}_j-h_j|}
	\leq \frac{3\bar{f}_X}{\underline{f}_X},
	\]
	and the desired result follows.
\end{proof}

\subsection{Proof of Lemma \ref{SA lem: spline transform}}

\begin{proof}
	For $s=0$, the result is trivial. For $0< s \leq p$, $\widehat{\bb}_{p,s}(x)$ is formally known as $B$-spline basis of order $p+1$ with knots $\{\hat{\tau}_1, \ldots, \hat{\tau}_{J-1}\}$ of multiplicities $(p-s+1, \ldots, p-s+1)$. See \citet[Definition 4.1]{Schumaker_2007_book}. Without loss of generality, suppose $\mathcal{X}=[0,1]$. Specifically, such a basis is constructed on an extended knot sequence $\{\xi_j\}_{j=1}^{2(p+1)+(p-s+1)(J-1)}$:
	\[
	\xi_1\leq\cdots\leq\xi_{p+1}\leq 0, \quad 1\leq\xi_{p+2+(p-s+1)(J-1)}\leq\cdots\leq \xi_{2(p+1)+(p-s+1)(J-1)}.
	\]
	and
	\[
	\xi_{p+2}\leq\cdots\leq \xi_{p+1+(p-s+1)(J-1)}=
	\underbrace{\hat{\tau}_1, \cdots, \hat{\tau}_1}_{p-s+1}, \cdots,
	\underbrace{\hat{\tau}_{J-1}, \cdots, \hat{\tau}_{J-1}}_{p-s+1}.
	\]
	
	By the well-known Recursive Relation of Splines, a typical function 
	$\widehat{b}_{p,s,\ell}(x)$ in $\widehat{\bb}_{p,s}(x)$ supported on $(\xi_{\ell}, \xi_{\ell+p+1})$ is expressed as
	\[
	\widehat{b}_{p,s,\ell}(x)=\sqrt{J}\sum_{j=\ell+1}^{\ell+p+1}
	C_{j}(x)
	\I(x\in[\xi_{j-1},\xi_{j})).
	\]
	where each $C_j(x)$ is a polynomial of degree $p$ as the sum of products of $p$ linear polynomials. See \citet[Section IX, Equation (19)]{DeBoor_1978_book}.
	Since $s\leq p$, we always have $\xi_{\ell}<\xi_{\ell+p+1}$. Thus, the support of such a basis function is well defined.
	Specifically, all $C_j(x)$s take the following form:
	\[
	C_j(x)=\sum_{\iota=1}^{M}
	\prod\limits_{(k,k')\in\mathcal{K}_\iota}
	\frac{(-1)^{c_{k,k'}}(x-\xi_k)}{\xi_k-\xi_{k'}}.
	\]
	Here, the convention is that ``$0/0=0$'', $M\leq 2^p$ is a constant denoting the number of summands, the cardinality of the  set $\mathcal{K}_s$ of index pairs is exactly $p$, and $c_{k,k'}$ is a constant used to change the sign of the summand. These indices may depend on $j$, which is omitted for notation simplicity. As explained previously, such a function is supported on at least one bin.
	
	We want to linearly represent $b_{p,s,\ell}(x)$ in terms of $\bb_{p,0}(x)$ with typical element 
	\begin{equation}\label{SA eq: piecewise power series}
		\varphi_{j,\alpha}(x)=\sqrt{J}\cdot\I_{\widehat{\mathcal{B}}_j}(x)\Big(\frac{x-\hat{\tau}_{j-1}}{\hat{h}_j} \Big)^\alpha,\quad
		0\leq\alpha\leq p,\quad 1\leq j\leq J.
	\end{equation} 
	Suppose without loss of generality, $\xi_{j-1}<\xi_j$ and $(\xi_{j-1}, \xi_j)$ is a cell within the support of $\widehat{b}_{p,s,\ell}(x)$.
	Let $c_{j,\alpha}$ be the coefficient of $\varphi_{j, \alpha}(x)$ in the linear representation of $\widehat{\bb}_{p,s}(x)$. Using the above results, it takes the following form
	\[
	c_{j,\alpha}=\sum_{\iota=1}^M 
	\frac{(\xi_{j}-\xi_{j-1})^\alpha
		\sum_{l_\iota=1}^{C_{p,\alpha}}
		\prod_{k=k_{l_\iota,1}}^{k_{l_\iota,p-\alpha}}(\xi_{j-1}-\xi_k)}
	{\prod\limits_{(k,k')\in\mathcal{K_\iota}}(-1)^{c_{k,k'}}(\xi_k-\xi_{k'})}.
	\]
	The quantities within the summation only depend on distance between knots, which is no greater than $(p+1)\max_{j}\hat{h}_j$ 
	since the support covers at most $(p+1)$ bins. Both denominator and numerator are products of $p$ such distances, and hence by Lemma \ref{SA lem: quantile partition}, $\sup_{j,\alpha}|c_{j,\alpha}|\lesssim_\P 1$. Then, $b_{p,s,\ell}(x)$ can be written as 
	\[
	b_{p,s,\ell}(x)=\sum_{j: \mathcal{B}_j\subset[\xi_\ell, \xi_{\ell+p+1}]}\sum_{\alpha=0}^pc_{j,\alpha}\psi_{j,\alpha}(x).
	\]
	The above expression gives the elements of the $\ell$th row of $\widehat{\bT}_s$.
	
	Since each row and each column of $\widehat{\mathbf{T}}_s$ only contain a finite number of nonzeros,	
	$\|\widehat{\mathbf{T}}_s\|_\infty\lesssim_\P 1$ and $\|\widehat{\bT}_s\|\lesssim_\P 1$. Using the fact $\max_{1\leq j\leq J}|\hat{h}_j-h_j|\lesssim_\P J^{-1}\sqrt{J\log J/n}$ given in the proof of Lemma \ref{SA lem: quantile partition}, and noticing the form of $c_{j,\alpha}$, $\max_{k,l}|(\widehat{\mathbf{T}}_s-\mathbf{T}_s)_{k,l}|\lesssim \sqrt{J\log J/n}$ where $(\widehat{\mathbf{T}}_s-\mathbf{T}_s)_{k,l}$ is $(k,l)$th element of $\widehat{\mathbf{T}}_s-\mathbf{T}_s$. Since  $(\widehat{\mathbf{T}}_s-\mathbf{T}_s)$ only has a finite number of nonzeros on every row and column, $\|\widehat{\mathbf{T}}_s-\mathbf{T}_s\|_\infty\lesssim_\P\sqrt{J\log J/n}$ and $\|\widehat{\mathbf{T}}_s-\mathbf{T}_s\|\lesssim_\P\sqrt{J\log J/n}$.
	
	Finally, we give an explicit expression of $c_{j,\alpha}$ for the case $s=p$, which may be of independent interest. In this case, $\mathbf{\bb}_{p,p}(x)$ is the usual $B$-spline basis with simple knots. Let $\widehat{b}_{p,p,\ell}(x)$ be a typical basis function supported on $[\hat{\tau}_\ell, \hat{\tau}_{\ell+p+1}]$. Then,
	using the recursive formula of $B$-splines, by induction we have
	\begin{equation} \label{SA eq: spline as truncated} 
		\widehat{b}_{p, p,\ell}(x)=(\hat{\tau}_{\ell+p+1}-\hat{\tau}_\ell)
		\sum_{j=\ell}^{\ell+p+1}\frac{(x-\hat{\tau}_j)^{p}_+}
		{\prod_{k=\ell\atop k\neq j}^{\ell+p+1}(\hat{\tau}_k-\hat{\tau}_j)},
	\end{equation}
	where $(z)_+$ equals to $z$ if $z\geq0$ and $0$ otherwise. Since $\widehat{b}_{p,p,\ell}(x)$ is zero outside of $(\hat{\tau}_\ell, \hat{\tau}_{\ell+p+1})$, $\widehat{b}_{p,p,\ell}(x)$ can be written as a linear combination of $\varphi_{j, \alpha}(x)$, $j=\ell+1,\ldots, \ell+p+1, \alpha=0, \ldots, p$:
	\begin{equation} \label{SA eq: spline as piecewise}
		\widehat{b}_{p,p,\ell}(x)=\sum_{\alpha=0}^{p}\sum_{j=\ell+1}^{\ell+p+1}
		c_{j,\alpha}\varphi_{j, \alpha}(x),\quad \text{for some} \;
		c_{j,\alpha}.
	\end{equation}
	For a generic cell $(\hat{\tau}_{j-1}, \hat{\tau}_{j})
	\subset(\hat{\tau}_\ell, \hat{\tau}_{\ell+p+1})$, all truncated polynomials $(x-\hat{\tau}_k)_+^{p}$ does not contribute to the coefficients of $\varphi_{j, \alpha}(x)$ if $k>j-1$. For any $\ell\leq k\leq j-1$, we can expand $(x-\hat{\tau}_{k})_+^{p}$ on $(\hat{\tau}_{j-1}, \hat{\tau}_{j})$ as
	\[
	(x-\hat{\tau}_k)^{p}=(x-\hat{\tau}_{j-1}+\hat{\tau}_{j-1}-\hat{\tau}_k)^{p}=\sum_{\alpha=0}^{p}\binom{p}{\alpha}
	\Big(\frac{x-\hat{\tau}_{j-1}}{\hat{\tau}_{j}-\hat{\tau}_{j-1}}\Big)^{\alpha}(\hat{\tau}_{j-1}-\hat{\tau}_k)^{p-\alpha}(\hat{\tau}_{j}-\hat{\tau}_{j-1})^\alpha.
	\]
	Thus, the contribution of $(x-\hat{\tau}_k)_+^{p}$ to the coefficients of $\varphi_{j, \alpha}(x)$ in Equation \eqref{SA eq: spline as piecewise}, combined with its coefficient in Equation \eqref{SA eq: spline as truncated}, is \[\binom{p}{\alpha}(\hat{\tau}_{j-1}-\hat{\tau}_k)^{p-\alpha}
	(\hat{\tau}_{j}-\hat{\tau}_{j-1})^\alpha(\hat{\tau}_{\ell+p+1}-\hat{\tau}_\ell)\Big(\prod_{k'=\ell\atop k'\neq k}^{\ell+p+1}(\hat{\tau}_{k'}-\hat{\tau}_k)\Big)^{-1}.
	\]
	Collecting all such coefficients contributed by $(x-\hat{\tau}_k)_+^{p}$, $k=\ell, \ldots, j-1$, we obtain
	\[ c_{j, \alpha}=\sum_{k=\ell}^{j-1}\binom{p}{\alpha}(\hat{\tau}_{j-1}-\hat{\tau}_k)^{p-\alpha}(\hat{\tau}_{j}-\hat{\tau}_{j-1})^\alpha
	(\hat{\tau}_{\ell+p+1}-\hat{\tau}_\ell)
	\Big(\prod_{k'=\ell\atop k'\neq k}^{\ell+p+1}(\hat{\tau}_{k'}-\hat{\tau}_k)\Big)^{-1}.
	\]
\end{proof}

\subsection{Proof of Lemma \ref{SA lem: local basis}}
\begin{proof}
		The sparsity of the basis follows by construction. 	To show the bound on $\|\widehat{\bb}_{p,s}^{(v)}(x)\|$, notice that when $s=0$, for any $x\in\mathcal{X}$ and any $j=1, \ldots, J(p+1)$, $0\leq \widehat{b}_{p,0,j}(x)\leq \sqrt{J}$.
		Define $\varphi_{j, \alpha}(x)$ as in Equation \eqref{SA eq: piecewise power series}. Since 
		\[
		\varphi_{j, \alpha}^{(v)}=\sqrt{J}\alpha(\alpha-1)\cdots(\alpha-v+1)\hat{h}_j^{-v}\I_{\widehat{\mathcal{B}}_j}(x)
		\Big(\frac{x-\hat{\tau}_{j-1}}{\hat{h}_j}\Big)^{\alpha-v}\lesssim \sqrt{J}\hat{h}_j^{-v},
		\]
		the bound on $\|\widehat{\bb}_{p,s}^{(v)}(x)\|$ simply follows from Lemma \ref{SA lem: quantile partition} and Lemma \ref{SA lem: spline transform}.
\end{proof}

\subsection{Proof of Lemma \ref{SA lem: uniform approx rate}}

\begin{proof}
	By Lemma \ref{SA lem: quantile partition}, it suffices to establish the approximation power of $\bb_{p,s}(x;\Delta)$ for all $\Delta\in\Pi$.
	For $v=0$, by Theorem 6.27 of \cite{Schumaker_2007_book},
	$\max_{\Delta\in\Pi}\min_{\bbeta\in\mathbb{R}^{K_{p,s}}}\sup_{x\in\mathcal{X}}
	|\mu_0(x)-\bb_{p,s}(x;\Delta)'\bbeta|\lesssim J^{-p-1}$.
	By \cite{Huang_2003_AoS} and Assumption \ref{SA Assumption DGP}, the Lebesgue factor of spline bases is bounded. Then, the bound on uniform approximation error coincides with that for $L_2$ projection error up to some universal constant. 
	
	For $v>0$, again, we only need to consider the case where $\Delta$ belongs to $\Pi$. For any $\Delta\in\Pi$, we can take the best $L_\infty$-approximation: for some $\bbeta_\infty(\Delta)\in\mathbb{R}^{K_{p,s}}$, $\|\mu_0(\cdot)-\bb_{p,s}(\cdot;\Delta)'\bbeta_\infty(\Delta)\|_\infty\lesssim J^{-p-1}$, and  
	$\|\mu_0^{(v)}(\cdot)-\bb_{p,s}^{(v)}(\cdot;\Delta)'\bbeta_\infty(\Delta)\|_\infty\lesssim J^{-p-1+v}$. Such a construction exists by Lemma SA-6.1 of \cite*{Cattaneo-Farrell-Feng_2020_AoS}. Then, 
	$\|\mu_0^{(v)}(\cdot)-\bb_{p,s}^{(v)}(\cdot;\Delta)'\bbeta_0(\Delta)\|_\infty
	\lesssim
	\|\mu_0^{(v)}(\cdot)-\bb_{p,s}^{(v)}(\cdot;\Delta)'\bbeta_\infty(\Delta)\|_\infty+
	\|\bb_{p,s}^{(v)}(\cdot;\Delta)'(\bbeta_\infty(\Delta)-\bbeta_0(\Delta))\|_\infty
	\lesssim J^{-p-1+v}+\|\bb_{p,s}^{(v)}(\cdot;\Delta)'(\bbeta_\infty(\Delta)-\bbeta_0(\Delta))\|_\infty$.
	By definition of $\bbeta_0(\Delta)$,
	\[
	\bbeta_0(\Delta)-\bbeta_\infty(\Delta)=\E[\bb_{p,s}(x_i;\Delta)\bb_{p,s}(x_i;\Delta)']^{-1}\E[\bb_{p,s}(x_i;\Delta)r_\infty(x_i;\Delta)],
	\]
	where $r_\infty(x_i;\Delta)=\mu_0(x_i)-\bb_{p,s}(x_i;\Delta)'\bbeta_\infty(\Delta)$. By the argument given later in the proof of Lemma \ref{SA lem: Gram, LS} in Section \ref{SA section: LS}, we have $\|\E[\bb_{p,s}(x_i;\Delta)\bb_{p,s}(x_i;\Delta)']^{-1}\|_\infty
	\lesssim 1$ uniformly over $\Delta\in\Pi$. Since $\bb_{p,s}(x_i;\Delta)$ is supported on a finite number of bins, 
	$\|\E[\bb_{p,s}(x_i;\Delta)r_\infty(x_i;\Delta)]\|_\infty\lesssim J^{-p-1-1/2}$. 
	Then the desired result follows.
\end{proof}

\subsection{Proof of Lemma \ref{SA lem: Gram, LS}}

\begin{proof}
    The upper bound on the maximum eigenvalue of $\bQ_0$ follows from Lemma \ref{SA lem: spline transform} and the quasi-uniformity property of population quantiles shown in the proof of Lemma \ref{SA lem: quantile partition}. Also, 
	in view of Lemma \ref{SA lem: quantile partition}, the lower bound on the minimum eigenvalue of $\bQ_0$ follows from Theorem 4.41 of \cite{Schumaker_2007_book}, by which the minimum eigenvalue of  $\bQ_0/J$ (the scaling factor dropped) is bounded by $\min_{1\leq j\leq J}h_j$ up to some universal constant.
	
	Now, we prove the convergence of $\widehat{\bQ}$.
	In view of Lemma \ref{SA lem: spline transform}, it suffices to show the convergence of $\widehat{\bQ}$ when $s=0$, i.e., $\|\E_n[\widehat{\bb}_{p,0}(x_i)\widehat{\bb}_{p,0}(x_i)']-
	\E[\bb_{p,0}(x_i)\bb_{p,0}(x_i)']\|\lesssim_\P\sqrt{J\log J/n}$.	
	By Lemma \ref{SA lem: quantile partition}, with probability approaching one, $\widehat{\Delta}\in\Pi$. Let $\mathcal{A}_n$ denote the event on which $\widehat{\Delta}\in\Pi$. Thus, $\P(\mathcal{A}_n^c)=o(1)$. On $\mathcal{A}_n$,
	\[
	\begin{split}
	&\Big\|\E_n[\widehat{\bb}_{p,0}(x_i)\widehat{\bb}_{p,0}(x_i)']-
	\E_{\widehat{\Delta}}[\widehat{\bb}_{p,0}(x_i)\widehat{\bb}_{p,0}(x_i)']\Big\|\\
	\leq\,&
	\sup_{\Delta\in\Pi}\Big\|\E_n[\bb_{p,0}(x_i;\Delta)\bb_{p,0}(x_i;\Delta)']-
	\E[\bb_{p,0}(x_i;\Delta)\bb_{p,0}(x_i;\Delta)']\Big\|.
	\end{split}
	\]
	By the relation between matrix norms, the right-hand-side of the above inequality is further bounded by
	$\sup_{\Delta\in\Pi}\|\E_n[\bb_{p,0}(x_i;\Delta)\bb_{p,0}(x_i;\Delta)']-
	\E[\bb_{p,0}(x_i;\Delta)\bb_{p,0}(x_i;\Delta)']\|_\infty$.
	Let $a_{kl}$ be a generic $(k,l)$th entry of the matrix inside $\|\cdot\|_\infty$. Then,
	\begin{equation*}
		|a_{kl}|=\Big|\E_n[b_{p,0,k}(x_{i}; \Delta)b_{p,0,l}(x_{i};\Delta)']-
		\E[b_{p,0,k}(x_{i};\Delta)b_{p,0,l}(x_{i};\Delta)']\Big|.
	\end{equation*} 
	If $b_{p,0,k}(\cdot \,;\Delta)$ and $b_{p,0,l}(\cdot \,;\Delta)$ are basis functions with different supports, $a_{kl}$ is zero. Now, define the following function class
	\[\mathcal{G}=\Big\{x\mapsto b_{p,0,k}(x;\Delta)b_{p,0,l}(x;\Delta):
	1\leq k, l\leq J(p+1), \Delta\in\Pi\Big\}.
	\]
	For this class of functions,
	$
	\sup_{g\in\mathcal{G}}|g|_\infty\lesssim J$ and
	$
	\sup_{g\in\mathcal{G}}\V[g]\leq 
	\sup_{g\in\mathcal{G}}\E[g^2]\lesssim J
	$
	where the second result follows from the fact that the size of the supports of $b_{0,k}(\cdot;\Delta)$ and $b_{0,l}(\cdot;\Delta)$ shrinks at the rate of $J^{-1}$. In addition, 
	each function in $\mathcal{G}$ is simply a dilation and translation of a polynomial function supported on $[0,1]$, plus a zero function, and the number of polynomial degree is finite. Then,
	by Proposition 3.6.12 of \citet*{Gine-Nickl_2016_book}, the collection $\mathcal{G}$ of such functions is of VC type, i.e., there exists some constant $C_z$ and $z>6$ such that
	\[ N(\mathcal{G}, L_2(\mathbb{Q}), \varepsilon\|\bar{G}\|_{L_2(\mathbb{Q})})\leq \Big(\frac{C_z}{\varepsilon}\Big)^{2z},
	\]
	for $\varepsilon$ small enough where we take $\bar{G}=CJ$ for some constant $C>0$ large enough. Theorem 6.1 of \citet*{Belloni-Chernozhukov-Chetverikov-Kato_2015_JoE},
	\begin{equation*}
		\E\Big[\sup_{g\in\mathcal{G}} \Big|\sum_{i=1}^{n}g(x_i)-\sum_{i=1}^{n}\E[g(x_i)]
		\Big|\Big]
		\lesssim  \sqrt{nJ\log J} + J\log J ,
	\end{equation*}
	implying that 
	$$\sup_{g\in\mathcal{G}}\Big|\frac{1}{n}\sum_{i=1}^{n}g(x_i)-\E[g(x_i)]\Big|
	\lesssim_\P \sqrt{J\log J/n}.
	$$
	Since any row or column of the matrix $n^{-1/2}\cdot \G_n[\bb_{p,0}(x_i;\Delta)\bb_{p,0}(x_i;\Delta)']$ only contains a finite number of nonzero entries, only depending on $p$, the above result suffices to show that 
	\[
	\Big\|\E_n[\widehat{\bb}_{p,0}(x_i)\widehat{\bb}_{p,0}(x_i)']-
	\E_{\widehat{\Delta}}[\widehat{\bb}_{p,0}(x_i)\widehat{\bb}_{p,0}(x_i)']\Big\|
	\lesssim_\P \sqrt{J\log J/n}.
	\]
	
	Next,
	let $\alpha_{kl}$ be a generic $(k,l)$th entry of
	$
	\E_{\widehat{\Delta}}[\widehat{\bb}_{p,0}(x_{i})\widehat{\bb}_{p,0}(x_{i})']/J - \E[\bb_{p,0}(x_{i})\bb_{p,0}(x_{i})']/J
	$,
	where by dividing the matrix by $J$, we drop the normalizing constant for notation simplicity. By definition, it is either equal to zero or can be rewritten as
	\begin{align}
		\alpha_{kl}
		=&\int_{\widehat{\mathcal{B}}_j}\Big(\frac{x-\hat{\tau}_j}{\hat{h}_j}\Big)^\ell f_X(x)dx 
		-\int_{\widehat{\mathcal{B}}_j}\Big(\frac{x-\tau_j}{h_j}\Big)^\ell f_X(x)dx \nonumber\\
		=&\hat{h}_j\int_0^1z^\ell f_X(z\hat{h}_j+\hat{\tau}_j)dz -
		h_j\int_0^1 z^\ell f_X(zh_j+\tau_j)dz \nonumber\\
		=&(\hat{h}_j-h_j)\int_0^1z^\ell f_X(z\hat{h}_j+\hat{\tau}_j)dz+
		h_j\int_0^1z^\ell\Big(f_X(z\hat{h}_j+\hat{\tau}_j)-f_X(zh_j+\tau_j)\Big)dz \label{SA eq: Gram derivation}
	\end{align} 
	for some $1\leq j\leq J$ and $0\leq\ell\leq 2p$.
	By Assumption \ref{SA Assumption DGP} and Lemma SA2 of \citet*{Calonico-Cattaneo-Titiunik_2015_JASA}, $\max_{1\leq j\leq J}f_X(\hat{\tau}_j)\lesssim 1$ and
	$\max_{1\leq j\leq J}|\hat{h}_j-h_j|\lesssim_\P J^{-1}\sqrt{J\log J/n}$. Also, Lemma SA2 of \citet*{Calonico-Cattaneo-Titiunik_2015_JASA} implies that 
	$$\sup_{z\in[0,1]}\max_{1\leq j\leq J}|\hat{\tau}_j+z\hat{h}_j-
	(\tau_j+zh_j)|\lesssim_\P\sqrt{J\log J/n}.$$
	Since $f_X(\cdot)$ is uniformly continuous on $\mathcal{X}$, 
	the second term in \eqref{SA eq: Gram derivation} is also $O_\P(J^{-1}\sqrt{J\log J/n})$. 
	Again, using the sparsity structure of the matrix $\E_{\widehat{\Delta}}[\widehat{\bb}_{p,0}(x_{i})\widehat{\bb}_{p,0}(x_{i})']/J - \E[\bb_{p,0}(x_{i})\bb_{p,0}(x_{i})']/J$, the above result suffices to show that $\|\E_{\widehat{\Delta}}[\widehat{\bb}_{p,0}(x_i)\widehat{\bb}_{p,0}(x_i)']-\bQ_0\|\lesssim_\P \sqrt{J\log J/n}$. 
	
	Given the above fact, it follows that $\|\widehat{\bQ}^{-1}\|\lesssim_\P 1$. Notice that $\widehat{\bQ}$ and $\bQ_0$ are banded matrices with finite band width. Then the bounds on $\|\widehat{\bQ}\|_\infty$ and $\|\widehat{\bQ}^{-1}-\bQ_0^{-1}\|_\infty$ hold by Theorem 2.2 of \cite{Demko_1977_SIAM}. This completes the proof.
\end{proof}

\subsection{Proof of Lemma \ref{SA lem: asymp variance, LS}}

\begin{proof}
	Since $\E[\epsilon_{i}^2|x_i=x]$ is bounded and bounded away from zero uniformly over $x\in\mathcal{X}$, we have 
	$\widehat{\bQ}\lesssim \bar{\bSigma}\lesssim\widehat{\bQ}$.
	Then, by Lemma \ref{SA lem: Gram, LS}, $1\lesssim_\P\lambda_{\min}(\bar{\bSigma})\lesssim \lambda_{\max}(\bar{\bSigma})\lesssim_\P 1$. The upper bound on $\bar{\Omega}(x)$ immediately follows by Lemmas \ref{SA lem: local basis} and \ref{SA lem: Gram, LS}.
	
	To establish the lower bound, it suffices to show $\inf_{x\in\mathcal{X}}\|\widehat{\bb}_{p,s}^{(v)}(x)\|\gtrsim_\P J^{1/2+v}$.
	For $s=0$, such a bound is trivial by construction. For other $s>0$, we only need to consider the case in which $\widehat{\Delta}\in\Pi$. Introduce an auxiliary function $\varrho(x)=(x-x_0)^v/h_{x_0}^v$ for any arbitrary point $x_0\in\mathcal{X}$, and $h_{x_0}$ is the length of $\mathcal{B}_{x_0}$, the bin containing $x_0$ in any given partition $\Delta\in\Pi$. Let $\{\varphi_j\}_{j=1}^{K_{p,s}}$ be the dual basis for $B$-splines $\breve{\bb}_{p,s}(x):=\bb_{p,s}(x;\Delta)/\sqrt{J}$, which is constructed as in Theorem 4.41 of \cite{Schumaker_2007_book}. The scaling factor $\sqrt{J}$ is dropped temporarily so that the definition of $\breve{\bb}_{p,s}(x)$ is consistent with that theorem. Since the $B$-spline basis reproduces polynomials,
	\[
	J^v\lesssim\varrho^{(v)}(x_0)=\sum_{j=1}^{K_{p,s}}(\varphi_j\varrho)\breve{b}_{p,s,j}^{(v)}(x_0).
	\]
	For any $x_0\in\mathcal{X}$, there are only a finite number of basis functions in $\breve{\bb}_{p,s}(x)$ supported on $\mathcal{B}_{x_0}$. By 
	Theorem 4.41 of \cite{Schumaker_2007_book}, for each $\breve{b}_{p,s,j}(x)$, $j=1, \cdots, K_{p,s}$, we have $|\varphi_j\varrho|\lesssim \|\varrho\|_{L_\infty[\mathcal{I}_j]}$ where $\mathcal{I}_j$ denotes the support of $\breve{b}_{p,s,j}(x)$ and $\|\cdot\|_{L_\infty[\mathcal{I}_j]}$ denotes the sup-norm on $\mathcal{I}_j$. All points within such $\mathcal{I}_j$ should be no greater than $(p+1)\max_{1\leq j\leq J}h_j(\Delta)$ away from $x_0$ where $h_j(\Delta)$ denotes the length of the $j$th bin in $\Delta$. Hence,  $\|\varrho\|_{L_\infty[\mathcal{I}_j]}\lesssim 1$. The desired lower bound follows.	
	The bound on $\Omega(x)$ can be established similarly.
\end{proof}

\subsection{Proof of Lemma \ref{SA lem: uniform converge var part, LS}}

\begin{proof}
	By Lemmas \ref{SA lem: spline transform}, \ref{SA lem: local basis} and \ref{SA lem: Gram, LS},  $\sup_{x\in\mathcal{X}}\|\widehat{\bb}_{p,s}^{(v)}(x)\|_1\lesssim_\P J^{1/2+v}$, $\|\widehat{\bQ}^{-1}\|_\infty\lesssim_\P 1$ and $\|\widehat{\bT}_s\|_\infty\lesssim_\P 1$. Define a function class
	\[
	\mathcal{G}=\Big\{(x_1, \epsilon_1)\mapsto b_{p,0,l}(x_1;\Delta)\epsilon_1:1\leq l\leq J(p+1), \Delta\in\Pi
	\Big\}.
	\]
	Then, $\sup_{g\in\mathcal{G}}|g|\lesssim \sqrt{J}|\epsilon_1|$, and hence take an envelop $\bar{G}=C\sqrt{J}|\epsilon_1|$ for some $C$ large enough. Moreover, $\sup_{g\in\mathcal{G}}\V[g]\lesssim 1$ and, as in the proof of Lemma \ref{SA lem: Gram, LS}, $\mathcal{G}$ is of VC-type. By Proposition 6.1 of \citet*{Belloni-Chernozhukov-Chetverikov-Kato_2015_JoE},
	\[\sup_{g\in\mathcal{G}}\Big|\frac{1}{n}\sum_{i=1}^{n}g(x_i, \epsilon_i)\Big|\lesssim_\P \sqrt{\frac{\log J}{n}}+\frac{J^{\frac{\nu}{2(\nu-2)}}\log J}{n}\lesssim\sqrt{\frac{\log J}{n}},
	\]
	and the desired result follows.
\end{proof}

\subsection{Proof of Lemma \ref{SA lem: proj approx error, LS}}

\begin{proof}
	Note that $\widehat{\bb}_{p,s}^{(v)}(x)'\widehat{\bQ}^{-1}\E_n[\widehat{\bb}_{p,s}(x_i)\widehat{r}_0(x_i)] = A_{1}(x) + A_{2}(x)$, with $A_{1}(x) := \widehat{\bb}_{p,s}^{(v)}(x)'(\widehat\bQ^{-1}-\bQ_0^{-1})\E_n[\widehat{\bb}_{p,s}(x_i)\widehat{r}_0(x_{i})]$ and $A_{2}(x) := \widehat{\bb}_{p,s}^{(v)}(x)'\bQ_0^{-1}\E_n[\widehat{\bb}_{p,s}(x_i)\widehat{r}_0(x_i)]$. By definition of $\widehat{r}_0(\cdot)$, we have $\E_{\widehat{\Delta}}[\widehat{\bb}_{p,s}(x_i)\widehat{r}_0(x_i)]=0$. 
	Define the following function class
	\[\mathcal{G}:=\Big\{x\mapsto b_{p,s,l}(x;\Delta)r_0(x;\Delta):1\leq l\leq K_{p,s}, \Delta\in\Pi \Big\}.
	\]
	By Lemma \ref{SA lem: uniform approx rate},   
	$\sup_{\Delta\in\Pi}|r_0(x;\Delta)|_\infty\lesssim J^{-p-1}$.
	Then, 
	$\sup_{g\in\mathcal{G}}|g|_\infty\lesssim J^{-p-1+1/2}$, and $\sup_{g\in\mathcal{G}}\V[g]\lesssim J^{-2(p+1)}$. In addition,  any function $g\in\mathcal{G}$ can be rewritten as 
	\[
	g(x)=b_{p,s,l}(x;\Delta)\Big(\mu_0(x)-\bb_{p,s}(x;\Delta)'\bbeta_0(\Delta)\Big)=
	b_{p,s,l}(x;\Delta)\mu_0(x)-\sum_{k=\underline{k}}^{\underline{k}+p}b_{p,s,l}(x;\Delta)b_{p,s,k}(x;\Delta)\beta_{0,k}(\Delta)
	\]
	for some $1\leq l, \underline{k}\leq K_{p,s}$ where $\beta_{0,k}(\Delta)$ denotes the $k$-th element of  $\bbeta_0(\Delta)$. Here we use the sparsity property of the partitioning basis: the summand in the second term is nonzero only if $b_{p,s,l}(x;\Delta)$ and $b_{p,s,k}(x;\Delta)$ have overlapping supports. For each $l$, there are at most $(p+1)$ such basis functions $b_{p,s,k}(x;\Delta)$s. 
	Also, the first term and  every summand in the second term are bounded by $\sqrt{J}$ up to some constant. 
	Then, using the same argument given in the proof of Lemma \ref{SA lem: Gram, LS}, 
	\[N(\mathcal{G}, L_2(\mathbb{Q}), \varepsilon\|\bar{G}\|_{L_2(\mathbb{Q})})\leq\Big(\frac{J^l}{\varepsilon}\Big)^z\]
	for some finite $l$ and $z$ and the envelop $\bar{G}=CJ^{-p-1+1/2}$ for $C>0$ large enough. By Theorem 6.1 of \cite*{Belloni-Chernozhukov-Chetverikov-Kato_2015_JoE},
	\[\sup_{g\in\mathcal{G}}\Big|\frac{1}{n}\sum_{i=1}^{n}g(x_i)\Big|\lesssim
	J^{-p-1}\sqrt{\frac{\log J}{n}}+\frac{J^{-p-1+1/2}\log J}{n},	
	\]
	and, by Lemma \ref{SA lem: Gram, LS}, $\|\widehat{\bQ}^{-1}-\bQ_0^{-1}\|_\infty\lesssim_\P\sqrt{J\log J/n}$.
	Then, using the bound on the basis given in Lemma \ref{SA lem: local basis},
	\begin{align*}
		&\sup_{x\in\mathcal{X}}|A_{1}(x)|\lesssim_\P J^{v}\sqrt{J}\sqrt{\frac{J\log J}{n}}J^{-p-1}\sqrt{\frac{\log J}{n}}=J^{-p-1+v}\frac{J\log J}{n}, \quad
		\text{and}\\
		&\sup_{x\in\mathcal{X}}|A_{2}(x)|\lesssim_\P J^v\sqrt{J}J^{-p-1}\sqrt{\frac{\log J}{n}}=J^{-p-1+v}\sqrt{\frac{J\log J}{n}}.
	\end{align*}
	These results complete the proof.
\end{proof}

\subsection{Proof of Lemma \ref{SA lem: parametric part}}

\begin{proof}
	We first show the convergence of $\widehat{\bgamma}$. 
	We denote the $(i,j)$th element of $\bM_{\bB}$ by $M_{ij}$. Then,
	\[ \widehat{\bgamma}-\bgamma_0=\Big(\frac{1}{n}\sum_{i=1}^{n}\sum_{j=1}^{n}M_{ij}\bw_i\bw_j'\Big)^{-1} \Big(\frac{1}{n}\sum_{i=1}^{n}\sum_{j=1}^{n}\bw_i M_{ij}(\mu_0(x_j)+\epsilon_{j})\Big).
	\]
	
	Define $\bV=\bW-\E[\bW|\bX]$ and $\bH=\E[\bW|\bX]$. Then, 
	\begin{equation*}
	\frac{\bW'\bM_{\bB}\bW}{n}=\frac{\bV'\bM_{\bB}\bV}{n}+\frac{\bH'\bM_{\bB}\bH}{n}+\frac{\bH'\bM_{\bB}\bV}{n}+\frac{\bV'\bM_{\bB}\bH}{n}.
	\end{equation*}
	We have
	\begin{equation*}
	\frac{\bV'\bM_{\bB}\bV}{n}=\frac{1}{n}\sum_{i=1}^{n}M_{ii}\bv_i\bv_i'+
	\frac{1}{n}\sum_{i=1}^{n}\sum_{j\neq i}M_{ij}\bv_i\bv_j'
	=\frac{1}{n}\sum_{i=1}^{n}M_{ii}\E[\bv_i\bv_i'|\bX]+O_\P\Big(\frac{1}{n}\Big)\gtrsim_\P 1,
	\end{equation*}
	where the penultimate equality holds by Lemma SA-1 of \cite*{Cattaneo-Jansson-Newey_2018_JASA} and the last by $\frac{1}{n}\sum_{i=1}^{n}M_{ii}=\frac{n-K_{p,s}}{n}\gtrsim 1$.
	Moreover, $\frac{\bH'\bM_{\bB}\bH}{n}\geq 0$,
	and
	$\frac{\bH'\bM_{\bB}\bV}{n}$ has mean zero conditional on $\bX$ and by Lemma SA-1 of \cite*{Cattaneo-Jansson-Newey_2018_JASA}, 
	\[ \Bigg\|\frac{\bH'\bM_{\bB}\bV}{n}\Bigg\|_{F}\lesssim_\P \frac{1}{\sqrt{n}}
	\Big(\tr\Big(\frac{\bH'\bH}{n}\Big)\Big)^{1/2}=o_\P(1),
	\]
	where $\|\cdot\|_F$ denotes the Frobenius norm for matrices.
	Therefore, we conclude that $\frac{\bW'\bM_{\bB}\bW}{n}\gtrsim_\P 1$.
	
	On the other hand, $\frac{1}{n}\sum_{i=1}^{n}\sum_{j=1}^{n}\bw_iM_{ij}\epsilon_j$ has mean zero with variance of order $O(1/n)$ by Lemma SA-2 of \cite*{Cattaneo-Jansson-Newey_2018_JASA}.
	In addition, as in Lemma 2 of \citet*{Cattaneo-Jansson-Newey_2018_ET}, let $\bG=(\mu_0(x_1), \ldots, \mu_0(x_n))'$ and note that
	\begin{align*}
	\frac{\bW'\bM_{\bB}\bG}{n}&=\frac{\bH'\bM_{\bB}\bG}{n}+\frac{\bV'\bM_{\bB}\bG}{n}\\
	&\lesssim
	\sqrt{\tr\Big(\frac{\bH'\bM_{\bB}\bH}{n}\Big)}\sqrt{\tr\Big(\frac{\bG'\bM_{\bB}\bG'}{n}\Big)}
	+\frac{1}{\sqrt{n}}\Big(\frac{\bG'\bM_{\bB}\bG}{n}\Big)^{1/2}\\
	&\lesssim_\P J^{-(\varsigma_w\wedge (p+1))}J^{-p-1}+\frac{J^{-p-1}}{\sqrt{n}}.
	\end{align*}
	Then, the first result follows from the rate restrictions imposed.
	
	To show the second result, by Lemmas \ref{SA lem: spline transform}, \ref{SA lem: local basis} and \ref{SA lem: Gram, LS}, $\sup_{x\in\mathcal{X}}
    \|\widehat{\bb}_{p,s}^{(v)}(x)\|_1\lesssim_\P J^{1/2+v}$,  $\|\widehat{\bQ}^{-1}\|_\infty \lesssim_\P 1$ and $\|\widehat{\bT}_s\|_\infty\lesssim_\P 1$.
	$\E_n[\widehat{\bb}_{p,0}(x_i)\bw_i']$ is a $J(p+1)\times d$ matrix and can be decomposed as follows:
	\begin{equation*}
	\E_n[\widehat{\bb}_{0}(x_i)\bw_{i}']
	=\E_n\Big[\widehat{\bb}_{0}(x_i)\E[\bw_{i}'|x_i]\Big]+
	 \E_n\Big[\widehat{\bb}_{0}(x_i)(\bw_{i}'-\E[\bw_{i}'|x_i])\Big].
	\end{equation*}
	By the argument in the proof of Lemma \ref{SA lem: Gram, LS} and the conditions that $\sup_{x\in\mathcal{X}}\|\E[\bw_{i}|x_i=x]\|\lesssim 1$ and $\frac{J\log J}{n}=o(1)$, 
	$\|\E_n[\widehat{\bb}_{0}(x_i)\E[\bw_{i}'|x_i]]\|_\infty\lesssim_\P J^{-1/2}$.
	Regarding the second term, note that it is a mean zero sequence, and for the $l$th covariate in $\bw$, $l=1, \ldots, d$,   
	\begin{align*}
	&\V\Big[\widehat{\bb}_{p,s}^{(v)}(x)'\widehat{\bQ}^{-1}\E_n[\widehat{\bb}_{s}(x_i)
	(w_{i,l}-\E[w_{i,l}|x_i])]\Big|\bX\Big]\\
	\lesssim\, &\frac{1}{n}\widehat{\bb}_{p,s}^{(v)}(x)'\widehat{\bQ}^{-1}
	\E_n[\widehat{\bb}_{s}(x_i)\widehat{\bb}_{s}(x_i)'\V[w_{i,l}|x_i]]
	\widehat{\bQ}^{-1}\widehat{\bb}_{p,s}^{(v)}(x)
	\lesssim \frac{J^{1+2v}}{n}.
	\end{align*}
	Thus the second result follows by Markov's inequality.
	
	Now suppose $\frac{J^{\frac{\nu}{\nu-2}}\log J}{n}\lesssim 1$ also holds.
	Using the argument given in Lemma \ref{SA lem: uniform converge var part, LS} and the assumption that $\sup_{x\in\mathcal{X}}\E[|w_{i,l}|^\nu|x_i=x]\lesssim 1$ for all $l$, we have  $\|\E_n[\widehat{\bb}_{s}(x_i)(w_{i,l}-\E[w_{i,l}|x_i])]\|_\infty\lesssim_\P \sqrt{\log J/n}$. Thus, the last result follows.	
\end{proof}

\subsection{Proof of Theorem \ref{SA thm: Stochastic Linear Approximation, LS}}
\begin{proof}
	The result follows by Lemmas \ref{SA lem: uniform approx rate}, \ref{SA lem: proj approx error, LS} and \ref{SA lem: parametric part}.
\end{proof}

\subsection{Proof of Corollary \ref{SA coro: uniform convergence, LS}}
\begin{proof}
     The result follows by Theorem \ref{SA thm: Stochastic Linear Approximation, LS} 
     and Lemma \ref{SA lem: uniform converge var part, LS}.
\end{proof}

\subsection{Proof of Theorem \ref{SA thm: meat matrix, LS}}
\begin{proof}
	Since $\widehat{\epsilon}_{i}:=y_{i}-\widehat{\bb}_{p,s}(x_{i})'\widehat{\bbeta}-\bw_{i}'\widehat{\bgamma}=\epsilon_{i}+\mu_0(x_{i})-\widehat{\bb}_{p,s}(x_{i})'\widehat{\bbeta}-\bw_{i}'(\widehat{\bgamma}-\bgamma_0)=:\epsilon_{i}+u_{i}$, we can write
	\begin{align*}
	&\E_n[\widehat{\bb}_{p,s}(x_{i})\widehat{\bb}_{p,s}(x_{i})'\widehat{\epsilon}_{i}^2]
	-\E[\bb_{p,s}(x_{i})\bb_{p,s}(x_{i})'\sigma^2(x_i)]\\
	=\;&\E_n[\widehat{\bb}_{p,s}(x_{i})
	\widehat{\bb}_{p,s}(x_{i})'u_i^2]
	+2\E_n[\widehat{\bb}_{p,s}(x_{i})
	\widehat{\bb}_{p,s}(x_{i})'u_i\epsilon_{i}]
	+\E_n[\widehat{\bb}_{p,s}(x_{i})
	\widehat{\bb}_{p,s}(x_{i})'(\epsilon_{i}^2-\sigma^2(x_i))]\\
	&+\Big(\E_n[\widehat{\bb}_{p,s}(x_{i})\widehat{\bb}_{p,s}(x_{i})'\sigma^2(x_i)]-
	\E[\bb_{p,s}(x_{i})\bb_{p,s}(x_{i})'\sigma^2(x_i)]\Big)\\
	=:&\bV_1+\bV_2+\bV_3+\bV_4.
	\end{align*}
	Now, we bound each term in the following.
	
	\textbf{Step 1:} For $\bV_1$, we further write $u_i=(\mu_0(x_{i})-\widehat{\bb}_{p,s}(x_{i})'\widehat{\bbeta})-\bw_{i}'(\widehat{\bgamma}-\bgamma_0)=:u_{i1}-u_{i2}$. Then
	\[
	\bV_1=\E_n[\widehat{\bb}_{p,s}(x_{i})
	\widehat{\bb}_{p,s}(x_{i})'(u_{i1}^2+u_{i2}^2-2u_{i1}u_{i2})]
	=:\bV_{11}+\bV_{12}-\bV_{13}.
	\]
    Since $\|2\E_n[\widehat{\bb}_{p,s}(x_{i})\widehat{\bb}_{p,s}(x_{i})'u_{i1}u_{i2}]\|
	\leq\|\E_n[\widehat{\bb}_{p,s}(x_{i})\widehat{\bb}_{p,s}(x_{i})'(u_{i1}^2+u_{i2}^2)]\|$, it suffices to bound $\bV_{11}$ and $\bV_{12}$. For $\bV_{11}$,
	\begin{equation*}
	\|\bV_{11}\|\leq\max_{1\leq i\leq n}|u_{i1}|^2
	\Big\|\E_n[\widehat{\bb}_{p,s}(x_{i})\widehat{\bb}_{p,s}(x_{i})']\Big\|
	\lesssim_\P \frac{J\log J}{n}+J^{-2(p+1)},
	\end{equation*}
	where the last inequality holds by Lemma \ref{SA lem: Gram, LS} and Corollary \ref{SA coro: uniform convergence, LS}. On the other hand, let $\widehat{\gamma}_\ell$ and $\gamma_{0,\ell}$ denote the $\ell$th entry of $\widehat{\bgamma}$ and $\bgamma_0$. We have
	\begin{align*}
	\|\bV_{12}\|&=\Big\|\E_n\Big[
	\widehat{\bb}_{p,s}(x_{i})\widehat{\bb}_{p,s}(x_{i})'\Big(\sum_{\ell=1}^{d}w_{i,\ell}^2(\widehat{\gamma}_\ell-\gamma_{0,\ell})^2+\sum_{\ell\neq\ell'}w_{i,\ell}w_{i,\ell'}(\widehat{\gamma_\ell}-\gamma_{0,\ell})(\widehat{\gamma}_{\ell'}-\gamma_{0,\ell'})\Big)\Big]\Big\|\\
	&\lesssim\Big\|\E_n\Big[\widehat{\bb}_{p,s}(x_{i})\widehat{\bb}_{p,s}(x_{i})'\Big(\sum_{\ell=1}^{d}w_{i,\ell}^2(\widehat{\gamma}_\ell-\gamma_{0,\ell})^2\Big)\Big]\Big\|
	\end{align*}
	by CR-inequality. 
	By Lemma \ref{SA lem: parametric part},  $\|\widehat{\bgamma}-\bgamma_0\|^2=o_\P(J/n)$. Then it suffices to show that for every $\ell=1, \ldots, d$, $\|\E_n[\widehat{\bb}_{p,s}(x_{i})\widehat{\bb}_{p,s}(x_{i})'w_{i,\ell}^2]\|\lesssim_\P 1$. Under the conditions given in the theorem, this bound can be established using the argument that will be given in Step 3 and 4 and that in Lemma \ref{SA lem: Gram, LS}.
	
	\textbf{Step 2:} For $\bV_2$, we have
	$\bV_2=2\E_n[\widehat{\bb}_{p,s}(x_{i})
	\widehat{\bb}_{p,s}(x_{i})'\epsilon_{i}(u_{i1}-u_{i2})]=:\bV_{21}-\bV_{22}$.
	Then,
	\begin{equation*}
	\|\bV_{21}\|\leq \max_{1\leq i\leq n}|u_{i1}|\Big(\Big\|
	\E_n[\widehat{\bb}_{p,s}(x_{i})\widehat{\bb}_{p,s}(x_{i})']\Big\|
	+\Big\|\E_n[\widehat{\bb}_{p,s}(x_{i})\widehat{\bb}_{p,s}(x_{i})'\epsilon_{i}^2]\Big\|\Big)
	\lesssim_\P \Big(\frac{J\log J}{n}\Big)^{1/2}+J^{-p-1},
	\end{equation*}
	where the last step follows by Lemma \ref{SA lem: Gram, LS} and the result given in Step 3. In addition,
	\begin{equation*}
	\|\bV_{22}\|=\Big\|2\E_n[\widehat{\bb}_{p,s}(x_{i})\widehat{\bb}_{p,s}(x_{i})'\epsilon_{i}\sum_{\ell=1}^dw_{i,\ell}(\widehat\gamma_\ell-\gamma_{0,\ell})]\Big\|\lesssim_\P \frac{1}{\sqrt{n}}+J^{-p-1-(\varsigma_w\wedge(p+1))}.
	\end{equation*}
	Since
	$\|2\E_n[\widehat{\bb}_{p,s}(x_{i})\widehat{\bb}_{p,s}(x_{i})'\epsilon_{i}w_{i,\ell}]\|\leq
	\|\E_n[\widehat{\bb}_{p,s}(x_{i})\widehat{\bb}_{p,s}(x_{i})'(\epsilon_{i}^2+w_{i,\ell}^2)]\|$,
	this bound on $\|\bV_{22}\|$ can be established using Lemma \ref{SA lem: parametric part} and the strategy given in Step 3 and Step 4 and that in Lemma \ref{SA lem: Gram, LS}.
	
	\textbf{Step 3:}
	For $\bV_3$, in view of Lemma \ref{SA lem: quantile partition} and \ref{SA lem: spline transform}, it suffices to show that
	\[\sup_{\Delta\in\Pi}\Big\|\E_n[\bb_{p,0}(x_{i};\Delta)
	\bb_{p,0}(x_{i};\Delta)'(\epsilon_{i}^2-\sigma^2(x_i))]\Big\|\lesssim_\P 
	\Big(\frac{J\log J}{n^{\frac{\nu-2}{\nu}}}\Big)^{1/2}.
	\]
	For notational simplicity, we write $\varphi_{i}=\epsilon_{i}^2-\sigma^2(x_i)$, 
	$\varphi_{i}^-=\varphi_{i}\I(|\varphi_{i}|\leq M)-\E[\varphi_{i}\I(|\varphi_{i}|\leq M)|x_i]$,
	$\varphi_{i}^+=\varphi_{i}\I(|\varphi_{i}|> M)-\E[\varphi_{i}\I(|\varphi_{i}|> M)|x_i]$ for some $M>0$ to be specified later. Since $\E[\varphi_{i}|x_i]=0$, $\varphi_{i}=\varphi_{i}^-+\varphi_{i}^+$. Then define a function class
	\[\mathcal{G}=\Big\{(x_{1}, \varphi_{1})\mapsto b_{p,0,l}(x_{1};\Delta)b_{p,0,k}(x_{1};\Delta)\varphi_{1}:1\leq l\leq J(p+1), 1\leq k\leq J(p+1), \Delta\in\Pi\Big\}.
	\]
	Then for $g\in\mathcal{G}$, $\sum_{i=1}^{n}g(x_{1}, \varphi_{1})
	=\sum_{i=1}^{n}g(x_{1}, \varphi_{1}^+)+\sum_{i=1}^{n}g(x_{1}, \varphi_{1}^-)$.
	
	Now, for the truncated piece, we have
	$\sup_{g\in\mathcal{G}}|g(x_{1}, \varphi_{1}^-)|\lesssim JM$,
	and
	\begin{align*}
	\sup_{g\in\mathcal{G}}\V[g(x_{1}, \varphi_{1}^-)]
	&\lesssim 
	\sup_{x\in\mathcal{X}}\E[(\varphi_{1}^-)^2|x_1=x]\sup_{\Delta\in\Pi}\sup_{1\leq l,k\leq J(p+1)}\E[b_{p,0,l}^2(x_{1};\Delta)b_{p,0,k}^2(x_{1};\Delta)] \\
	&\lesssim  JM \sup_{x\in\mathcal{X}}
	\E\Big[|\varphi_{1}|\Big|x_i=x\Big]\lesssim JM.
	\end{align*}
	The VC condition holds by the same argument given in the proof of Lemma \ref{SA lem: Gram, LS}. Then, by Proposition 6.1 of \citet*{Belloni-Chernozhukov-Chetverikov-Kato_2015_JoE},
	\[
	\E\Big[\sup_{g\in\mathcal{G}}\Big|\E_n[ g(x_{i}, \varphi_{i}^-)]\Big|\Big]
	\lesssim \Big(\frac{JM\log (JM)}{n}\Big)^{1/2}+\frac{JM\log(JM)}{n}.
	\]
	
	Regarding the tail, we apply Theorem 2.14.1 of \citet*{vandevarrt-Wellner_1996_book} and obtain
	\begin{align*}
	\E\Big[\sup_{g\in\mathcal{G}}\Big|\E_n[g(x_{i}, \varphi_{i}^+)]\Big|\Big]
	&\lesssim \frac{1}{\sqrt{n}}J
	\E\Big[\sqrt{\E_n[|\varphi_{i}^+|^2]}\Big]\\
	&\leq \frac{1}{\sqrt{n}}J
	(\E[\max_{1\leq i\leq n}|\varphi_{i}^+|])^{1/2}(\E[\E_n[|\varphi_{i}^+|])^{1/2}\\
	&\lesssim \frac{J}{\sqrt{n}}\cdot \frac{n^{\frac{1}{\nu}}}{M^{(\nu-2)/4}},
	\end{align*}
	where the second line follows by Cauchy-Schwarz inequality and the third line uses the fact that
	\[
	\E[\max_{1\leq i\leq n}|\varphi_{i}^+|]\lesssim
	\E[\max_{1\leq i\leq n}\epsilon_{i}^2]\lesssim
	n^{2/\nu}, \quad \text{and}\quad
	\E[\E_n[|\varphi_{i}^+|]]\leq \E[|\varphi_{1}|^+|]\lesssim \frac{\E[|\epsilon_1|^{\nu}]}{M^{(\nu-2)/2}}.
	\]
	Then the desired result follows simply by setting $M=J^{\frac{2}{\nu-2}}$ and the sparsity of the basis.
	
	\textbf{Step 4:} For $\bV_4$, since by Assumption \ref{SA Assumption LS},
	$\sup_{x\in\mathcal{X}}\E[\epsilon_{i}^2|x_i=x]\lesssim 1$. Then, by the same argument given in the proof of Lemma \ref{SA lem: Gram, LS},  
	\begin{align*}
	&\sup_{\Delta\in\Pi}\Big\|\E_n[\bb_{p,s}(x_{i};\Delta)\bb_{p,s}(x_{i};\Delta)'\sigma^2(x_i)]
	-\E\Big[\bb_{p,s}(x_{i};\Delta)\bb_{p,s}(x_{i};\Delta)'\epsilon_{i}^2\Big]\Big\|
	\lesssim_\P \sqrt{J\log J/n}, \quad\text{and}\\
	&\Big\|\E_{\widehat{\Delta}}\Big[\widehat{\bb}_{p,s}(x_{i})\widehat{\bb}_{p,s}(x_{i})'\epsilon_{i}^2\Big]-
	\E\Big[\bb_{p,s}(x_{i})\bb_{p,s}(x_{i})'\epsilon_{i}^2\Big]\Big\|\lesssim_\P \sqrt{J\log J/n}.
	\end{align*}
	Then the proof is complete.
\end{proof}

\subsection{Proof of Theorem \ref{SA thm: pointwise normality, LS}}
\begin{proof}
	We first show that for each fixed $x\in\mathcal{X}$,
	\[
	\bar{\Omega}(x)^{-1/2}\widehat{\bb}_{p,s}^{(v)}(x)'\widehat{\bQ}^{-1}\G_n[\widehat{\bb}_{p,s}(x_i)\epsilon_i]=:\G_n[a_{i}\epsilon_{i}]
	\] 
	is asymptotically normal.	
	Conditional on $\bX$, it is a mean zero independent sequence over $i$ with variance equal to $1$.
	Then by Berry-Esseen inequality,
	\begin{align*}
		\sup_{u\in\mathbb{R}}
		\Big|\P(\G_n[a_i\epsilon_i]\leq u|\bX)-\Phi(u)\Big|
		\leq \min\bigg(1, \frac{\sum_{i=1}^{n}\E[|a_{i}\epsilon_{i}|^3|\bX]}{n^{3/2}}\bigg).
	\end{align*}
	Now, using Lemmas \ref{SA lem: local basis}, \ref{SA lem: Gram, LS} and \ref{SA lem: asymp variance, LS},
	\begin{align*}
		\quad \frac{1}{n^{3/2}}\sum_{i=1}^{n}\E\Big[|a_{i}\epsilon_{i}|^3\Big|\bX\Big]
		&\lesssim \bar{\Omega}(x)^{-3/2}\frac{1}{n^{3/2}}\sum_{i=1}^{n}\E\Big[
		|\widehat{\bb}_{p,s}^{(v)}(x)'\widehat{\bQ}^{-1}\widehat{\bb}_{p,s}(x_{i})\epsilon_{i}|^3\Big|\bX\Big]\\
		&\lesssim \bar{\Omega}(x)^{-3/2}\frac{1}{n^{3/2}}\sum_{i=1}^n|\widehat{\bb}_{p,s}^{(v)}(x)'\widehat{\bQ}^{-1}\widehat{\bb}_{p,s}(x_{i})|^3\\
		&\leq\bar{\Omega}(x)^{-3/2}\frac{\sup_{x\in\mathcal{X}}\sup_{z\in\mathcal{X}}
			|\widehat{\bb}_{p,s}^{(v)}(x)'\widehat{\bQ}^{-1}\widehat{\bb}_{p,s}(z)|}{n^{3/2}}
		\sum_{i=1}^{n}
		|\widehat{\bb}_{p,s}^{(v)}(x)'\widehat{\bQ}^{-1}\widehat{\bb}_{p,s}(x_{i})|^2\\
		&\lesssim_\P\frac{1}{J^{3/2+3v}}\cdot\frac{J^{1+v}}{\sqrt{n}}\cdot J^{1+2v}
		\rightarrow 0
	\end{align*}
	since $J/n=o(1)$.
	By Theorem \ref{SA thm: meat matrix, LS}, the above weak convergence still holds if $\bar{\Omega}(x)$ is replaced by $\widehat{\Omega}(x)$. Now, the desired result follows by Lemmas \ref{SA lem: uniform approx rate}, \ref{SA lem: proj approx error, LS} and \ref{SA lem: parametric part}.
\end{proof}

\subsection{Proof of Theorem \ref{SA thm: IMSE, LS}}
\begin{proof}
       Since $\widehat\Upsilon(x,\widehat\bw)$ differs from $\widehat\mu(x)$ only when $v=0$, we will first focus on the IMSE of $\widehat\mu^{(v)}(x)$. 
	We rely on the following decomposition:
	\begin{equation} \label{SA eq: decomp}
		\begin{split}
			\widehat{\mu}^{(v)}(x)-\mu_0^{(v)}(x)=&\,
			\widehat{\bb}_{p,s}^{(v)}(x)'\widehat{\bQ}^{-1}\E_n[\widehat{\bb}_{p,s}(x_{i})\epsilon_i]
			+\widehat{\bb}_{p,s}^{(v)}(x)'\widehat{\bQ}^{-1}\E_n[\widehat{\bb}_{p,s}(x_{i})\widehat{r}_0(x_{i})] + \\
			&\Big(\widehat{\bb}_{p,s}^{(v)}(x)'\widehat{\bbeta}_0-\mu_0^{(v)}(x)\Big)
			-\widehat{\bb}_{p,s}^{(v)}(x)'\widehat{\bQ}^{-1}\E_n[\widehat{\bb}_{p,s}(x_i)\bw_i'](\widehat{\bgamma}-\bgamma_0).
		\end{split}
	\end{equation}
	The proof is divided into several steps.
	
	\textbf{Step 1:}
	By Lemma \ref{SA lem: parametric part}, the variance of the last term is of smaller order, and thus it suffices to characterize the conditional variance of 
	$A(x):=\widehat{\bb}_{p,s}^{(v)}(x)'\widehat{\bQ}^{-1}\E_n[\widehat{\bb}_{p,s}\epsilon_i]$. By Lemma \ref{SA lem: Gram, LS},
	\begin{align*}
		\int_{\mathcal{X}}\V[A(x)|\bX]\omega(x)dx&=\frac{1}{n}\tr\Big(
		\bQ_0^{-1}\bSigma_0\bQ_0^{-1}
		\int_{\mathcal{X}}\widehat{\bb}_{p,s}^{(v)}(x)\widehat{\bb}_{p,s}^{(v)}(x)'\omega(x)dx\Big)+o_\P\Big(\frac{J^{1+2v}}{n}\Big).
	\end{align*}
	In fact, using the argument given in the proof of Lemma \ref{SA lem: local basis}, we also have 
	\[
	\bigg\|\int_{\mathcal{X}}\widehat{\bb}_{p,s}^{(v)}(x)\widehat{\bb}_{p,s}^{(v)}(x)'\omega(x)dx
	- \int_{\mathcal{X}}\bb_{p,s}^{(v)}(x)\bb_{p,s}^{(v)}(x)'\omega(x)dx
	\bigg\| =o_\P(J^{2v}),
	\]
	and since $\sigma^2(x)$ and $\omega(x)$ are bounded and bounded away from zero,
	\[
	\mathscr{V}_n(p,s,v)=J^{-(1+2v)}\tr\Big(
	\bQ_0^{-1}\bSigma_0\bQ_0^{-1}
	\int_{\mathcal{X}}\bb_{p,s}^{(v)}(x)\bb_{p,s}^{(v)}(x)'\omega(x)dx\Big)
	\asymp 1.
	\]
	
	\textbf{Step 2:}
	By decomposition \eqref{SA eq: decomp},
	\[
	\begin{split}
		\E[\widehat{\mu}^{(v)}(x)|\bX, \bW]-\mu_0^{(v)}(x)&=
		\widehat{\bb}^{(v)}_{p,s}(x)'\widehat{\bQ}^{-1}\E_n[\widehat{\bb}_{p,s}(x_i)\widehat{r}_0(x_i)]+
		\Big(\widehat{\bb}^{(v)}_{p,s}(x)'\widehat{\bbeta}_0-\mu_0^{(v)}(x)\Big)\\
		&\quad-\widehat{\bb}_{p,s}^{(v)}(x)'\widehat{\bQ}^{-1}\E_n[\widehat{\bb}_{p,s}(x_i)\bw_i']
		\E[(\widehat{\bgamma}-\bgamma_0)|\bX, \bW]\\
		&=:\mathfrak{B}_1(x)+\mathfrak{B}_2(x)+\mathfrak{B}_3(x).
	\end{split}
	\]
	
	By Lemma \ref{SA lem: proj approx error, LS}, $\int_{\mathcal{X}}\mathfrak{B}_1(x)^2\omega(x)dx=o_\P(J^{-2p-2+2v})$. 
	By Lemma \ref{SA lem: parametric part}, $\int_{\mathcal{X}}\mathfrak{B}_3(x)^2\omega(x)dx=o_\P (J^{-2p-2+2v})$.
	By Lemma \ref{SA lem: uniform approx rate},
	$\int_{\mathcal{X}}\mathfrak{B}_2(x)^2\omega(x)dx\lesssim_\P J^{-2p-2+2v}$.
	By Cauchy-Schwarz inequality, the integrals of those cross-product terms is of higher-order in the IMSE expansion, and the leading term in the integrated squared bias is 
	\[
	J^{2p+2-2v}\int_{\mathcal{X}}\Big(\widehat{\bb}_{p,s}^{(v)}(x)'\widehat{\bbeta}_0-\mu_0^{(v)}(x)\Big)^2\omega(x)dx\lesssim_\P 1.
	\]
	Then, by Lemma SA-6.1 of \cite*{Cattaneo-Farrell-Feng_2020_AoS}, for $s=p$,
	\begin{equation}\label{SA eq: uniform approx}
		\sup_{x\in\mathcal{X}}
		\bigg|\mu_0^{(v)}(x)-
		\widehat{\bb}_{p,p}^{(v)}(x)'\bbeta_\infty(\widehat{\Delta})-
		\frac{\mu^{(p+1)}(x)}{(p+1-v)!}\hat{h}_x^{p+1-v}\mathscr{E}_{p+1-v}
		\Big(\frac{x-\hat{\tau}_x^{\mathtt{L}}}{\hat{h}_x}\Big)\bigg|=o_\P(J^{-(p+1-v)}),
	\end{equation}
	where for each $m\in\mathbb{Z}_+$, $\mathscr{E}_m(\cdot)$ is the $m$th Bernoulli polynomial, $\hat{\tau}_x^{\mathtt{L}}$ is the start of the (random) interval in $\widehat{\Delta}$ containing $x$ and $\hat{h}_x$ denotes its length. When $s<p$, $\widehat{\bb}_{p,p}(x)'\bbeta_\infty$ is still an element in the space spanned by $\widehat{\bb}_{p,s}(x)$. In other words, it provides a valid approximation of $\mu_0^{(v)}(x)$ in the larger space in terms of sup-norm. Then it follows that
	\begin{align}
		&\widehat{\bb}_{p,s}^{(v)}(x)'\widehat{\bbeta}_0-\mu_0^{(v)}(x) \nonumber \\
		=\;&
		\widehat{\bb}_{p,s}^{(v)}(x)'\Big(\E_{\widehat{\Delta}}[\widehat{\bb}_{p,s}(x_i)\widehat{\bb}_{p,s}(x_i)']\Big)^{-1}\E_{\widehat{\Delta}}[\widehat{\bb}_{p,s}(x_i)\mu_0(x_i)]-\mu_0^{(v)}(x) \nonumber \\
		=\;&\widehat{\bb}_{p,s}^{(v)}(x)'\Big(\E_{\widehat{\Delta}}[\widehat{\bb}_{p,s}(x_i)\widehat{\bb}_{p,s}(x_i)']\Big)^{-1}
		\E_{\widehat{\Delta}}\bigg[\widehat{\bb}_{p,s}(x_i)\frac{\mu_0^{(p+1)}(x_i)}{(p+1)!}\hat{h}_{x_i}^{p+1}\mathscr{E}_{p+1}
		\Big(\frac{x_i-\hat{\tau}_{x_i}^{\mathtt{L}}}{\hat{h}_{x_i}}\Big)\bigg] \nonumber\\
		&\qquad-\frac{\mu_0^{(p+1)}(x)}{(p+1-v)!}\hat{h}_x^{p+1-v}\mathscr{E}_{p+1-v}
		\Big(\frac{x-\hat{\tau}_x^{\mathtt{L}}}{\hat{h}_x}\Big) +o_\P(J^{-p-1+v}) \nonumber\\
		=\;&J^{-p-1}\widehat{\bb}_{p,s}^{(v)}(x)'\bQ_0^{-1}\bT_s
		\E_{\widehat{\Delta}}\bigg[\widehat{\bb}_{p,0}(x_i)\frac{\mu_0^{(p+1)}(x_i)}{(p+1)!f_X(x_i)^{p+1}}\mathscr{E}_{p+1}
		\Big(\frac{x_i-\hat{\tau}_{x_i}^{\mathtt{L}}}{\hat{h}_{x_i}}\Big)\bigg]  \nonumber\\
		&\qquad-\frac{J^{-p-1+v}\mu_0^{(p+1)}(x)}{(p+1-v)!f_X(x)^{p+1-v}}\mathscr{E}_{p+1-v}
		\Big(\frac{x-\hat{\tau}_x^{\mathtt{L}}}{\hat{h}_x}\Big) +o_\P(J^{-p-1+v}), \label{SA eq: orth approx error}
	\end{align}
	where the last step uses Lemmas \ref{SA lem: quantile partition}-\ref{SA lem: local basis} and \ref{SA lem: Gram, LS}, and $o_\P(\cdot)$ holds uniformly over $x\in\mathcal{X}$.
	Taking integral of the squared bias and using Assumption \ref{SA Assumption DGP} and Lemmas \ref{SA lem: quantile partition}--\ref{SA lem: local basis} and  \ref{SA lem: Gram, LS} again, we have three leading terms:
	\begin{align*}
		M_1(x):=&\,\int_{\mathcal{X}}\bigg(\frac{J^{-p-1+v}\mu_0^{(p+1)}(x)}{(p+1-v)!f_X(x)^{p+1-v}}\mathscr{E}_{p+1-v}
		\Big(\frac{x-\hat{\tau}_x^{\mathtt{L}}}{\hat{h}_x}\Big)\bigg)^2\omega(x) dx \\
		=&\,\frac{J^{-2p-2+2v}|\mathscr{E}_{2p+2-2v}|}{(2p+2-2v)!}
		\int_{\mathcal{X}}\Big[\frac{\mu_0^{(p+1)}(x)}{f_X(x)^{p+1-v}}\Big]^2\omega(x)dx + o_\P(J^{-2p-2+2v}),\\
		M_2(x):=&J^{-2p-2}\int_{\mathcal{X}}\bigg(\widehat{\bb}_{p,s}^{(v)}(x)'\bQ_0^{-1}\bT_s
		\E_{\widehat{\Delta}}\Big[\widehat{\bb}_{p,0}(x_i)\frac{\mu_0^{(p+1)}(x_i)}{(p+1)!f_X(x_i)^{p+1}}\mathscr{E}_{p+1}
		\Big(\frac{x_i-\hat{\tau}_{x_i}^{\mathtt{L}}}{\hat{h}_{x_i}}\Big)\Big]\bigg)^2 \omega(x)dx \\
		=&J^{-2p-2}\boldsymbol{\xi}_{0,f}'\bT_s'\bQ_0^{-1}
		\Big(\int_{\mathcal{X}}\bb_{s}^{(v)}(x)\bb_{s}^{(v)}(x)'\omega(x)dx\Big)\,
		\bQ_0^{-1}\bT_s\boldsymbol{\xi}_{0,f}+o_\P(J^{-2p-2+2v}), \\
		M_3(x):=&J^{-2p-2+v}\int_{\mathcal{X}}\bigg\{\bigg(\widehat{\bb}_{p,s}^{(v)}(x)'\bQ_0^{-1}\bT_s
		\E_{\widehat{\Delta}}\Big[\widehat{\bb}_{p,0}(x_i)\frac{\mu_0^{(p+1)}(x_i)}{(p+1)!f_X(x_i)^{p+1}}\mathscr{E}_{p+1}
		\Big(\frac{x_i-\hat{\tau}_{x_i}^{\mathtt{L}}}{\hat{h}_{x_i}}\Big)\Big]\bigg)\\
		&\hspace{1in}\times \frac{\mu_0^{(p+1)}(x)}{(p+1-v)!f_X(x)^{p+1-v}}\mathscr{E}_{p+1-v}
		\Big(\frac{x-\hat{\tau}_x^{\mathtt{L}}}{\hat{h}_x}\Big)\bigg\}
		\omega(x)dx\\
		=&J^{-2p-2+v}\boldsymbol{\xi}_{0,f}'\bT_s'\bQ_0^{-1}\bT_s\boldsymbol{\xi}_{v,\omega}+o_\P(J^{-2p-2+2v}),
	\end{align*}
	where $\mathscr{E}_{2p+2-2v}$ is the $(2p+2-2v)$th Bernoulli number, and for a weighting function $\lambda(\cdot)$ (which can be replaced by $f_X(\cdot)$ and $\omega(\cdot)$ respectively), we define
	\[
	\boldsymbol{\xi}_{v,\lambda}=\int_{\mathcal{X}}
	\bb_{p,0}^{(v)}(x)\frac{\mu_0^{(p+1)}(x)}{(p+1-v)!f_X(x)^{p+1-v}}\mathscr{E}_{p+1-v}
	\Big(\frac{x-\tau_{x}^{\mathtt{L}}}{h_{x}}\Big)
	\lambda(x)dx.
	\]
	$\tau_{x}$ and $h_{x}$ are defined the same way as $\hat{\tau}_{x}$ and $\hat{h}_x$, but are based on $\Delta_0$, the partition using population quantiles.
	Therefore, the leading terms now only rely on the non-random partition $\Delta_0$ as well as other deterministic functions, which are simply equivalent to the leading bias if we repeat the above derivation but set $\widehat{\Delta}=\Delta_0$. 
	
 \textbf{Step 3:} For $v=0$, we will have two additional terms $\widehat{\bw}'(\widehat{\bgamma}-\bgamma_0)$ and $(\widehat{\bw}-\bw)'\bgamma_0$ in the decomposition of $\widehat\Upsilon(x,\widehat\bw)-\Upsilon_0(x,\bw)$. By Assumption, $\widehat{\bw}-\bw=o_\P(\sqrt{J/n}+J^{-p-1})$, and thus $(\widehat{\bw}-\bw)'\bgamma_0$ as a (conditional) bias term is of higher order. The term  
$\widehat{\bw}'(\widehat{\bgamma}-\bgamma_0)$ can be treated the same way as we analyze  $\widehat{\bb}_{p,s}(x)'\widehat{\bQ}^{-1}\E_n[\widehat{\bb}_{p,s}(x_i)\bw_i'](\widehat{\bgamma}-\bgamma_0)$. By Lemma \ref{SA lem: parametric part}, it is also of higher order. Then, the proof is complete.
\end{proof}

\subsection{Proof of Corollary \ref{SA coro: IMSE, LS}}
\begin{proof}
	The proof is divided into two steps.
	
	\textbf{Step 1:}
	Consider the special case in which $s=0$. $\mathscr{V}_n(p,0,v)$ depends on three matrices: $\bQ_0$, $\bSigma_0$ and $\int_{\mathcal{X}}\bb_{p,0}^{(v)}(x)\bb_{p,0}^{(v)}(x)'\omega(x)dx$.
	Importantly, they are block diagonal with finite block sizes, and the basis functions that form these matrices have local supports. By continuity of $\omega(x)$, $f_X(x)$ and $\sigma^2(x)$, these matrices can be further approximated:
	\begin{equation*}
		\bQ_0=\breve{\bQ}\mathfrak{D}_f+o_\P(1), \;
		\bSigma_0=\breve{\bQ}\mathfrak{D}_{\sigma^2f}+o_\P(1),\;\text{and}
		\;
		\int_{\mathcal{X}}\bb_{p,0}^{(v)}(x)\bb_{p,0}^{(v)}(x)'\omega(x)dx=
		\breve{\bQ}_{v}\mathfrak{D}_{\omega}+o_\P(J^{2v}),
	\end{equation*}
	where
	\begin{align*}
		&\check{\bQ}=\int_{\mathcal{X}}\bb_{p,0}(x)\bb_{p,0}(x)'dx, \;
		\check{\bQ}_{v}=\int_{\mathcal{X}}\bb_{p,0}^{(v)}(x)\bb_{p,0}^{(v)}(x)'dx,\;
		\mathfrak{D}_f=\diag\{f_X(\check{x}_1), \cdots, f_X(\check{x}_{J(p+1)})\},\\
		&\mathfrak{D}_{\sigma^2f}=\diag\{\sigma^2(\check{x}_1)f_X(\check{x}_1), \cdots, \sigma^2(\check{x}_{J(p+1)})f_X(\check{x}_{J(p+1)})\},\;\text{ and }\;
		\mathfrak{D}_{\omega}=\diag\{\omega(\check{x}_1), \ldots, \omega(\check{x}_{J(p+1)})\}.
	\end{align*}
	``$o_\P(\cdot)$'' in the above equations means the operator norm of the remainder  is $o_\P(\cdot)$, and for $l=1, \ldots, J(p+1)$, each $\check{x}_l$ is an arbitrary point in the support of $b_{p,0,l}(x)$. For simplicity, we choose these points such that $x_l=x_{l'}$ if $b_{p,0,l}(\cdot)$ and $b_{p,0,l'}(\cdot)$ have the same support. Therefore, we have
	\[
	\int_{\mathcal{X}}\V[A(x)|\bX]\omega(x)dx=
	\frac{1}{n}\tr\Big(\mathfrak{D}_{\sigma^2\omega/f}\breve{\bQ}^{-1}\breve{\bQ}_{v}\Big)+o_\P\Big(\frac{J^{1+2v}}{n}\Big),
	\]
	where $\mathfrak{D}_{\sigma^2\omega/f}=\diag\{\sigma^2(\check{x}_1)\omega(\check{x}_1)/f_X(\check{x}_1), \ldots, \sigma^2(\check{x}_{J(p+1)})\omega(\check{x}_{J(p+1)})/f_X(\check{x}_{J(p+1)})\}$.
	
	Finally, by change of variables, we can rewrite $\breve{\bQ}^{-1}\breve{\bQ}_{v}$ as a block diagonal matrix $\diag\{\tilde{\bQ}_1, \cdots, \widetilde{\bQ}_{J}\}$ where the $l$th block $\widetilde{\bQ}_l$, $l=1, \ldots, j$, can be written as
	\[
	\widetilde{\bQ}_l=h_l^{-2v}\Big(\int_{0}^{1}\bm{\varphi}(z)\bm{\varphi}(z)'dz\Big)^{-1}\int_0^1\bm{\varphi}^{(v)}(z)\bm{\varphi}^{(v)}(z)'dz
	\]
	for $\bm{\varphi}(z)=(1, z, \ldots, z^{p})$. Employing Lemma \ref{SA lem: quantile partition} and letting the trace converge to the Riemann integral, we conclude that
	\[
	\int_{\mathcal{X}}\V[A(x)|\bX]\omega(x)dx=
	\frac{J^{1+2v}}{n}\mathscr{V}(p,0,v)
	+o_\P\Big(\frac{J^{1+2v}}{n}\Big),
	\]
	where $\mathscr{V}(p,0,v):=\tr\Big\{\Big(\int_{0}^{1}\bm{\varphi}(z)\bm{\varphi}(z)'dz\Big)^{-1}\int_0^1\bm{\varphi}^{(v)}(z)\bm{\varphi}^{(v)}(z)'dz\Big\}
	\int_{\mathcal{X}}\sigma^2(x)f_X(x)^{2v}\omega(x)dx$.
	
	\textbf{Step 2:}
	Now, consider the special case in which $s=0$.
	By Lemma A.3 of \citet*{Cattaneo-Farrell-Feng_2020_AoS}, we can construct an $L_\infty$ approximation error
	\[r_\infty^{(v)}(x;\widehat{\Delta}):=\mu_0^{(v)}(x)-\widehat{\bb}_{p,0}^{(v)}(x)'\bbeta_\infty(\widehat{\Delta})=
	\frac{\mu_0^{(p+1)}(x)}{(p+1-v)!}\hat{h}_x^{p+1-v}\mathscr{B}_{p+1-v}
	\Big(\frac{x-\hat{\tau}_x^{\mathtt{L}}}{\hat{h}_x}\Big)+o_\P(J^{-(p+1-v)}),
	\]
	where for each $m\in\mathbb{Z}_+$, $\binom{2m}{m}\mathscr{B}_m(\cdot)$ is the $m$th shifted Legendre polynomial on $[0,1]$, $\hat{\tau}_x^{\mathtt{L}}$ is the start of the (random) interval in $\widehat{\Delta}$ containing $x$ and $\hat{h}_x$ denotes its length. In addition,
	\begin{align*}
		&\quad\max_{1\leq j\leq J(p+1)} |\E_{\widehat{\Delta}}[\widehat{b}_{p,0,j}(x)r_\infty(x;\widehat{\Delta})]|\\
		&=\max_{1\leq j\leq J(p+1)}
		\Big|\int_{\mathcal{X}}\widehat{b}_{p,0,j}(x)r_\infty(x;\widehat{\Delta})f_X(x)dx\Big|\\
		&=\max_{1\leq j\leq J(p+1)}
		\Big|\int_{\hat{\tau}_x^{\mathtt{L}}}^{\hat{\tau}_x^{\mathtt{L}}+\hat{h}_x}\widehat{b}_{p,0,j}(x)r_\infty(x;\widehat{\Delta})f_X(\hat{\tau}_x^{\mathtt{L}})dx\Big|+o_\P(J^{-p-1-1/2})\\
		&=\max_{1\leq j\leq J(p+1)}
		\Big|f_X(\hat{\tau}_x^{\mathtt{L}})\frac{\mu_0^{(p+1)}(x)J^{-p-1}}{(p+1)!}
		\int_{\hat{\tau}_x^{\mathtt{L}}}^{\hat{\tau}_x^{\mathtt{L}}+\hat{h}_x}
		\widehat{b}_{p,0,j}(x)
		\mathscr{B}_{p+1}\Big(\frac{x-\hat{\tau}_x^{\mathtt{L}}}{\hat{h}_x}\Big)dx\Big|+o_\P(J^{-p-1-1/2})\\
		&=o_\P(J^{-p-1-1/2}),
	\end{align*}
	where the last line follows by change of variables and the orthogonality of Legendre polynomials. Thus, $r_\infty(x;\widehat{\Delta})$ is approximately orthogonal to the space spanned by $\widehat{\bb}_{p,0}(x)$. Immediately, we have
	\[
	\|\E_{\widehat{\Delta}}[\bb(x;\widehat{\Delta})r_\infty(x;\widehat{\Delta})]\|
	=o_\P(J^{-p-1}).
	\]
	Since $\E_{\widehat{\Delta}}[\widehat{\bb}_{p,0}(x)r_0(x;\widehat{\Delta})]=0$,
	\[
	\|\E_{\widehat{\Delta}}[\widehat{\bb}_{p,0}(x)(r_0(x;\widehat{\Delta})-r_\infty(x;\widehat{\Delta}))]\|
	=\|\E_{\widehat{\Delta}}[\widehat{\bb}_{p,0}(x)\widehat{\bb}_{p,0}(x)'
	(\bbeta_\infty(\widehat{\Delta})-\bbeta_0(\widehat{\Delta}))]\|
	=o_\P(J^{-p-1}).
	\]
	By Lemma \ref{SA lem: Gram, LS}, $\lambda_{\min}(\E_{\widehat{\Delta}}[\widehat{\bb}_{p,0}(x_i)\widehat{\bb}_{p,0}(x_i)]')\gtrsim_\P 1$, and thus $\|\bbeta_\infty(\widehat{\Delta})-\bbeta_0(\widehat{\Delta})\|=o_\P(J^{-p-1})$.
	Then,
	\begin{align*}
		&\int_{\mathcal{X}}\Big(\widehat{\bb}_{p,0}^{(v)}(x)'(\bbeta_0(\widehat{\Delta})-\bbeta_\infty(\widehat{\Delta}))\Big)^2 \omega(x)dx\\
		\leq&\, \lambda_{\max}\Big(\int_{\mathcal{X}}\widehat{\bb}_{p,0}^{(v)}(x)\widehat{\bb}_{p,0}^{(v)}(x)'\omega(x)dx\Big)\|\bbeta_0(\widehat{\Delta})-\bbeta_\infty(\widehat{\Delta})\|^2
		=o_\P(J^{-2p-2+2v}).
	\end{align*} 
	Therefore, we can represent the leading term in the integrated squared bias by $L_\infty$ approximation error:  $\int_{\mathcal{X}}\mathfrak{B}_2(x)^2\omega(x)dx=\int_{\mathcal{X}}(\mu_0^{(v)}(x)-\widehat{\bb}_{p,0}^{(v)}(x)'\bbeta_\infty(\widehat{\Delta}))^2\omega(x)dx+o_\P(J^{-2p-2+2v})$. Finally, using the results given in Lemma \ref{SA lem: quantile partition}, change of variables and the definition of Riemann integral, we conclude that
	\[
	\int_{\mathcal{X}}\Big(\E[\widehat{\mu}^{(v)}(x)|\bX,\bW]-\mu_0^{(v)}(x)\Big)^2\omega(x)dx= J^{-2(p+1-v)}\mathscr{B}(p,0,v)+o_\P(J^{-2p-2+2v})
	\]
	where 
	\[
	\mathscr{B}(p,0,v)=\frac{\int_{0}^{1}[\mathscr{B}_{p+1-v}(z)]^2 dz}
	{((p+1-v)!)^2}\int_{\mathcal{X}}\frac{[\mu_0^{(p+1)}(x)]^2}{f_X(x)^{2p+2-2v}}\omega(x)dx.
	\]
	Then the proof is complete.
\end{proof}

\subsection{Proof of Theorem \ref{SA thm: strong approximation, LS}}
\begin{proof}
	The proof is divided into several steps.
	
	\textbf{Step 1:}	
	Note that
	\begin{align*}
		&\sup_{x\in\mathcal{X}}\bigg|\frac{\widehat{\mu}^{(v)}(x)-\mu_0^{(v)}(x)}{\sqrt{\widehat{\Omega}(x)/n}}-\frac{\widehat{\mu}^{(v)}(x)-\mu_0^{(v)}(x)}{\sqrt{\Omega(x)/n}}\bigg|\\
		\leq& \sup_{x\in\mathcal{X}}\bigg|\frac{\widehat{\mu}^{(v)}(x)-\mu_0^{(v)}(x)}{\sqrt{\Omega(x)/n}}\bigg|
		\sup_{x\in\mathcal{X}}\bigg|\frac{\widehat{\Omega}(x)^{1/2}-\Omega(x)^{1/2}}{\widehat{\Omega}(x)^{1/2}}\bigg|\\
		\lesssim&_\P \Big(\sqrt{\log J}+\sqrt{n}J^{-p-1-1/2}\Big)\Big(J^{-p-1}+\sqrt{\frac{J\log J}{n^{1-\frac{2}{\nu}}}}\Big)
	\end{align*}
	where the last step uses Lemma \ref{SA lem: asymp variance, LS}, Corollary  \ref{SA coro: uniform convergence, LS} and Theorem \ref{SA thm: meat matrix, LS}.
	Then, in view of Lemmas \ref{SA lem: uniform approx rate}, \ref{SA lem: proj approx error, LS}, \ref{SA lem: parametric part} and Theorem \ref{SA thm: meat matrix, LS} and the rate restriction given in the lemma, we have
	\[
	\sup_{x\in\mathcal{X}}\bigg|\frac{\widehat{\mu}^{(v)}(x)-\mu_0^{(v)}(x)}
	{\sqrt{\widehat{\Omega}(x)/n}} -
	\frac{\widehat{\bb}_{p,s}^{(v)}(x)'\widehat{\bQ}^{-1}}{\sqrt{\Omega(x)}}\G_n[\widehat{\bb}_{p,s}(x_i)\epsilon_i]\bigg|=o_\P(a_n^{-1}).
	\]

	\textbf{Step 2:}
	Let us write 
	$\mathscr{K}(x,x_i)=\Omega(x)^{-1/2}\widehat{\bb}_{p,s}^{(v)}(x)'\widehat{\bQ}^{-1}\bb_{p,s}(x_i)$. 
	Now we
	rearrange $\{x_i\}_{i=1}^n$ as a sequence of order statistics $\{x_{(i)}\}_{i=1}^n$, i.e., $x_{(1)}\leq\cdots\leq x_{(n)}$.
	Accordingly, $\{\epsilon_i\}_{i=1}^n$ and $\{\sigma^2(x_i)\}_{i=1}^n$ are ordered as concomitants $\{\epsilon_{[i]}\}_{i=1}^n$ and $\{\sigma^2_{[i]}\}_{i=1}^n$ where $\sigma^2_{[i]}=\sigma^2(x_{(i)})$. Clearly, conditional on $\bX$, $\{\epsilon_{[i]}\}_{i=1}^n$ is still an independent mean zero sequence. Then by Assumptions \ref{SA Assumption DGP}, \ref{SA Assumption LS} and the result of  \cite{Sakhanenko_1991_SAM}, there exists a sequence of i.i.d. standard normal random variables $\{\zeta_{[i]}\}_{i=1}^n$ such that
	\[
	\max_{1\leq \ell\leq n}|S_{\ell}|:=
	\max_{1\leq \ell \leq n}\Big|\sum_{i=1}^\ell\epsilon_{[i]} - \sum_{i=1}^\ell\sigma_{[i]}\zeta_{[i]}\Big|\lesssim_\P n^{\frac{1}{\nu}}.
	\]
	Then, using summation by parts,
	\begin{align*}
		&\,\sup_{x\in\mathcal{X}}\left|\sum_{i=1}^{n}\mathscr{K}(x, x_{(i)})(\epsilon_{[i]}-\sigma_{[i]}\zeta_{[i]})\right|\\
		=&\,\sup_{x\in\mathcal{X}}\left|\mathscr{K}(x, x_{(n)})S_{n}
		-\sum_{i=1}^{n-1}S_{i}\left(\mathscr{K}(x,x_{(i+1)})
		-\mathscr{K}(x,x_{(i)})\right)\right|\\
		\leq&\,\sup_{x\in\mathcal{X}}\max_{1\leq i\leq n}|\mathscr{K}(x,x_i)||S_{n}|
		+\sup_{x\in\mathcal{X}}\left|
		\frac{\widehat{\bb}_{p,s}^{(v)}(x)'\widehat{\bQ}^{-1}}{\sqrt{\Omega(x)}}
		\sum_{i=1}^{n-1}S_{i}\Big(\widehat{\bb}_{p,s}(x_{(i+1)}) -
		\widehat{\bb}_{p,s}(x_{(i)})\Big)\right|\\
		\leq&\,\sup_{x\in\mathcal{X}}\max_{1\leq i\leq n}|\mathscr{K}(x,x_i)||S_{n}|
		+\sup_{x\in\mathcal{X}}
		\Bigg\|\frac{\widehat{\bQ}^{-1}\widehat{\bb}_{p,s}^{(v)}(x)}{\sqrt{\Omega(x)}}\Bigg\|_1
		\left\|\sum_{i=1}^{n-1}S_{i}\Big(\widehat{\bb}_{p,s}(x_{(i+1)})-
		\widehat{\bb}_{p,s}(x_{(i)})\Big)\right\|_\infty.
	\end{align*}
	
	By Lemmas \ref{SA lem: local basis}, \ref{SA lem: Gram, LS} and \ref{SA lem: asymp variance, LS},
	$\sup_{x\in\mathcal{X}}\sup_{x_i\in\mathcal{X}}|\mathscr{K}(x,x_i)|\lesssim_\P \sqrt{J}$, and $$\sup_{x\in\mathcal{X}}\Bigg\|\frac{\widehat{\bQ}^{-1}\widehat{\bb}_{p,s}^{(v)}(x)}{\sqrt{\Omega(x)}}\Bigg\|_1\lesssim_\P 1.$$
	Then, notice that
	\begin{equation*}
		\max_{1\leq l \leq K_{p,s}}
		\Big|\sum_{i=1}^{n-1}\Big(\widehat{b}_{p,s,l}(x_{(i+1)})-\widehat{b}_{p,s,l}(x_{(i)})\Big)S_{l}\Big|
		\leq \max_{1\leq l \leq K_{p,s}}
		\sum_{i=1}^{n-1}\Big|\widehat{b}_{p,s,l}(x_{(i+1)})-\widehat{b}_{p,s,l}(x_{(i)})\Big|
		\max_{1\leq \ell \leq n}\Big|S_{\ell}\Big|.
	\end{equation*}
	By construction of the ordering, $\max_{1\leq l \leq K_{p,s}}
	\sum_{i=1}^{n-1}\Big|\widehat{b}_{p,s,l}(x_{(i+1)})-\widehat{b}_{p,s,l}(x_{(i)})\Big|\lesssim \sqrt{J}$. Under the rate restriction in the theorem, this suffices to show that for any $\xi>0$, $$\P\Big(\sup_{x\in\mathcal{X}}\,|\G_n[\mathscr{K}(x,x_i)(\epsilon_i-\sigma_i\zeta_i)]|>\xi a_n^{-1}\Big|\bX\Big)=o_\P(1),$$ 
	where we recover the original ordering.
	Since $\G_n[\widehat{\bb}(x_i)\zeta_i\sigma_{i}]=_{d|\bX}\mathbf{N}(0, \bar{\bSigma})$ ($=_{d|\bX}$ denotes ``equal in distribution conditional on $\bX$''), the above steps construct the following approximating process:
	\[ \bar{Z}_p(x):=\frac{\widehat{\bb}_{p,s}^{(v)}(x)'\widehat{\bQ}^{-1}}{\sqrt{\Omega(x)}}\bar{\bSigma}^{1/2}\bN_{K_{p,s}}.
	\]
	Then, it remains to show $\widehat{\bQ}^{-1}$ and $\bar{\bSigma}$ can be replaced by their population analogues without affecting the approximation, which is verified in the next step.	
	
	\textbf{Step 3:}	
	Note that
	\[
	\begin{split}
		\sup_{x\in\mathcal{X}}|\bar{Z}_p(x)-Z_p(x)|
		&\leq
		\sup_{x\in\mathcal{X}}\Big|\frac{\widehat{\bb}^{(v)}_{p,s}(x)'(\widehat{\bQ}^{-1}-\bQ_0^{-1})}{\sqrt{\Omega(x)}}\bar{\bSigma}^{1/2}\bN_{K_{p,s}}\Big|\\
		& \qquad + \sup_{x\in\mathcal{X}}\Big|\frac{\widehat{\bb}^{(v)}_{p,s}(x)'\bQ_0^{-1}}{\sqrt{\Omega(x)}}\Big(\bar{\bSigma}^{1/2}-\bSigma_0^{1/2}\Big)\bN_{K_{p,s}}\Big|\\
		& \qquad + \sup_{x\in\mathcal{X}}\Big|\frac{\widehat{\bb}_{p,0}^{(v)}(x)'
			(\widehat{\bT}_s-\bT_s)\bQ_0^{-1}}{\sqrt{\Omega(x)}}\bSigma_0^{1/2}\bN_{K_{p,s}}\Big|,
	\end{split}
	\]
	where each term on the right-hand side is a mean-zero Gaussian process conditional on $\bX$.
	By Lemmas \ref{SA lem: spline transform} and \ref{SA lem: Gram, LS}, $\|\widehat{\bQ}^{-1}-\bQ_0^{-1}\|\lesssim_\P\sqrt{J\log J/n}$ and $\|\widehat{\bT}_s-\bT_s\|\lesssim_\P\sqrt{J\log J/n}$. Also, using the argument in the proof of Lemma \ref{SA lem: Gram, LS} and Theorem X.3.8 of \citet*{Bhatia_2013_book}, $\|\bar{\bSigma}^{1/2}-\bSigma_0^{1/2}\|\lesssim_\P\sqrt{J\log J/n}$.
	By Gaussian Maximal Inequality \citep*[see, e.g.,][Corollary 2.2.8]{vandevarrt-Wellner_1996_book},
	\[\E\Big[\sup_{x\in\mathcal{X}}|\bar{Z}_p(x)-Z_p(x)|\Big|\bX \Big]\lesssim_\P \sqrt{\log J}
	\Big(\|\bar{\bSigma}^{1/2}-\bSigma_0^{1/2}\|+\|\widehat{\bQ}^{-1}-\bQ_0^{-1}\|+\|\widehat{\bT}_s-\bT_s\|\Big)=o_\P(a_n^{-1}),
	\]
	where the last line follows from the imposed rate restriction. 
	
	As a reminder, if we drop the third term on the right-hand side, we obtain the same strong approximation result except that the approximating process is
	\[
	\frac{\widehat{\bb}_{p,s}^{(v)}(\cdot)'\bQ_0^{-1}\bSigma_0^{1/2}}{\sqrt{\Omega(x)}}\bN_{K_{p,s}}.
	\]

 \textbf{Step 4:} The above steps have shown the desired result for $v>0$ already.
	For $v=0$,
	\[
	T_p(x)=\frac{\widehat{\Upsilon}(x, \widehat{\bw})-\Upsilon_0(x, \bw)}{\sqrt{\widehat{\Omega}(x)/n}}
	=\frac{\widehat{\mu}(x)-\mu_0(x)}{\sqrt{\widehat{\Omega}(x)/n}}+
	\frac{\widehat{\bw}'\widehat{\bgamma}-
		\bw'\bgamma_0}{\sqrt{\widehat{\Omega}(x)/n}},
	\]
	where
	\[
	\frac{\widehat{\bw}'\widehat{\bgamma}-\bw'\bgamma_0}{\sqrt{\widehat{\Omega}(x)/n}}=
	\frac{(\widehat{\bw}-\bw)'\widehat{\bgamma}}{\sqrt{\widehat{\Omega}(x)/n}}
	+\frac{\bw'(\widehat{\bgamma}-\bgamma_0)}{\sqrt{\widehat{\Omega}(x)/n}}=o_\P(a_n^{-1})
	\] 
	by Lemma \ref{SA lem: parametric part}, Theorem \ref{SA thm: meat matrix, LS} and the condition $\|\widehat{\bw}-\bw\|=o_\P(a_n^{-1}\sqrt{J/n})$. Therefore, the desired strong approximation for $\widehat{\Upsilon}(x,\widehat{\bw})$ follows from the previous steps. Then, the proof is complete. 	
\end{proof}

\subsection{Proof of Theorem \ref{SA thm: plug-in approx, LS}}
\begin{proof}
	This conclusion follows from Lemmas \ref{SA lem: local basis} and \ref{SA lem: Gram, LS}, Theorem \ref{SA thm: meat matrix, LS} and Gaussian Maximal Inequality as applied in Step 3 in the proof of Theorem \ref{SA thm: strong approximation, LS}.
\end{proof}


\subsection{Proof of Theorem \ref{SA thm: sup approx}}
\begin{proof}
	We first show that
	\[
	\sup_{u\in\mathbb{R}}\Big|\P\Big(\sup_{x\in\mathcal{X}}
	|T_p(x)|\leq u\Big) -
	\P\Big(\sup_{x\in\mathcal{X}}|Z_p(x)|\leq u\Big)\Big|=o(1).
	\]
	By Theorem \ref{SA thm: strong approximation, LS}, there exists a sequence of constants $\xi_n$ such that $\xi_n=o(1)$ and
	\[ \P\Big(\Big|\sup_{x\in\mathcal{X}}|T_p(x)|-\sup_{x\in\mathcal{X}}|Z_p(x)|\Big|>\xi_n/a_n\Big)=o(1).
	\]
	Then,
	\begin{align*}
	\P\Big(\sup_{x\in\mathcal{X}}|T_p(x)|\leq u\Big)&\leq
	\P\Big(\Big\{\sup_{x\in\mathcal{X}}|T_p(x)|\leq u\Big\}\cap\Big\{\Big|\sup_{x\in\mathcal{X}}|T_p(x)|-
	\sup_{x\in\mathcal{X}}|Z_p(x)|\Big|\leq\xi_n/a_n\Big\}\Big)+o(1)\\
	&\leq \P\Big(\sup_{x\in\mathcal{X}}|Z_p(x)|\leq u+\xi_n/a_n\Big)
	+o(1)\\
	&\leq \P\Big(\sup_{x\in\mathcal{X}}|Z_p(x)|\leq u\Big)+
	\sup_{u\in\mathbb{R}}\E\Big[\P\Big(\Big|\sup_{x\in\mathcal{X}}|Z_p(x)|-u\Big|\leq \xi_n/a_n\Big|\bX\Big)\Big]\\
	&\leq \P\Big(\sup_{x\in\mathcal{X}}|Z_p(x)|\leq u\Big)+
	\E\Big[\sup_{u\in\mathbb{R}}\P\Big(\Big|\sup_{x\in\mathcal{X}}
	|Z_p(x)|-u\Big|\leq \xi_n/a_n\Big|\bX\Big)\Big]
	+o(1).
	\end{align*}
	Now, apply the Anti-Concentration Inequality conditional on $\bX$ (see \citealp*{Chernozhukov-Chetverikov-Kato_2014b_AoS}) to the second term:
	\begin{align*}
	\sup_{u\in\mathbb{R}}\;
	\P\Big(\Big|\sup_{x\in\mathcal{X}}|Z_p(x)|-u\Big|\leq \xi_n/a_n\Big|\bX\Big)&\leq 4\xi_na_n^{-1}\E\Big[\sup_{x\in\mathcal{X}}|Z_p(x)|\Big|\bX\Big]+o(1)\\
	&\lesssim_\P \xi_n a_n^{-1}\sqrt{\log J}+o(1)\rightarrow 0
	\end{align*}
	where the last step uses Gaussian Maximal Inequality
	\citep*[see][Corollary 2.2.8]{vandevarrt-Wellner_1996_book}.
	By Dominated Convergence Theorem,
	\[\E\Big[
	\sup_{u\in\mathbb{R}}\P\Big(\Big|\sup_{x\in\mathcal{X}}|Z_p(x)|-u\Big|\leq \xi_n/a_n\Big|\bX\Big)\Big]=o(1).
	\]
	The other side of the inequality follows similarly.
	
	By similar argument, using Theorem \ref{SA thm: plug-in approx, LS}, we have
	\[
	\sup_{u\in\mathbb{R}}\Big|\P\Big(\sup_{x\in\mathcal{X}}|\widehat{Z}_p(x)|\leq u\Big|\bD\Big) -
	\P\Big(\sup_{x\in\mathcal{X}}|Z_p(x)|\leq u\Big|\bX\Big)\Big|=o_\P(1).
	\]
	Then it remains to show that
	\begin{equation} \label{eq: sup ks distance 1}
	\sup_{u\in\mathbb{R}}\Big|\P\Big(\sup_{x\in\mathcal{X}}|Z_p(x)|\leq u\Big)-
	\P\Big(\sup_{x\in\mathcal{X}}|Z_p(x)|\leq u|\bX\Big)\Big|=o_\P(1).
	\end{equation}
	Now, we can write
	\[ Z_p(x)=\frac{\widehat{\bb}_{p,0}^{(v)}(x)'}{\sqrt{\widehat{\bb}_{p,0}^{(v)}(x)'\bV_0\widehat{\bb}_{p,0}^{(v)}(x)}}\breve{\bN}_{K_{p,0}}
	\]
	where $\bV_0=\bT_s'\bQ_0^{-1}\bSigma_0\bQ_0^{-1}\bT_s$ and $\breve{\bN}_{K_{p,0}}:=\bT_s'\bQ_0^{-1}\bSigma_0^{1/2}\bN_{K_{p,s}}$ is a $K_{p,0}$-dimensional normal random vector. Importantly, by this construction, $\breve{\bN}_{K_{p,0}}$ and $\bV_0$ do not depend on $\widehat{\Delta}$ and $x$, and they are only determined by the deterministic partition $\Delta_0$.
	
	Now, first consider $v=0$. For any two partitions $\Delta_1, \Delta_2\in\Pi$, for any $x\in\mathcal{X}$, there exists $\check{x}\in\mathcal{X}$ such that
	\[\bb_{p,0}^{(0)}(x;\Delta_1)=\bb_{p,0}^{(0)}(\check{x};\Delta_2),\]
	and vice versa.
	Therefore, the following two events are equivalent: $\{\omega:\sup_{x\in\mathcal{X}}|Z_p(x;\Delta_1)|\leq u\}=
	\{\omega:\sup_{x\in\mathcal{X}}|Z_p(x;\Delta_2)|\leq u\}$ for any $u$.
	Thus, 
	$$\E\Big[\P\Big(\sup_{x\in\mathcal{X}}|Z_p(x)|\leq u
	\Big|\bX\Big)\Big]=\P\Big(\sup_{x\in\mathcal{X}}|Z_p(x)|\leq u
	\Big|\bX\Big)+o_\P(1).$$
	Then for $v=0$, the desired result follows.
	
	For $v>0$, simply notice that 
	$\widehat{\bb}_{p,0}^{(v)}(x)=\widehat{\mathfrak{T}}_v\widehat{\bb}_{p,0}(x)$ for some transformation matrix $\widehat{\mathfrak{T}}_v$. Clearly, $\widehat{\mathfrak{T}}_v$ takes a similar structure as $\widehat{\bT}_s$: each row and each column only have a finite number of nonzeros. Each nonzero element is simply $\hat{h}_j^{-v}$ up to some constants. By Lemma \ref{SA lem: quantile partition}, it can be shown that $\|\widehat{\mathfrak{T}}_v-\mathfrak{T}_v\|\lesssim J^v\sqrt{J\log J/n}$ where $\mathfrak{T}_v$ is the population analogue ($\hat{h}_j$ replaced by $h_j$). Repeating the argument given in, e.g., the proof of Theorems \ref{SA thm: strong approximation, LS} and \ref{SA thm: plug-in approx, LS}, we can replace $\widehat{\mathfrak{T}}_v$ in $Z_p(x)$ by $\mathfrak{T}_v$ without affecting the approximation rate. Then the desired result follows by repeating the argument given for $v=0$ above. 
\end{proof}

\subsection{Proof of Theorem \ref{SA thm: robust CB}}
\begin{proof}
	 Let $\xi_{1,n}=o(1)$, $\xi_{2,n}=o(1)$ and $\xi_{3,n}=o(1)$. Then,
	\begin{align*}
	\P\left[\sup_{x\in\mathcal{X}}|T_{p}(x)|\leq \cval
	\right]&\leq
	\P\left[\sup_{x\in\mathcal{X}}|Z_{p}(x)|\leq \cval+\xi_{1,n}/a_n
	\right]+o(1)\\
	&\leq 
	\P\left[\sup_{x\in\mathcal{X}}|Z_{p}(x)|\leq c^0(1-\alpha+\xi_{3,n})+(\xi_{1,n}+\xi_{2,n})/a_n
	\right]+o(1)\\
	&\leq \P\left[\sup_{x\in\mathcal{X}}|Z_{p}(x)|\leq c^0(1-\alpha+\xi_{3,n})\right]+o(1)\rightarrow 1-\alpha,
	\end{align*}
	where $c^0(1-\alpha+\xi_{3,n})$ denotes the $(1-\alpha+\xi_{3,n})$-quantile of $\sup_{x\in\mathcal{X}}|Z_{p}(x)|$ (given the partition), the first inequality holds by Theorem \ref{SA thm: strong approximation, LS}, the second by Lemma A.1 of \cite*{Belloni-Chernozhukov-Chetverikov-Kato_2015_JoE}, and the third by Anti-Concentration Inequality in \cite*{Chernozhukov-Chetverikov-Kato_2014b_AoS}. The other side of the bound follows similarly.
\end{proof}

\subsection{Proof of Theorem \ref{SA thm: testing specification}}
\begin{proof}
	Throughout this proof, we let $\xi_{1,n}=o(1)$, $\xi_{2,n}=o(1)$ and
	$\xi_{3,n}=o(1)$ be sequences of vanishing constants. Moreover, let $A_n$ be a sequence of diverging constants such that $\sqrt{\log J}A_n\lesssim \sqrt{\frac{n}{J^{1+2v}}}$. 	
	Note that under $\dot{\mathsf{H}}_0$,
	\[ \sup_{x\in\mathcal{X}}
	|\dot{T}_p(x)|\leq \sup_{x\in\mathcal{X}}
	\bigg|\frac{\widehat{\Upsilon}^{(v)}(x,\widehat{\bw})-
		\Upsilon_0^{(v)}(x,\bw)}
	{\sqrt{\widehat\Omega(x)/n}}\bigg|+
	\sup_{x\in\mathcal{X}}\bigg|
	\frac{\Upsilon_0^{(v)}(x,\bw)-
		M^{(v)}(x,\widehat{\bw};\widetilde{\btheta},\widetilde{\bgamma})}
	{\sqrt{\widehat{\Omega}(x)/n}}\bigg|.
	\]
	Therefore,
	\begin{align*}
	\P\Big[\sup_{x\in\mathcal{X}}|\dot{T}_p(x)|>\cval\Big]&
	\leq
	\P\bigg[\sup_{x\in\mathcal{X}}|T_p(x)|>\cval-
	\sup_{x\in\mathcal{X}}\bigg|\frac{\Upsilon_0^{(v)}(x,\bw)-
 M^{(v)}(x,\widehat{\bw};\widetilde{\btheta},\widetilde{\bgamma})}{\sqrt{\widehat{\Omega}(x)/n}}\bigg|\bigg] \\
	&\leq \P\bigg[\sup_{x\in\mathcal{X}}|Z_p(x)|>\cval-\xi_{1,n}/a_n - 
    \sup_{x\in\mathcal{X}}\bigg|\frac{\Upsilon_0^{(v)}(x,\bw)-M^{(v)}(x,\widehat{\bw};\widetilde{\btheta},\widetilde{\bgamma})}{\sqrt{\widehat{\Omega}(x)/n}}\bigg|\bigg] + o(1)\\
	&\leq \P\bigg[\sup_{x\in\mathcal{X}}|Z_p(x)|>c^0(1-\alpha-\xi_{3,n})-(\xi_{1,n}+\xi_{2,n})/a_n - \\
	&\qquad \sup_{x\in\mathcal{X}}\bigg|\frac{\Upsilon_0^{(v)}(x,\bw)-M^{(v)}(x,\widehat{\bw};\widetilde{\btheta},\widetilde{\bgamma})}{\sqrt{\widehat{\Omega}(x)/n}}\bigg|\bigg] + o(1)\\
	&\leq \P\Big[\sup_{x\in\mathcal{X}}|Z_p(x)|>c^0(1-\alpha-\xi_{3,n})\Big]+o(1)\\
	&=\alpha+o(1)
	\end{align*}
	where $c^0(1-\alpha-\xi_{3,n})$ denotes the $(1-\alpha-\xi_{3,n})$-quantile of $\sup_{x\in\mathcal{X}}|Z_p(x)|$ (given the partition), the second inequality holds by Theorem \ref{SA thm: strong approximation, LS}, the third by Lemma A.1 of \cite*{Belloni-Chernozhukov-Chetverikov-Kato_2015_JoE}, the fourth by the fact that $\sup_{x\in\mathcal{X}}\big|\frac{\Upsilon_0^{(v)}(x,\bw)-M^{(v)}(x,\widehat{\bw};\widetilde{\btheta},\widetilde{\bgamma})}{\sqrt{\widehat{\Omega}(x)/n}}\big|=o_\P(\frac{1}{\sqrt{\log J}})$ and Anti-Concentration Inequality in \cite*{Chernozhukov-Chetverikov-Kato_2014b_AoS}. The other side of the bound follows similarly.
	
	On the other hand, under $\dot{\mathsf{H}}_\text{A}$,
	\begin{align*}
	&\P\Big[\sup_{x\in\mathcal{X}}|\dot{T}_p(x)|>\cval\Big] \\
	=\,&\P\Big[\sup_{x\in\mathcal{X}}\Big|T_p(x)+
	\frac{\Upsilon_0^{(v)}(x,\bw)-M^{(v)}(x,\bw;\bar{\btheta},\bar{\bgamma})}{\sqrt{\widehat{\Omega}(x)/n}}+
	\frac{M^{(v)}(x,\bw;\bar{\btheta},\bar{\bgamma})-M^{(v)}(x,\widehat{\bw};\widetilde{\btheta},\widetilde{\bgamma})}{\sqrt{\widehat{\Omega}(x)/n}}\Big|>\cval\Big]\\
	\geq\,& \P\bigg[\sup_{x\in\mathcal{X}}|T_p(x)|< \sup_{x\in\mathcal{X}}\bigg|\frac{\Upsilon_0^{(v)}(x,\bw)-M^{(v)}(x,\bw;\bar{\btheta},\bar{\bgamma})}
	{\sqrt{\widehat{\Omega}(x)/n}} +
	\frac{M^{(v)}(x,\bw;\bar{\btheta},\bar{\bgamma})-M^{(v)}(x,\widehat{\bw};\widetilde{\btheta},\widetilde{\bgamma})}{\sqrt{\widehat{\Omega}(x)/n}}\bigg| 
	-\cval\bigg] \\
	\geq\,& \P\Big[\sup_{x\in\mathcal{X}} |Z_p(x)|\leq \sqrt{\log J}A_n-\xi_{1,n}/a_n\Big]-o(1)\\
	\geq\,& 1-o(1).
	\end{align*}
	where the fourth line holds by Lemma \ref{SA lem: asymp variance, LS}, Theorem \ref{SA thm: meat matrix, LS}, Theorem \ref{SA thm: strong approximation, LS}, the condition that $J^v\sqrt{J\log J/n}=o(1)$ and the definition of $A_n$, and the last by the Talagrand-Samorodnitsky Concentration Inequality \cite*[Proposition A.2.7]{vandevarrt-Wellner_1996_book}.
\end{proof}

\subsection{Proof of Theorem \ref{SA thm: testing shape restriction}}
\begin{proof}
	The definitions of $A_n$, $\xi_{1,n}, \xi_{2,n}$ and $\xi_{3,n}$ are the same as in the proof of Theorem \ref{SA thm: testing specification}. 
	Note that under $\ddot{\mathsf{H}}_0$,
	\[
	\sup_{x\in\mathcal{X}}\ddot{T}_p(x)
	\leq\sup_{x\in\mathcal{X}}T_p(x)+
	\sup_{x\in\mathcal{X}}\frac{|M^{(v)}(x,\bw;\bar{\btheta},\bar{\bgamma})-
		M^{(v)}(x,\widehat{\bw};\widetilde{\btheta},\widetilde{\bgamma})|}
	{\sqrt{\widehat{\Omega}(x)/n}}.
	\]
	Then,
	\begin{align*}
	\P\Big[ \sup_{x\in\mathcal{X}} \ddot{T}_p(x)> \cval\Big]
	&\leq \P\bigg[ \sup_{x\in\mathcal{X}} T_p(x)> \cval
	- \sup_{x\in\mathcal{X}}\frac{|M^{(v)}(x,\bw;
		\bar{\btheta},\bar{\bgamma})-
		M^{(v)}(x,\widehat{\bw};\widetilde{\btheta},\widetilde{\bgamma})|}
	{\sqrt{\widehat{\Omega}(x)/n}}
	\bigg] \\
	&\leq \P\Big[ \sup_{x\in\mathcal{X}} Z_p(x)> \cval-\xi_{1,n}/a_n\Big] + o(1)\\
	&\leq \P\Big[ \sup_{x\in\mathcal{X}} Z_p(x)> c^0(1-\alpha-\xi_{3,n})-(\xi_{1,n}+\xi_{2,n})/a_n\Big] + o(1)\\ 
	&\leq \P\Big[ \sup_{x\in\mathcal{X}} Z_p(x)> c^0(1-\alpha-\xi_{3,n})\Big] + o(1)\\
	&=\alpha+o(1)
	\end{align*}
	where $c^0(1-\alpha-\xi_{3,n})$ denotes the $(1-\alpha-\xi_{3,n})$-quantile of $\sup_{x\in\mathcal{X}}Z_p(x)$ (given the partition), the second line holds by Theorem \ref{SA thm: strong approximation, LS}, the third by Lemma A.1 of \cite*{Belloni-Chernozhukov-Chetverikov-Kato_2015_JoE}, the fourth by Anti-Concentration Inequality in \cite*{Chernozhukov-Chetverikov-Kato_2014b_AoS}.
	
	On the other hand, under $\ddot{\mathsf{H}}_\text{A}$,
	\begin{align*}
	\P\Big[ \sup_{x\in\mathcal{X}} \ddot{T}_p(x)> \cval\Big]
	&=\P\bigg[ \sup_{x\in\mathcal{X}}\Big( T_p(x)+
	\frac{\Upsilon_0^{(v)}(x,\bw)-
		M^{(v)}(x,\widehat{\bw};\widetilde{\btheta},\widetilde{\bgamma})}
	{\sqrt{\widehat{\Omega}(x)/n}}-\cval\Big)>0\bigg]\\
	&\geq
	\P\Big[ \sup_{x\in\mathcal{X}} |T_p(x)|< \sup_{x\in\mathcal{X}}\frac{
		\Upsilon_0^{(v)}(x,\bw)-
		M^{(v)}(x,\widehat{\bw};\widetilde{\btheta},\widetilde{\bgamma})}
	{\sqrt{\widehat{\Omega}(x)/n}}-\cval, \\
	&\hspace{2.5em}\sup_{x\in\mathcal{X}}\frac{\Upsilon_0^{(v)}(x,\bw)-
		M^{(v)}(x,\widehat{\bw};\widetilde{\btheta},\widetilde{\bgamma})}
	{\sqrt{\widehat{\Omega}(x)/n}}>\cval\Big]\\
	&\geq \P\Big[ \sup_{x\in\mathcal{X}} |T_p(x)|< \sup_{x\in\mathcal{X}}\frac{\Upsilon_0^{(v)}(x,\bw)-
		M^{(v)}(x,\widehat{\bw};\widetilde{\btheta},\widetilde{\bgamma})}{\sqrt{\widehat{\Omega}(x)/n}}
	-\cval\Big]-o(1)\\
	&\geq \P\Big[ \sup_{x\in\mathcal{X}} |T_p(x)|< \sqrt{\log J}A_n\Big]-o(1)\\
	&\geq \P\Big[ \sup_{x\in\mathcal{X}} |Z_p(x)|<\sqrt{\log J}A_n-\xi_{1,n}/a_n\Big] - o(1) \\
	&\geq 1 - o(1) 
	\end{align*}
	where the fourth line holds by Lemma \ref{SA lem: asymp variance, LS},  Theorem \ref{SA thm: meat matrix, LS}, Lemma A.1 of \cite*{Belloni-Chernozhukov-Chetverikov-Kato_2015_JoE}, the assumptions that $J^v\sqrt{J\log J/n}=o(1)$ and $\sup_{x\in\mathcal{X}}\\|M^{(v)}(x,\widehat{\bw};\widetilde{\btheta},\widetilde{\bgamma})-
	M^{(v)}(x,\bw;\bar{\btheta},\bar{\bgamma})|=o_\P(1)$, the fifth by definition of $A_n$, and the sixth by Theorem \ref{SA thm: strong approximation, LS}, and the last by Proposition A.2.7 in \cite*{vandevarrt-Wellner_1996_book}.
	
\end{proof}


\section*{References} 
\begingroup
\renewcommand{\section}[2]{}	
\bibliography{CCFF_2024_AER--bib}
\bibliographystyle{aer}
\endgroup